\documentclass[a4paper,11pt]{article}
\pdfoutput=1 \usepackage{amssymb}
\usepackage{amsmath}
\usepackage{braket}
\usepackage{mathtools}
\usepackage[usenames,dvipsnames,svgnames,table]{xcolor}
\usepackage [utf8] {inputenc}
\usepackage{color}
\usepackage{cite}
\usepackage{subfigure}
\usepackage{lmodern}
\usepackage{footnote}
\usepackage{glossaries-extra}
\setabbreviationstyle[acronym]{long-short}
\glssetcategoryattribute{acronym}{nohyperfirst}{true}

\usepackage{slashed}
\usepackage[pdftex] {graphicx}
\usepackage{multirow}
\usepackage{jheppub} 
\usepackage{here}
\usepackage{hyperref}
\usepackage{verbatim}
\usepackage{simplewick}
\usepackage[justification=centering, singlelinecheck=false]{caption}

\newcommand{\cK}[0]{\mathcal K}

\newcommand{\cM}[0]{\mathcal M}
\newcommand{\cO}[0]{\mathcal O}

\newcommand{\cS}[0]{\mathcal S}

\newcommand{\wt}[0]{\widetilde}
\newcommand{\mc}[0]{\mathcal}

\newcommand{\df}[0]{\mathrm{df}}
\newcommand{\iso}[0]{{\rm iso}}

\newcommand{\Kiso}[0]{{\cK_{\df,3}^{\iso}}}

\newcommand{\Kisoone}[0]{{\cK_{\df,3}^{\iso,1}}}
\newcommand{\Kdf}[0]{{\cK_{\df,3}}}

\newcommand{\KdfX}[1]{{\cK_{\df,3}^{#1}}}

\newcommand{\PV}[0]{{\mathrm{PV}}}

\newcommand{\K}[0]{\mathcal K}

\newcommand{\bm}[0]{\boldsymbol}

\newcommand{\Kisotwo}[0]{{\cK^{\iso,2}_{\df,3}}}
\newcommand{\KA}[0]{{\cK^{(2,A)}_{\df,3}}}
\newcommand{\KB}[0]{{\cK^{(2,B)}_{\df,3}}}
\newcommand{\KAS}[0]{{\cK^{\text{AS}}_{\df,3}}}
\newcommand{\KASone}[0]{{\cK^{\text{AS,1}}_{\df,3}}}
\newcommand{\KAStwo}[0]{{\cK^{\text{AS,2}}_{\df,3}}}

\newcommand{\LtoK}[0]{Hansen:2014eka}
\newcommand{\KtoM}[0]{Hansen:2015zga}
\newcommand{\KSS}[0]{Kim:2005gf}

\newcommand{\BHSQC}[0]{Briceno:2017tce}
\newcommand{\BHSnum}[0]{Briceno:2018mlh}
\newcommand{\BHSK}[0]{Briceno:2018aml}
\newcommand{\HSQCa}[0]{Hansen:2014eka}
\newcommand{\HSQCb}[0]{Hansen:2015zga}
\newcommand{\dwave}[0]{Blanton:2019igq}

\newcommand{\Akakia}[0]{Hammer:2017uqm}
\newcommand{\Akakib}[0]{Hammer:2017kms}

\newcommand{\MDpi}[0]{Mai:2018djl}

\newcommand{\HSrev}[0]{Hansen:2019nir}

\newcommand{\HH}[0]{Horz:2019rrn}

\newcommand{\largera}[0]{Romero-Lopez:2019qrt}
\newcommand{\MD}[0]{Mai:2017bge}
\newcommand{\Akakinum}[0]{Doring:2018xxx}
\newcommand{\RaulSpin}[0]{Briceno:2015csa}

\newacronym{CMF}{CMF}{center-of-momentum frame}
\newcommand{\CMF}[0]{\gls{CMF}}

\preprint{CERN-TH-2020-045}

\title{Generalizing the relativistic quantization condition to include all three-pion isospin channels}
\author[1]{Maxwell T. Hansen}
\affiliation[1]{Theoretical Physics Department, CERN, 1211 Geneva 23, Switzerland}
\author[2]{, Fernando Romero-L\'opez}
\affiliation[2]{IFIC, CSIC-Universitat de Val\`encia, 46980 Paterna, Spain}
\author[3]{, and Stephen R. Sharpe}
\affiliation[3]{Physics Department, University of Washington, Seattle, WA 98195-1560, USA}

\emailAdd{maxwell.hansen@cern.ch}
\emailAdd{fernando.romero@uv.es}
\emailAdd{srsharpe@uw.edu}

\abstract{
We present a generalization of the relativistic, finite-volume, three-particle quantization condition for non-identical pions in isosymmetric QCD.
The resulting formalism allows one to use discrete finite-volume energies, determined using lattice QCD, to constrain scattering amplitudes for all possible values of two- and three-pion isospin. As for the case of identical pions considered previously, the result splits into two steps: The first defines a non-perturbative function with roots equal to the allowed energies, $E_n(L)$, 
in a given cubic volume with side-length $L$. 
This function depends on an intermediate three-body quantity, denoted $\Kdf$, 
which can thus be constrained from lattice QCD input. 
The second step is a set of integral equations relating $\Kdf$ to the physical scattering amplitude, $\mathcal M_3$. 
Both of the key relations, $E_n(L) \leftrightarrow \Kdf$ and $\Kdf \leftrightarrow \mathcal M_3$, are shown to be block-diagonal in the basis of definite three-pion isospin, $I_{\pi \pi \pi}$, 
so that one in fact recovers four independent relations, 
corresponding to $I_{\pi \pi \pi}=0,1,2,3$.
We also provide the generalized threshold expansion of $\Kdf$ for all channels, as well as parameterizations for all 
three-pion resonances present for $I_{\pi\pi\pi}=0$ and $I_{\pi\pi\pi}=1$.
As an example of the utility of the generalized formalism,
 we present a toy implementation of the quantization condition for $I_{\pi\pi\pi}=0$, 
 focusing on the quantum numbers of the $\omega$ and $h_1$ resonances.
}

\allowdisplaybreaks
\begin{document}

\maketitle
\flushbottom
\clearpage

\section{Introduction}

The computation of scattering amplitudes using lattice quantum chromodinamics (LQCD) has seen enormous progress in the last few years. The majority of calculations are based on the finite-volume formalism of L\"uscher \cite{Luscher:1986n2}, which relates discrete finite-volume energies in a cubic, periodic, spatial volume of side-length $L$, to the scattering amplitude of two identical spin-zero particles. This relation is exact up to corrections scaling as $e^{- m L}$, with $m$ the pion mass, but holds only for energies in the regime of elastic scattering, i.e.~below the lowest-lying three- or four-particle threshold.
The formalism has since been extended to generic two-particle systems \cite{Luscher:1991n1,Kari:1995, Kim:2005gf, He:2005ey, Bernard:2010fp, Hansen:2012tf, Briceno:2012yi, Briceno:2014oea, Romero-Lopez:2018zyy,Woss:2020cmp}, for which, however, the same restrictions apply.
At unphysically heavy pion masses, many resonances satisfy this restriction, leading to a recent explosion of LQCD resonant studies as reviewed, for example, in ref.~\cite{Briceno:2017max}. However, for physical masses, many experimentally observed resonances have significant branching fractions to modes containing three (or more) particles. Thus, the development of a multi-particle formalism is essential in order to gain insight into the nature of these states.

In the last few years, significant theoretical effort has been devoted to extensions and alternatives to the two-particle Lüscher formalism for more-than-two-particle systems. In particular, a three-particle quantization condition for identical (pseudo)scalars has been derived following three different approaches:\footnote{See also refs.~\cite{Polejaeva:2012ut,Guo:2017ism,Klos:2018sen,Guo:2018ibd}.} 
(i) generic relativistic effective field theory (RFT)~\cite{\HSQCa,\HSQCb,\BHSQC,\BHSnum,\BHSK,\dwave,\largera,Blanton:2019vdk}, 
(ii) nonrelativistic effective field theory (NREFT)~\cite{\Akakia,\Akakib,\Akakinum,Pang:2019dfe}, and (iii) (relativistic) finite volume unitarity (FVU) \cite{\MD,\MDpi,Mai:2019fba}.
(See ref.~\cite{\HSrev} for a review of the three approaches.) At this stage, only the RFT formalism has been explicitly worked out including higher partial waves.
 
These theoretical developments have been accompanied by significant progress in lattice calculations. In previous work, the three-particle coupling was extracted using the ground state energy in QCD \cite{Beane:2007es, Detmold:2008fn,Mai:2018djl}, and also in $\varphi^4$ theory \cite{Romero-Lopez:2018rcb}. Going beyond this, the determination of complete spectra with quantum numbers of three pions has been achieved by multiple groups in the last two years \cite{\HH,Culver:2019vvu,Woss:2018irj}. In fact, very recently, a large number of three-$\pi^+$ levels (including those in moving frames) has been combined with the RFT formalism to constrain the three-particle scattering amplitude from first principles QCD \cite{Blanton:2019vdk}.

As the present quantization conditions are only valid for identical particles, their use is limited to three pions (or kaons or heavy mesons) at maximal isospin, and thus only for weakly interacting channels with no resonances. Motivated by this, in the present paper we provide the generalization of the RFT approach to include nonidentical, mass-degenerate (pseudo)scalar particles. Specifically, we focus on a general three-pion state in QCD with exact isospin symmetry (and thus exact G parity, preventing two-to-three transitions).

A feature of all three-particle approaches 
is that the extraction of scattering amplitudes proceeds via an intermediate 
three-particle scattering quantity, denoted in the RFT approach by $\Kdf$. 
In particular, the RFT quantization condition provides,
for each finite-volume three-particle energy, $E_n(L)$,
a combined constraint  on $\Kdf$ and the two-particle scattering amplitude, $\mathcal M_2$.
Additional constraints on $\mathcal M_2$ are provided by the two-particle spectrum using
the L\"uscher formalism.
Then, in a second step, infinite-volume integral equations are used to relate $\Kdf$ to the physical scattering amplitude, $\mathcal M_3$.
To implement these steps in practice, one requires
a physically motivated parametrization of $\Kdf$ that includes, for example, a truncation
in the angular momentum of two-particle subsystems.

Our work generalizes all aspects of this work flow to three-pion scattering for all allowed values of two- and three-pion isospin. 
In section~\ref{sec:derivation} we derive the generalized formalism.
We first review the results of refs.~\cite{\LtoK,\KtoM} for identical particles [section \ref{sec:der:identical}], before providing the extensions to non-identical pions, first of the relation between $E_n(L)$ to $\Kdf$ [section \ref{sec:der:LtoKfbasis}] and then of the integral equations relating $\Kdf$ to $\mathcal M_3$ [section \ref{sec:genKtoM}]. 
These are presented for states with definite individual pion flavors.
The change of basis to definite total isospin is given 
in sections \ref{sec:blockdiag} and \ref{sec:blockdiagM3}.
An important consequence of projecting to total isospin is that the results block diagonalize into four separate relations, one for each of the allowed values of the total three-pion isospin: 
$I_{\pi \pi \pi}=0,1,2,3$.

With the formalism in hand, in section~\ref{sec:param} we describe strategies 
to parametrize $\Kdf$. We determine the form of the  threshold expansion 
for all choices of $I_{\pi\pi\pi}$,
and provide expressions for $\Kdf$ that produce three-particle resonant behavior for each
of the choices of $I_{\pi\pi\pi}$ and $J^P$ for which such behavior is experimentally observed.

To illustrate the utility of the generalized formalism, we present a numerical implementation
for the $I_{\pi\pi\pi}=0$ channel in section~\ref{sec:toy}.
  We do so using forms of $\Kdf$ that lead to both vector and axial-vector resonances, mimicking the experimentally observed $\omega$ and $h_1$. The finite-volume energies exhibit avoided level crossings associated with the allowed cascading resonant decays, e.g.~$h_1 \to \rho \pi \to \pi \pi \pi$. 

This completes the main text, following which 
section~\ref{sec:conc} gives a brief summary of the work and a
discussion of the future outlook.
We include four appendices to address various technical details. 
First, in appendix \ref{app:derivation}, 
we provide further discussion of the derivation of the generalized quantization condition. 
Second, in appendix \ref{app:QC}, 
we collect the definitions of the building blocks entering the quantization condition. 
Third, appendix \ref{app:A} describes the different bases we use for three-pion states.
Finally, appendix \ref{app:B} summarizes some group theoretical results that are relevant
to the implementation of the quantization condition.

\section{Derivation} 
\label{sec:derivation}

In this section we derive the quantization condition for general three-pion states. Following the approach of refs.~\cite{\KSS,\LtoK}, we first introduce a matrix of correlation functions
\begin{equation}
C_{L;jk}(P) \equiv \int d x^0 \int_{L^3} d^3 x \ e^{- i \boldsymbol P \cdot \boldsymbol x + i E t} \ \langle \text{T} \mathcal O_{j}(x) \mathcal O^\dagger_{k}(0) \rangle_L \,.
\end{equation}
Here $\mathcal O^\dagger_{k}$ are $\mathcal O_{j}$ are operators that, respectively,
create and destroy three-pion states, with quantum numbers and
additional information specified by the indices $j,k$. 
In the following paragraphs we give a concrete choice for these operators that is particularly convenient for the present derivation. The correlator is defined in the context of a generic, isospin-symmetric effective theory of pions. The underlying fields are denoted by $\pi_{+}(x), \ \pi_-(x)$ and $\pi_0(x)$, and are normalized such that
\begin{equation}
\label{eq:pi_norm}
\langle 0 \vert \pi_{q}(x)  \vert \pi, q, \boldsymbol p \rangle = e^{-i p \cdot x} \,,
\end{equation}
where $\vert \pi, q, \boldsymbol p \rangle$ is a state with mass $m$ and charge $q$,
and $p^0=\omega_p = \sqrt{\boldsymbol p^2+m^2}$.
We use Minkowski four-vectors,
adopting the metric convention $p \cdot x = p^0 x^0 - \boldsymbol p \cdot \boldsymbol x$.
The finite volume is implemented by requiring that all fields satisfy periodic boundary conditions in each of the spatial directions.,
$\pi(x) = \pi(x + L \boldsymbol{e}_i)$.

In the derivation of refs.~\cite{\LtoK,\KtoM}, the analysis was simplified
by assuming that the interactions
of the identical scalar particles satisfied a $Z_2$ symmetry that led to
particle number conservation modulo two.\footnote{This is not a fundamental limitation on the derivation; the generalization
without a $Z_2$ symmetry is derived in ref.~\cite{\BHSQC}.}
This implied, for example, that there were no intermediate four-pion states in
the correlator $C_L$.
This simplification carries over to the present analysis because
we are assuming exact isospin symmetry, so that G parity is exactly conserved,
and serves as the $Z_2$ symmetry.

For a given choice of total momentum $\boldsymbol P$, which is constrained by
the boundary conditions to take one of the values $2\pi \boldsymbol n /L$, with
$\boldsymbol n$ a vector of integers, the correlator $C_{L,ij}(E,\boldsymbol P)$ 
has poles in $E$ at the positions of the finite-volume eigenstates. Our aim is
to derive a quantization condition whose solutions give the energies of these eigenstates.

\begin{figure}
\begin{center}
\vspace{-10pt}
\includegraphics[width=\textwidth]{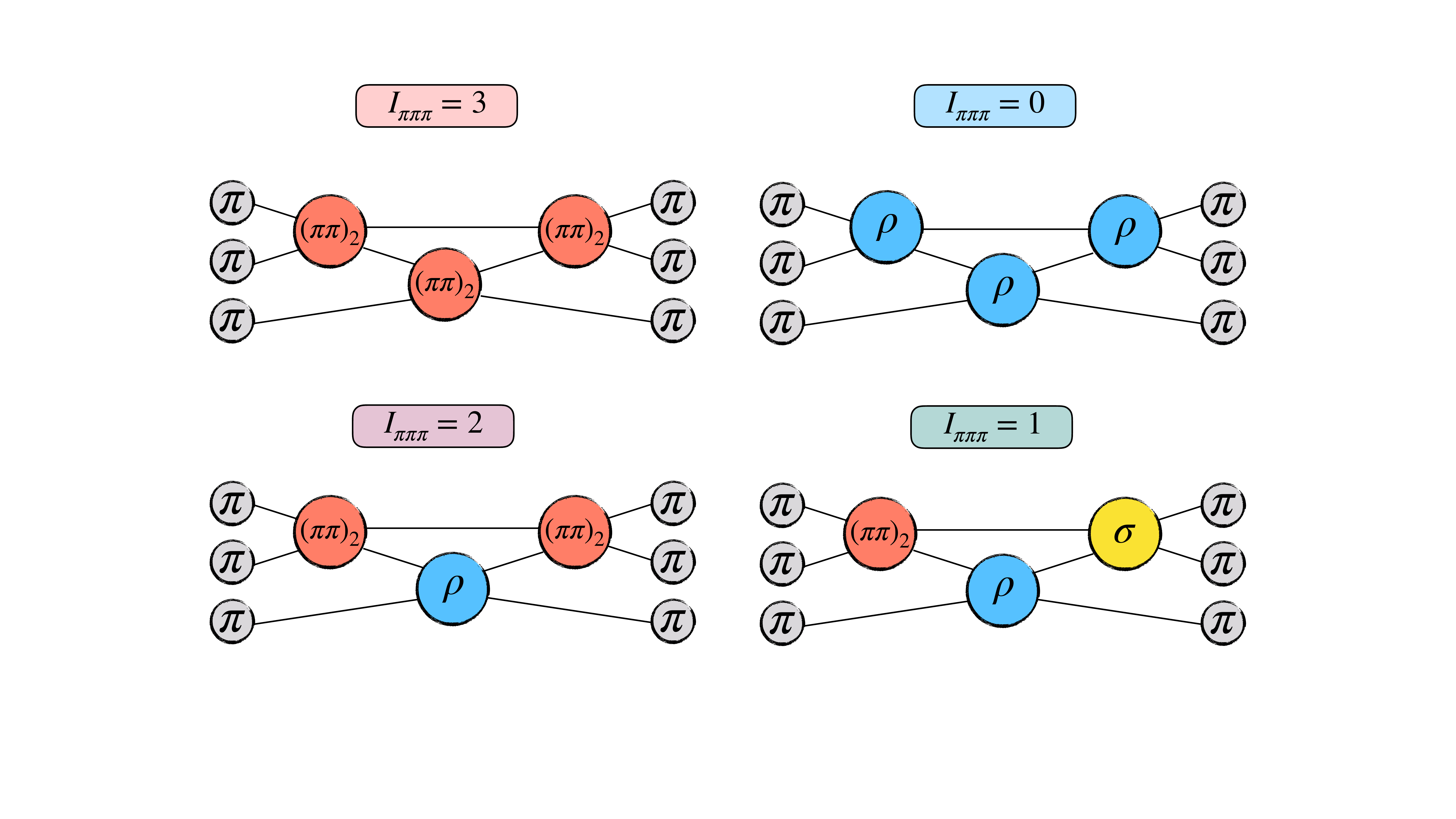}
\vspace{-60pt}
\caption{Sketch of subchannels for pairwise interactions present in each three-pion system with fixed overall isospin, $I_{\pi\pi\pi}$. For  $I_{\pi\pi\pi}=0$ and $3$, only one subchannel is present, having $I_{\pi\pi}=1$ and $I_{\pi\pi}=2$, respectively. For $I_{\pi\pi\pi}=2$, two subchannels are present,  with $I_{\pi\pi} = 1$ and $2$, implying that the three-particle quantization condition lives in a two-dimensional flavor space. For $I_{\pi\pi\pi} = 1$, all three two-pion subchannels contribute ($I_{\pi\pi}=0,1, $ and $2$), leading to a three-dimensional flavor space. For convenience, we use the shorthand notation $(I_{\pi\pi} =  0) \equiv ``\sigma"$, $(I_{\pi\pi} =  1) \equiv ``\rho"$, and  $(I_{\pi\pi} =  2) \equiv ``(\pi\pi)_2"$, 
in which we label  (when possible) the two-pion subchannels by the renonances present in them.
\label{fig:pions}}
\end{center}
\end{figure}

There are 27 distinct combinations of three-pion fields, assuming that we distinguish identical fields with position labels, $x_1,x_2,x_3$. 
It is useful to understand this multiplicity from the viewpoint of 
combining three objects with isospin 1. This leads to seven irreducible representations
(irreps)
\begin{equation}
\bm 1 \otimes \bm 1 \otimes \bm 1 = (\bm0 \oplus \bm1 \oplus \bm2) \otimes \bm1
= (\bm 1) \oplus (\bm0  \oplus \bm1 \oplus \bm2) \oplus
(\bm1 \oplus \bm2 \oplus \bm3)\,.
\end{equation}
We see that the total three-pion isospin can have values
$I_{\pi \pi \pi} = 0, 1, 2, 3$,  with respective multiplicities $1,3,2,1$.
The multiplicities correspond to the number of possible values of the two-pion
isospin, $I_{\pi\pi}$, that can appear: all three values for $I_{\pi\pi\pi}=1$, 
 two values, $I_{\pi\pi}=1,2$, for $I_{\pi\pi\pi}=2$, and only one value each for
 $I_{\pi\pi\pi}=0$ and $3$, namely $I_{\pi\pi}=1$ and $2$, respectively.
The situation is summarized in fig.~\ref{fig:pions}.

Since we are treating isospin as an exact symmetry, we need only consider
one choice of $I_z$ (or, equivalently, one choice of electric charge) 
from each of the seven irreps.
A convenient choice is to use the combination with vanishing electric charge,
since this appears once in each irrep. Thus, henceforth we focus on the space of
the seven neutral operators:
\begin{equation}
\label{eq:opdefA}
\widetilde { \mathcal O}(a,b,k) \equiv  \left(
\begin{array}{c}
 \widetilde{\pi}_{-}( a) \  \widetilde{\pi}_{0}( b) \  \widetilde{\pi}_{+}( k)  \\[5pt]
 \widetilde{\pi}_{0}( a) \  \widetilde{\pi}_{-}( b) \  \widetilde{\pi}_{+}( k)  \\[5pt]
 \widetilde{\pi}_{-}( a) \  \widetilde{\pi}_{+}( b) \  \widetilde{\pi}_{0}( k)  \\[5pt]
 \widetilde{\pi}_{0}( a) \  \widetilde{\pi}_{0}( b) \  \widetilde{\pi}_{0}( k)  \\[5pt]
 \widetilde{\pi}_{+}( a) \  \widetilde{\pi}_{-}( b) \  \widetilde{\pi}_{0}( k)  \\[5pt]
 \widetilde{\pi}_{0}( a) \  \widetilde{\pi}_{+}( b) \  \widetilde{\pi}_{-}( k)  \\[5pt]
 \widetilde{\pi}_{+}( a) \  \widetilde{\pi}_{0}( b) \  \widetilde{\pi}_{-}( k) 
\end{array}
\right) \,.
\end{equation}
Here we have written the fields in momentum space as this will prove  convenient
below.
These operators are related to $\mathcal O_j(x)$ via
\begin{equation}
\label{eq:opdefB}
\mathcal O_j(x) \equiv \int_{a,b,k} \, f(a,b,k) \, e^{- i (a+b+k) \cdot x} \, \widetilde { \mathcal O}_j(a,b,k) \,,
\end{equation}
where $\int_k \equiv \int d k^0/(2 \pi) \sum_{\bm k}$, with the sum over 
$\bm k$ being over the finite-volume set introduced above for $\bm P$. 
We also adopt the convention here, and below, that the factor of $1/L^3$ accompanying each sum is left implicit.
$f(a,b,k)$ is a smooth function that specifies the detailed form of $\mathcal O_j$.
As we discuss further below, it is crucial that $f(a,b,k)$ is {\em not} symmetric with respect to permutations of its arguments. More precisely, $f(a,b,k)$ is defined such that all seven operators defining $\mathcal O_j (x)$ are in fact distinct, which is necessary to ensure that all definite-isospin operators are non-zero.

Having defined the column of operators, $\mathcal O_j$, we are now in position to derive a skeleton expansion for $C_{L;ij}$, exactly as was done in ref.~\cite{\LtoK}. The only distinction compared to the earlier work is that the operators, appearing on the far left and far right of every diagram, now represent a column (on the left) and row (on the right), so that each Feynman diagram encodes a $7 \times 7$ matrix, defining a contribution to the matrix of correlators, $C_{L;ij}$. As we discuss in the following, this matrix structure naturally propagates through all steps of the derivation so that the final result appears identical to that of ref.~\cite{\LtoK}, but with the additional flavor channel assigned to each of the building blocks. The final step is to perform a change of basis into states with definite two- and three-pion isospin. This block diagonalizes $C_{L;ij}$, as expected, and one recovers four distinct quantization conditions, one each for $I_{\pi \pi \pi}=0,1,2,3$. While the $I_{\pi \pi \pi} = 0$ and $3$ conditions are one-dimensional in the flavor index,  $I_{\pi \pi \pi} = 1$ and $2$ are 3 and 2 dimensional, respectively, 
encoding the coupled-channel scattering of the various allowed $I_{\pi \pi}$ 
subchannels.

\subsection{Formalism for identical (pseudo-)scalars}
\label{sec:der:identical}

In this subsection we review the results of refs.~\cite{\LtoK,\KtoM} for 
the case of three identical particles, which apply here for the $I_{\pi\pi\pi}=3$
channel. These results will serve as stepping stones  to the generalization
for other values of $I_{\pi\pi\pi}$.
In ref.~\cite{\LtoK}, it was shown that the finite-volume correlator  
for three identical (pseudo-)scalars can be written
\begin{equation}
\label{eq:CLdecom}
C_L(P) = C_\infty(P) + i A_3' F_3 \frac{1}{1 + \mathcal K_{\df,3} F_3} A_3 \,,
\end{equation}
where
\begin{equation}
\label{eq:F3def}
2 \omega L^3 \times F_3 \equiv \frac{F}{3} - F   \frac{1}{1 + \mathcal M_{2,L} G } \mathcal M_{2,L}  F \,, \ \ \ \ \ \  \mathcal M_{2,L} \equiv \frac{1}{\mathcal K_2^{-1} + F} \,.
\end{equation}
This result holds for $m^2 < E^2 - \boldsymbol P^2 < (5m)^2$ and neglects $L$ dependence of the form $e^{- m L}$, while keeping all power-like scaling. The intuitive picture behind its derivation is that only three-pion states can go on shell for the kinematics considered, and only these on-shell states can propagate large distances to feel the periodicity and induce $1/L^n$ corrections.
The quantities $\omega, F, G, \mathcal K_2, \mathcal K_{\df,3}, A_3', A_3$ 
and $C_\infty$ are each defined in detail in ref.~\cite{\LtoK}, 
as is the matrix space on which all quantities act.\footnote{{The quantities we call $A_3$ and $A'_3$ here are denoted $A$ and $A'$ in
refs.~\cite{\LtoK,\KtoM}.}
}
Here we only give a brief summary of the most important details, with some additional definitions provided in appendix \ref{app:QC}. 
All objects besides $C_L$ and $C_\infty$ are defined on an index space denoted by $k, \ell, m$ where $k$ represents the three-momentum for the spectator particle,
 i.e.~is shorthand for a finite-volume momentum $\bm k$,
 and $\ell, m$ give the angular-momentum of the non-spectator pair. A cutoff on the $k$ index is built into all matrices so that this index space is always finite. To intuitively understand the appearance of the cutoff function, note that, for fixed total energy  $E$
 and momentum $\boldsymbol P$, if the spectator carries 
 $k^\mu = (\omega_k, \boldsymbol k)$
  then the squared invariant mass of the non-spectator pair is
\begin{equation}
E_{2,k}^{\star 2} \equiv (E - \omega_k)^2 - (\boldsymbol P - \boldsymbol k)^2 \,.
\end{equation}
This becomes negative for sufficiently large $\boldsymbol k^2$ implying that the state cannot go on the mass shell and therefore does not induce power-like $L$ dependence. Thus it is possible to absorb the deep subthreshold behavior into the definitions 
of $\mathcal K_2, \mathcal K_{\df,3}, A_3', A_3$  and $C_\infty$ 
and to cut off the matrix space.

The objects $\omega$, $F$, $G$, $\mathcal K_2$ and $\mathcal K_{\df,3}$ are all matrices on the $k, \ell, m$ space, e.g.~$F = F_{k' \ell' m', k \ell m}$, 
whereas $A_3'$ and $A_3$ are row and column vectors respectively, 
e.g.~$A_3 = A_{3; k \ell m}$. In this way all indices in eqs.~\eqref{eq:CLdecom} and \eqref{eq:F3def} are fully contracted, with adjacent factors multiplied according to usual matrix multiplication.
The $L$-dependence in these results enters both through the index space, $k$, and through explicit dependence inside of $F$ and $G$, which are defined in
 eqs.~\eqref{eq:Ft1} and \eqref{eq:defG}, respectively. 
 The simplest object entering eq.~\eqref{eq:F3def} is the diagonal kinematic matrix
\begin{equation}
\omega_{k' \ell' m', k \ell m} \equiv  \delta_{k' k} \delta_{\ell' \ell} \delta_{m' m} \sqrt{\boldsymbol k^2 + m^2} \,.
\end{equation}

This leaves only two quantities to define: the two- and three-particle K matrices, 
$\mathcal K_2$ and $\mathcal K_{\df,3}$, respectively. 
The former is given in eq.~(\ref{eq:K2mat}).
It depends on the two-to-two scattering phase shift, $\delta_{\ell}$,
 in each angular momentum channel, for two-particle energies ranging from $0$
  (well below the threshold at $2m$) up to $E^\star - m$.
Here we have introduced the notation $E^\star = \sqrt{E^2 - \boldsymbol P^2}$, for the three-particle \CMF~energy. In practice, one must choose a value $\ell_{\text{max}}$ above which the phase shift is assumed negligible, in order to render $\mathcal K_2$ finite-dimensional. Then it can be determined using the two-particle quantization condition, together with finite-volume energies from a numerical lattice calculation.
  
The remaining object, $\mathcal K_{\df,3}$, encodes the short-distance part of the three-particle amplitude.
 We close this subsection by explaining, first, how this quantity can be constrained from finite-volume three-particle energies and, second,
 how it is related to the physical observable, the three-particle scattering amplitude.

The utility of eq.~\eqref{eq:CLdecom} is that it allows one to identify the poles in $C_L(P)$ as a function of $E$, corresponding to the three-body finite-volume spectrum for fixed values of $L$ and $\bm P$.
These pole locations, denoted $E_n(L)$ for $n=0,1,2, \hdots \,$, occur at energies for which
\begin{equation}
\label{eq:identQC}
\text{det}_{k,\ell,m} \big [1 + \mathcal K_{\df,3}(E^\star) F_3(E, \boldsymbol P, L) \big ] =0 \,,
\end{equation}
where we have made the kinematic dependence explicit. Thus, given many values of $E_n(L)$, ideally for different $\boldsymbol P$ and $L$, one can identify parameterizations of $\K_{\df,3}(E^\star)$ that describe the system and fix the values of the parameters. As with $\mathcal K_2$, also here a value of $\ell_{\text{max}}$ must be set to render $ \mathcal K_{\df,3}(E^\star) $ finite-dimensional. Indeed, the angular momentum cutoffs in the two- and three-particle sectors must be performed in a self consistent way, as is described in ref.~\cite{\dwave}.

Now, taking $\K_{\df,3}(E^\star)$ as known, we present its relation to the three-particle scattering amplitude, $\mathcal M_3$, first derived in ref.~\cite{\KtoM}. As is explained in that work, one can relate $C_L(P)$ to a new finite-volume correlator, $\mathcal M_{3,L}(P)$, in a two-step procedure. First we take only the second term of eq.~\eqref{eq:CLdecom},
 multiply by $i$, and amputate $A_3' F [2 \omega L^3]^{-1}$ on the left and $[2 \omega L^3]^{-1}F A_3$ on the right to reach
\begin{align}
C'_L(P) & \equiv  - \bigg[ \frac{F}{2 \omega L^3} \bigg ]^{-1} F_3  \frac{1}{1 + \mathcal K_{\df,3} F_3} \bigg[ \frac{F}{2 \omega L^3} \bigg ]^{-1} \,, \\
& =   \mathcal D^{(u,u)}_{\text{disc}}  +   \mathcal D^{(u,u)} +   \mathcal L^{(u)}_L    \frac{1}{1 + \mathcal K_{\df,3} F_3}     \mathcal K_{\df,3}    \mathcal R^{(u)}_L \,,  
\end{align}
where in the second step we have introduced 
\begin{align}
\label{eq:Ddisc}
\mathcal D^{(u,u)}_{\text{disc}} & \equiv   - \bigg[ \frac{F}{2 \omega L^3} \bigg ]^{-1}  \bigg [ \frac{F}{6 \omega L^3} - \frac{ F     \mathcal M_{2,L}  F }{2 \omega L^3} \,   \bigg ] \bigg[ \frac{F}{2 \omega L^3} \bigg ]^{-1} \,, \\
\mathcal D^{(u,u)}  & \equiv -\bigg[ \frac{F}{2 \omega L^3} \bigg ]^{-1}  F_3 \bigg[ \frac{F}{2 \omega L^3} \bigg ]^{-1} - \mathcal D^{(u,u)}_{\text{disc}} \,, \\
  \mathcal L^{(u)}_L & \equiv \bigg[ \frac{F}{2 \omega L^3} \bigg ]^{-1} F_3 \,, \\
   \mathcal R^{(u)}_L & \equiv F_3 \bigg[ \frac{F}{2 \omega L^3} \bigg ]^{-1}\,.
\end{align}
Note that $\mathcal D^{(u,u)}$, $ \mathcal L^{(u)}_L $ and $ \mathcal R^{(u)}_L$ are closely related to $F_3$, differing only by the amputation factors and, in the case of $\mathcal D^{(u,u)}$, by the subtraction of $\mathcal D^{(u,u)}_{\text{disc}}$. The latter is labeled with the subscript ``disc'' for disconnected, referring to the fact that these terms arise from diagrams in which one of the three-particles does not interact with the other two. The second step towards defining $\mathcal M_{3,L}(P)$ is to drop $\mathcal D^{(u,u)}_{\text{disc}} $ and to symmetrize the resulting function with respect to the exchange of pion momenta. The result is \begin{align}
\label{eq:M3Ldef}
 \mathcal M_{3,L}(P) &\equiv \mathcal S \Big [  \mathcal M_{3,L}^{(u,u)}(P)   \Big ] \,, \\
  \mathcal M_{3,L}^{(u,u)}(P) & \equiv  \mathcal D^{(u,u)} +   \mathcal L^{(u)}_L    \frac{1}{1 + \mathcal K_{\df,3} F_3}     \mathcal K_{\df,3}    \mathcal R^{(u)}_L \,,
\end{align}
where $\mathcal S$ indicates the symmetrization.\footnote{%
The quantity $\mathcal M_{3,L}^{(u,u)}$ given here is actually slightly different
from the object with the same name defined in Ref.~\cite{\KtoM}. The distinction is that the $\mathcal M_{3,L}^{(u,u)}$ is this work has been
partially symmetrized, leading to small differences in $\mathcal L^{(u)}$
and $\mathcal R^{(u)}$. However, these differences have
 no impact on the fully symmetrized quantity, $\mathcal M_{3,L}$, 
 which is identical to that in Ref.~\cite{\KtoM}.
}
 This is explained in detail in section~\ref{sec:genKtoM} below, in the context of the generic isospin system.

The motivation for these seemingly \emph{ad hoc} redefinitions is that the new correlator, $ \mathcal M_{3,L}(P)$, is closely related to the physical, fully connected three-to-three scattering amplitude. Substituting $P = (E, \boldsymbol P)$, the connection is given by 
\begin{equation}
\label{eq:MLinfty}
\mathcal M_3(E, \boldsymbol P) = \lim_{\epsilon \to 0^+} \lim_{L \to \infty}  \mathcal M_{3,L}(E + i \epsilon, \boldsymbol P) \,.
\end{equation} 
This ordered double limit can be evaluated analytically to produce an integral equation relating $\mathcal K_{\df,3}$ to the $\mathcal M_3$. This completes the complicated mapping from the finite-volume spectrum to infinite-volume amplitudes.
Again, we point the reader to ref.~\cite{\KtoM} for a full derivation and for the explicit forms of the integral equations. 

\subsection{Generalized quantization condition}
\label{sec:der:LtoKfbasis}

\begin{figure}
\begin{center}
\includegraphics[width=\textwidth]{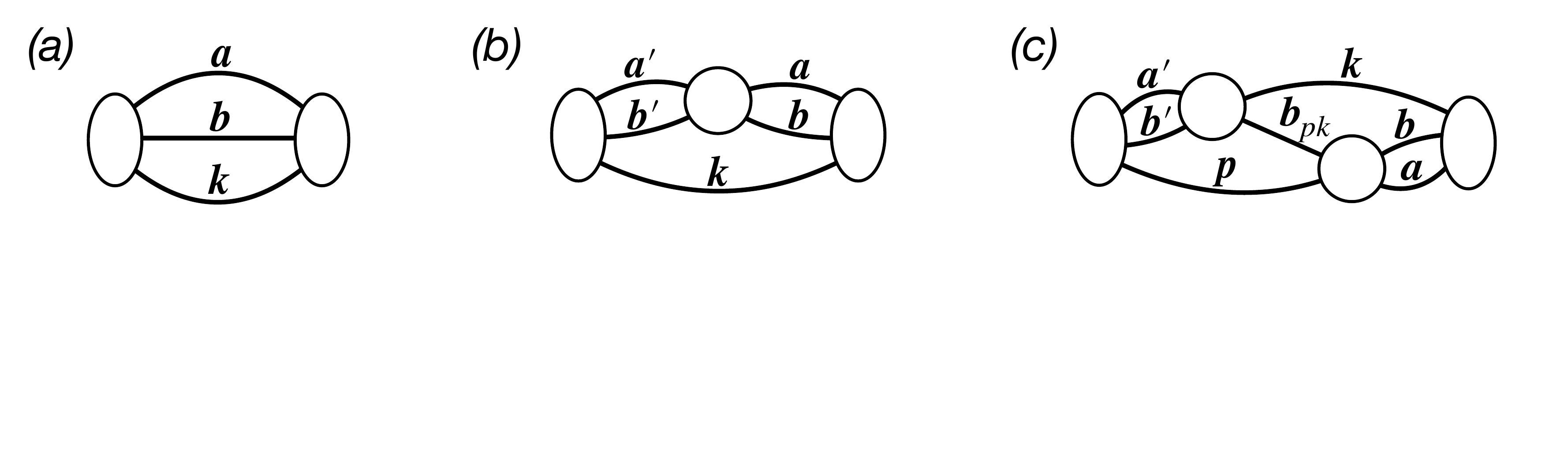}
\vspace{-70pt}
\caption{Three Feynman diagram topologies required to illustrate the extension to generic isospin. 
\label{fig:diagexamples}}
\end{center}
\end{figure}

In this subsection we generalize the derivation of the quantization condition [eq.~(\ref{eq:identQC})] to the system of three pions with any allowed total isospin.
The relation of the generalized $\mathcal K_{\df,3}$ to the corresponding generalized
scattering amplitude is discussed in the next subsection.

As explained above, the finite-volume correlator, $C_{L, ij}$, becomes a $7 \times 7$ matrix on the space of all possible neutral three-pion configurations. 
We find that, to generalize the quantization condition,
we also need to extend all the objects in the correlator decomposition [eq.~\eqref{eq:CLdecom}], the quantization condition [eq.~\eqref{eq:identQC}] and the relation to $\mathcal M_3$ [eqs.~\eqref{eq:M3Ldef} and \eqref{eq:MLinfty}] to be matrices on the seven-dimensional flavor space. We stress that all objects, including $C_\infty$, $A_3$ and $A_3'$ become flavor matrices, even though the latter are defined as either scalars or vectors in the $k \ell m$ indices.

In the original derivation of ref.~\cite{\LtoK}, the first step was to identify a skeleton expansion that expressed $C_{L}$ in terms of generalized Bethe-Salpeter kernels and fully dressed propagators. Cutting rules were then applied to write each diagram as a sum of various contributions, and summing over all possibilities lead to eq.~\eqref{eq:CLdecom}. A key feature that will simplify the present generalization is that the new matrix space can be completely implemented already at the level of Bethe-Salpeter kernels and fully dressed propagators, i.e.~before the steps of decomposition and summation. These final steps, which lead to the main complications in the earlier work, can then be copied over with the new index space passing in a straightforward way into $F$, $G$, $\mathcal K_2$ and the other matrices entering the final results.

To illustrate this we carefully consider the three diagrams of fig.~\ref{fig:diagexamples}. We give expressions for each of these in turn, first for the case of identical particles and then for the general isospin extensions. In this way, all building blocks are defined for the new quantization condition, which is then given in eq.~\eqref{eq:genIsoQC} below.

Beginning with fig.~\ref{fig:diagexamples}(a), the expression in the case of three identical particles is
\begin{equation}
\label{eq:diagA}
C^{[\ref{fig:diagexamples}(a)]}_L(P) = \frac{1}{6} \sum_{\boldsymbol k, \boldsymbol a} \int \frac{d a^0}{2 \pi}  \int \frac{d k^0}{2 \pi} i \sigma(k,a)  \,  \Delta(a)  \Delta(b) \Delta(k)  \,    i \sigma^\dagger(k,a) \,,
\end{equation}
where $\sigma(k,a)$ and $\sigma^\dagger(k,a)$ are endcap factors encoding the coupling of the operator to a three-particle state and $\Delta(a)$ is a fully dressed propagator. 
As explained in ref.~\cite{\LtoK}, this can be rewritten as
\begin{equation}
\label{eq:diagAcut}
C^{[\ref{fig:diagexamples}(a)]}_L(P) = C^{[\ref{fig:diagexamples}(a)]}_\infty(P) + i \sigma \frac{i F}{6 \omega L^3} i \sigma^\dagger \,,
\end{equation}
where the first term on the right-hand side is the contribution from the diagram of fig.~\ref{fig:diagexamples}(a) to the infinite-volume correlation function. In the second term we have introduced $\sigma$ and $\sigma^\dagger$ as row and column vectors, respectively, on the $k \ell m$ space. These are ultimately combined with other terms to define $A_3'$ and $A_3$, respectively.

In the extension to general three-pion isospin, eq.~\eqref{eq:diagA} is replaced with  
\begin{equation}
\label{eq:diagAgenIso}
C^{[\ref{fig:diagexamples}(a)]}_{L,j l }(P) = \frac{1}{6} \sum_{n,n'} \sum_{\boldsymbol k, \boldsymbol a} \int \frac{d a^0}{2 \pi}  \int \frac{d k^0}{2 \pi} i \sigma_{jn}(k,a)  \,  [\Delta(a)  \Delta(b) \Delta(k)]_{nn'}  \,    i \sigma^\dagger_{n' l }(k,a) \,,
\end{equation}
where $b=P-k-a$.
 Here $[\Delta(a) \Delta(b) \Delta(k)]_{nn'}$ is a diagonal matrix of propagator triplets, in which each entry is built from charged and neutral pion propagators according to eq.~\eqref{eq:opdefA}. We repeat the pion content of each entry here for convenience:
 \begin{multline}
[\Delta(a) \Delta(b) \Delta(k)] = \textrm{diag}\Big(
{  [-][\,0\,][+]},\,
{  [\,0\,][-][+]}, \,
{  [-][+][\,0\,]},\,
{  [\,0\,][\,0\,][\,0\,]},\, \\
{  [+][-][\,0\,]},\,
{  [\,0\,][+][-]},\,
{  [+][\,0\,][-]} \Big) \,,
\end{multline}
where $[-][\,0\,][+] = \Delta_{-}(a) \Delta_{0}(b) \Delta_{+}(k)$, etc.,
 the subscript indicating the pion field at the sink of the two-point function 
 defining the fully-dressed propagator.
 In fact, in the iso-symmetric theory, the propagators are all equal as functions,
 $\Delta_{-}(a)= \Delta_{0}(a)= \Delta_{+}(a)$.
 Nonetheless, it is useful to treat these objects as distinct, 
 in order to better identify the patterns arising in our 
 matrix representation of the Feynman rules.

The endcap matrices, $\sigma_{j l }(k,a)$ and $\sigma^\dagger_{j l }(k,a)$, are built from the function $f(a,b,k)$, introduced in eq.~\eqref{eq:opdefB}, that encodes how the fundamental fields, $\pi_0$, $\pi_+$ and $\pi_-$, are used to build up the annihilation operators $\mathcal O_j(x)$. The exact relation is
\begin{equation}
\sigma(k,a) = \begin{pmatrix}
 f(a,b,k) & f(b,a,k) & f(a,k,b) & 0 & f(b,k,a) & f(k,a,b) & f(k,b,a) \\
 f(b,a,k) & f(a,b,k) & f(k,a,b) & 0 & f(k,b,a) & f(a,k,b) & f(b,k,a) \\
 f(a,k,b) & f(b,k,a) & f(a,b,k) & 0 & f(b,a,k) & f(k,b,a) & f(k,a,b) \\
 0 & 0 & 0 & g(a,b,k) & 0 & 0 & 0 \\
 f(k,a,b) & f(k,b,a) & f(b,a,k) & 0 & f(a,b,k) & f(b,k,a) & f(a,k,b) \\
 f(b,k,a) & f(a,k,b) & f(k,b,a) & 0 & f(k,a,b) & f(a,b,k) & f(b,a,k) \\
 f(k,b,a) & f(k,a,b) & f(b,k,a) & 0 & f(a,k,b) & f(b,a,k) & f(a,b,k) \\
\end{pmatrix} \,,
\end{equation}
where
\begin{equation}
 g(a,b,k) \equiv f(a,b,k) + f(b,a,k) + f(a,k,b) + f(b,k,a) + f(k,a,b) + f(k,b,a)     
\end{equation}
is the symmetrized version of $f(a,b,k)$.
Here the $(i,j)$ entry of the matrix can be understood as the non-interacting overlap of the operator $\mathcal O_i(0)$ with the $j$th state. The latter is defined with the convention of eq.~\eqref{eq:opdefA}. So, for example, the (1,3) entry follows from
\begin{align}
\big \langle 0 \big \vert \mathcal O_1(0) \big \vert   \pi \pi \pi , j\!=\!3 \big \rangle & =
 \bcontraction[1ex]{        \int_{a',b',k'} \big \langle 0 \big \vert f(a',b',k')        }{        \, \widetilde \pi_{-} (a')        }{        \, \widetilde \pi_0(b') \,  \widetilde \pi_+(k') \big \vert          }{         \pi_-(a)        }
\acontraction[1ex]{        \int_{a',b',k'} \big \langle 0 \big \vert f(a',b',k') \,\widetilde \pi_{-}(a') \,        }{       \widetilde \pi_0(b')         }{        \,  \widetilde \pi_+(k') \big \vert \pi_-(a) \pi_+(b)        }{        \pi_0(k)        }
\bcontraction[2ex]{        \int_{a',b',k'} \big \langle 0 \big \vert f(a',b',k') \,\widetilde \pi_{-}(a') \, \widetilde \pi_0(b') \,          }{         \widetilde \pi_+(k')       }{       \big \vert \pi_-(a)          }{       \pi_+(b)         }
\int_{a',b',k'} \big \langle 0 \big \vert f(a',b',k') \,\widetilde \pi_{-}(a') \, \widetilde \pi_0(b') \,  \widetilde \pi_+(k') \big \vert \pi_-(a) \pi_+(b) \pi_0(k) \big \rangle \,,
\nonumber  \\
 & = f(a,k,b) \,,
\end{align}
where $\vert   \pi \pi \pi , j\!=\!3 \big \rangle$ represents the non-interacting state with momentum assignment given by the index. The $6$ different terms in the (4,4) entry arise from the $3!$ contractions of the neutral operator with the neutral state.

This complicated matrix structure in the case of the non-interacting diagram, fig.~\ref{fig:diagexamples}(a), may seem surprising. The structure arises simply because six of the seven entries in the column $\mathcal O_j(x)$ (all entries besides $j=4$) are built from $\pi^-$, $\pi^0$ and $\pi^+$, distinguished only by the momentum assignments as shown in eqs.~\eqref{eq:opdefA} and \eqref{eq:opdefB}. Thus, even when all interactions are turned off, 
 $C_{L,j l }$ is still nonzero for any combination of $j, l  \neq 4$.
  
In more detail, the definition of $\sigma_{ij}(k,a)$ ensures that eq.~\eqref{eq:diagAgenIso} 
gives the correct expression for $C_{L,j l }^{[\ref{fig:diagexamples}(a)]}$, 
for all choices of $j$ and $ l $. 
Here one must consider three distinct cases.
First for $j=4,  l  \neq 4$, as well as $j \neq 4,  l =4$, the correlator vanishes,
as expected for the non-interacting contribution connecting a 
$[-][\,0\,][+]$ channel with a $[\,0\,][\,0\,][\,0\,]$. 
Second, if both $j, l  \neq 4$ then one recovers a non-zero contribution 
with a factor of $ 6$ arising from the contracted matrix indices. 
For example for $j=1,l=2$ one finds
\begin{align}
\begin{split}
\sum_k \sigma_{1k}(k,a) \sigma^\dagger_{k2}(k,a) & =  
f(a,b,k) f^*(b,a,k) + f(b,a,k)f^*(a,b,k) + f(a,k,b)f^*(k,a,b) \\[-5pt]
& \hspace{-40pt}  + f(b,k,a)f^*(k,b,a) + f(k,a,b)f^*(a,k,b) + f(k,b,a)f^*(b,k,a)
     \nonumber
    \end{split} \\[7pt]
&  \hspace{-42pt}  \to 6 f(a,b,k) f^*(b,a,k)     \,,
\end{align}
where the arrow represents a replacement in the integral that is justified as all other factors are exchange symmetric with respect to $a$, $b$, and $k$.
This compensates the $1/6$ pre-factor, leading to the correct expression 
for a diagram with three distinguishable particles. 
Finally, $j= l =4$ yields the diagram with three neutral particles and in this case 
the $1/6$ survives and correctly gives the symmetry factor for identical particles.

Having demonstrated that eq.~\eqref{eq:diagAgenIso} gives the correct generalization of eq.~\eqref{eq:diagA}, it is now very straightforward to generalize the decomposition, eq.~\eqref{eq:diagAcut}. We find
\begin{align}
\label{eq:diagAgenIsocut}
\textbf C^{[\ref{fig:diagexamples}(a)]}_L(P) & = \textbf C^{[\ref{fig:diagexamples}(a)]}_\infty(P) + \frac13  \boldsymbol \sigma \, \textbf F \, \boldsymbol \sigma^\dagger \,, \\
\label{eq:Fgendef}
\left[\textbf F\right]_{j l }  & \equiv \frac{i F}{2 \omega L^3} \delta_{j l } \,,
\end{align}
where $\delta_{j l }$ is the identity matrix on the seven-dimensional flavor space. 
Here we find it convenient to absorb various factors of $i$, $\omega$ and $L$ into the boldface definitions. Specifically, we use 
\begin{equation}
\left[\boldsymbol \sigma\right]_{j l } = i \sigma_{j l }\,,\quad
\big[\boldsymbol \sigma^\dagger\big]_{j l } = i \big[\sigma^\dagger\big]_{j l }\,,
\ \ {\rm and}\ \
\left[\textbf C_L(P)\right]_{j l } = C_{L,j l }(P)\,,
\end{equation} 
In the following we generally follow the convention of using bold-faced symbols 
whenever flavor-space indices are suppressed. 

\bigskip

We turn now to the diagram shown in fig.~\ref{fig:diagexamples}(b). In the case of a single channel of identical particles the corresponding expression is
\begin{multline}
\label{eq:diagB}
C^{[\ref{fig:diagexamples}(b)]}_L(P) = \frac{1}{4} \sum_{\boldsymbol k, \boldsymbol a, \boldsymbol a'} \int \frac{d a^0}{2 \pi}  \int \frac{d a'^0}{2 \pi}  \int \frac{d k^0}{2 \pi} i \sigma(k,a')  \,  \Delta(a')  \Delta(b')  \\ \times i \mathcal B(a', b'; a, b) \,  \Delta(a) \Delta(b) \, \Delta(k)  \,    i \sigma^\dagger(k,a) \,,
\end{multline}
where $\mathcal B$ is the infinite-volume Bethe-Salpeter kernel. As we demonstrate in ref.~\cite{\LtoK} this leads to a contribution of the form 
\begin{equation}
\label{eq:diagBcut}
C^{[\ref{fig:diagexamples}(b)]}_L(P) =  i \sigma \,
\frac{i F}{2 \omega L^2} \, i \mathcal K_2 \,  i F \,
   i \sigma^\dagger + \cdots \,,
\end{equation}
where $\mathcal K_2$ is the two-particle K matrix, up to some subtleties in the sub-threshold definition, as discussed in refs.~\cite{\LtoK,\KtoM}. 
The ellipsis in eq.~\eqref{eq:diagBcut} indicates that additional terms arise containing
less than two factors of $F$. 
Indeed, many of the complications in ref.~\cite{\LtoK} arise in the demonstration that these terms can be reabsorbed into redefinitions of $C_\infty$, $\sigma$ 
and $\sigma^\dagger$, in a consistent way that generalizes to all orders. It is this patterm of absorbing higher-order terms that leads to the conversion of $\mathcal B$ into the K matrix.

Following the pattern established above, 
our next step is to give the isospin generalization of eq.~\eqref{eq:diagB}
\begin{multline}
\label{eq:diagBgenIso}
C^{[\ref{fig:diagexamples}(b)]}_{L,j l }(P) = \frac{1}{4} \sum_{\boldsymbol k, \boldsymbol a, \boldsymbol a'} \int \frac{d a^0}{2 \pi}  \int \frac{d a'^0}{2 \pi}  \int \frac{d k^0}{2 \pi} i \sigma_{jn}(k,a')  \,  [\Delta(a')  \Delta(b') \Delta(k)]_{nn'}  \\ \times [\Delta(k)^{-1} i \mathcal B(a', b'; a, b)]_{n' n''} \,  [ \Delta(a) \Delta(b) \Delta(k)  ]_{n'' n'''} \,    i \sigma^\dagger_{n'''  l }(k,a) \,.
\end{multline}
All quantities here have been defined, with the exception of $[\Delta(k)^{-1} i \mathcal B(a', b'; a, b)]_{n' n''}$. This object is a matrix on the flavor space, with non-zero entries only when the third particles of the $n'$ and $n''$ states coincide, see again eq.~\eqref{eq:opdefA}. In the case where $n'$ and $n''$ do have a common spectator, the entry is defined by setting $\Delta(k)^{-1}$ to the spectator species and taking $\mathcal B$ as the Bethe-Salpeter kernel for the  scattering of the $n'$ and $n''$ non-spectator pairs. 
 We give a concrete expression of this matrix structure
  (in the context of $\mathcal K_2$) in eqs.~\eqref{eq:K2def1}-\eqref{eq:K2def4} below.

As with eq.~\eqref{eq:diagAgenIso}, it is straightforward to show that 
\eqref{eq:diagBgenIso} 
 gives the correct result for the correlator
for all choices of $j$ and $ l $. 
For example, if $j=4$ and $ l \ne 4$, then
the left-hand loop (containing momenta $a'$ and $b'$) consists of three $\pi_0$s,
and the expression then forces the spectator in the right-hand loop
(that with momenta $a$ and $b$) to also be a $\pi_0$. 
There are then two options for $n''=n'''$ available, namely $n''=3$ and $5$ 
($n''=4$ being disallowed since $ l \ne 4$).
These two options correspond to the scattering process in the Bethe-Salpeter
kernel being $\pi_0(a') \pi_0(b') \leftarrow \pi_+(a) \pi_-(b)$ and 
$\pi_0(a') \pi_0(b') \leftarrow \pi_-(a) \pi_+(b)$, respectively.
These give equal contributions because in the loop sums/integrals we can
freely interchange the dummy labels $a$ and $b$. 
This redundancy
cancels the prefactor of $1/2$ for right-hand loop, while leaving it for the left-hand
loop, as required for a diagram with only one exchange-symmetric two-particle loop.

We are now ready to present the isospin generalization
of eq.~\eqref{eq:diagBcut},
\begin{equation}
\label{eq:diagBgenIsocut}
\textbf C^{[\ref{fig:diagexamples}(b)]}_L(P) =  \boldsymbol \sigma \,
\textbf F  \,  \textbf K_2 \,   \textbf F \,
   \boldsymbol \sigma^\dagger + \cdots \,,
\end{equation}
where all objects have been defined above besides
\begin{equation}
\label{eq:K2def1}
\textbf K_2 \equiv i [2 \omega L^3] 
\begin{pmatrix} 
\mathcal K_{+}  & & \\  
  & \mathcal K_{0 } & \\ 
  & & \mathcal K_{-} 
   \end{pmatrix} \,.
\end{equation}
Here our notation indicates a block-diagonal matrix,
in which the subscript on each block denotes the charge of the spectator.
The blocks are given explicitly by
\begin{align}
\mathcal K_{+} & \equiv 
 \begin{pmatrix}
  [ \pi_- \pi_0      \leftarrow  \pi_- \pi_0 ] \  \ &\  \    [ \pi_- \pi_0      \leftarrow  \pi_0 \, \pi_-  ]     \\ 
  [ \pi_0 \, \pi_-      \leftarrow  \pi_- \pi_0 ] \  \ &\  \    [ \pi_0 \, \pi_-      \leftarrow  \pi_0 \, \pi_- ]      \
 \end{pmatrix} \,, \\[5pt]
  \mathcal K_{0 } & \equiv 
 \begin{pmatrix}
    {     [\pi_- \pi_+     \leftarrow    \pi_- \pi_+]     } \  \ 
&\  \  {  [\pi_- \pi_+     \leftarrow   \pi_0 \, \pi_0]     } \   \ 
&\  \  {  [\pi_- \pi_+     \leftarrow  \pi_+ \pi_-]    } \     \\ 
     {     [\pi_0 \,  \pi_0     \leftarrow    \pi_- \pi_+]      } \  \ 
&\  \  {   [\pi_0 \,  \pi_0     \leftarrow   \pi_0 \,  \pi_0]     } \   \ 
&\  \  {   [\pi_0 \,  \pi_0     \leftarrow  \pi_+ \pi_-]     } \     \\ 
      {     [\pi_+ \pi_-     \leftarrow   \pi_- \pi_+]     } \  \ 
&\  \  {    [\pi_+ \pi_-     \leftarrow   \pi_0 \,  \pi_0]     } \   \ 
&\  \  {    [\pi_+ \pi_-     \leftarrow  \pi_+ \pi_-]     } \    
   \end{pmatrix} \,, \\[5pt]
   \mathcal K_{-} & \equiv 
 \begin{pmatrix}
   {     [\pi_0 \, \pi_+     \leftarrow  \pi_0 \, \pi_+]    } \  \ 
&\  \  { [\pi_0 \, \pi_+     \leftarrow   \pi_+ \pi_0]     } \     \\ 
    {    [\pi_+ \pi_0    \leftarrow  \pi_0 \, \pi_+]     } \  \ 
&\  \  { [\pi_+ \pi_0    \leftarrow   \pi_+ \pi_0] \   } \  
 \end{pmatrix} \,,
 \label{eq:K2def4}
\end{align} 
where each scattering process in square brackets
 indicates the corresponding two-particle K matrix. 
We stress that many entries  in these K matrices are trivially related, e.g.
\begin{equation}
    {     [\pi_- (a') \pi_+(b')     \leftarrow    \pi_- (a) \pi_+ (b)]    }
    =
    {     [\pi_- (a') \pi_+(b')     \leftarrow    \pi_+ (b) \pi_- (a)]    }\,.
\end{equation}
This completes the discussion of fig.~\ref{fig:diagexamples}(b).

To conclude the extension of the quantization condition, it remains only to consider fig.~\ref{fig:diagexamples}(c). Here we immediately give the isospin-generalized expression
\begin{multline}
\label{eq:diagCgenIso}
C^{[\ref{fig:diagexamples}(c)]}_{L,j l }(P) = \frac{1}{4} \sum_{\boldsymbol k, \boldsymbol a, \boldsymbol a'} \int \frac{d a^0}{2 \pi}  \int \frac{d a'^0}{2 \pi}  \int \frac{d k^0}{2 \pi} i \sigma_{jn}(k,a')  \,  [\Delta(a')  \Delta(b') \Delta(k)]_{nn'}  \\ \times [\Delta(k)^{-1} i \mathcal B(a', b'; p,b_{pk} )]_{n' n''} \,  [ \Delta(p)  \Delta(b_{pk}) \Delta(k)  ]^{\textbf G}_{n'' n'''} \, \\ \times  [\Delta(p)^{-1} i \mathcal B(b_{pk}, k ; a, b)]_{n''' m''}  [\Delta(a)  \Delta(b)  \Delta(p)]_{m'' m'} i \sigma^\dagger_{m'  l }(p,a) \,,
\end{multline}
where $b_{pk}=P-p - k$.
All quantities are defined above except for the propagator triplet with the $\textbf G$ superscript, which represents the contribution of the central cut in fig.~\ref{fig:diagexamples}(c). To give an explicit expression, we introduce the
matrix
\begin{equation}
T_G =
\begin{pmatrix}
  \square \,  & \, \square \,  & \, \square \,  & \, \square \,  & \, \square \,  & \, \square \,  & \, \blacksquare   \\
  \square \,  & \, \square \,  & \, \square \,  & \, \square \,  & \, \blacksquare \,  & \, \square \,  & \, \square   \\
  \square \,  & \, \square \,  & \, \square \,  & \, \square \,  & \, \square \,  & \, \blacksquare \,  & \, \square   \\
  \square \,  & \, \square \,  & \, \square \,  & \, \blacksquare \,  & \, \square \,  & \, \square \,  & \, \square   \\
  \square \,  & \, \blacksquare \,  & \, \square \,  & \, \square \,  & \, \square \,  & \, \square \,  & \, \square   \\
  \square \,  & \, \square \,  & \, \blacksquare \,  & \, \square \,  & \, \square \,  & \, \square \,  & \, \square   \\
  \blacksquare \,  & \, \square \,  & \, \square \,  & \, \square \,  & \, \square \,  & \, \square \,  & \, \square   \\
\end{pmatrix}
 \,, \qquad \qquad \square = 0 \,, \ \ \blacksquare = 1 \,.
 \label{eq:TGdef}
\end{equation}
(Here and below we use empty and filled squares to present matrices of 0s and 1s as we find this form more readable.) 
This corresponds to interchanging the first and last particles
in each channel, which is what is required by the ``switching'' of the spectator
particle in fig.~\ref{fig:diagexamples}(c). {Note that $T_G$ is a reducible representation of the element $(13)$ of the permutation group $\mathcal{S}_3$ in the notation of appendix \ref{app:A}.}
Using this matrix we then have
\begin{equation}
 [ \Delta(p)  \Delta(b_{pk}) \Delta(k)  ]^{\textbf G}_{n n''}
 =  
  [ \Delta(p)  \Delta(b_{pk}) \Delta(k)  ]_{n n'} [ T_G ]_{n' n''}\,.
\label{eq:DeltaG}
\end{equation}

In ref.~\cite{\LtoK} we demonstrated that such exchange propagators
gave rise to a new kind of finite-volume cut involving $G$.
We find that the isospin-generalized result is
\begin{equation}
\textbf C^{[\ref{fig:diagexamples}(c)]}_{L}(P) =  \boldsymbol \sigma \,
\textbf F  \,  \textbf K_2 \,  \textbf G \, \textbf K_2 \, \textbf F \,
   \boldsymbol \sigma^\dagger + \cdots \,,
\end{equation}
where
\begin{equation}
\label{eq:Ggendef}
\textbf G = i \frac{1}{2 \omega L^3}  G\, T_G\,.
\end{equation}
 We stress that, in contrast to $\mathcal K_2$ and $F$, the matrix $G$ does not commute with $1/[2 \omega L^3]$ on the $k, \ell, m$ index space. For this reason we have been careful to show the order of the product defining $\textbf G$.

At this point we have introduced the key quantities entering the generalized quantization condition: $\textbf F$, $\textbf K_2$ and $\textbf G$. With these objects defined, 
every step in the decompositions of refs.~\cite{\LtoK,\KtoM} naturally generalizes to flavor space, 
 with each equation carrying over essentially verbatim, but with extra flavor indices.
The only significant difference is that certain steps, related to symmetrization, require
additional justification when flavor is included. 
This is discussed in appendix~\ref{app:derivation}, where the additional arguments are
given. In the end,
one reaches a decomposition of the finite-volume correlator that is exactly analogous to eq.~\eqref{eq:CLdecom} above:
\begin{equation}
\label{eq:CLdecomgenIso}
\textbf C_L(P) =\textbf  C_\infty(P) - \textbf A_3' \textbf F_3 \frac{1}{1 - \textbf K_{\df,3} \textbf F_3} \textbf A_3 \,,
\end{equation}
where
\begin{equation}
\label{eq:F3defgenIso}
\textbf  F_3 \equiv \frac{\textbf F}{3} + \textbf F   \frac{1}{1 - \textbf M_{2,L} \textbf G } \textbf M_{2,L}  \textbf F \,, \ \ \ \ \ \  \textbf M_{2,L} \equiv \frac{1}{\textbf K_2^{-1} - \textbf F} \,.
\end{equation}
The sign changes in eqs.~(\ref{eq:CLdecomgenIso}) and 
(\ref{eq:F3defgenIso}) as compared to
eqs.~\eqref{eq:CLdecom} and \eqref{eq:F3def}
are due to the factors of $i$ that are
absorbed into the bold-faced quantities.\footnote{{For completeness, we note that $\textbf A_3$ and $\textbf A'_3$ include factors
of $i$: they are the flavor generalizations of $iA_3$ and $i A'_3$, respectively.
They are the generalized all-orders endcaps, whose leading terms are 
$\bm \sigma^\dagger$  and $\bm \sigma$, respectively.
Similarly $\textbf K_{\df,3}$ is the flavor generalization of $i \mathcal K_{\df,3}$.
 }
}

The endcap factors, $\textbf A_3'$ and $\textbf A_3$, are matrices on the seven-dimensional flavor space, describing the coupling of each of the seven operators [see eq.~\eqref{eq:opdefA}] to each of the seven interacting asymptotic states. The exact definitions are unimportant for this work and it suffices to know that these quantities, like $\textbf C_{\infty}(P)$, have only exponentially suppressed dependence on $L$, and do not contain the finite-volume poles that we are after. Thus, just as in the single channel case, the finite-volume spectrum is given by all divergences of the matrix appearing between $\textbf A_3'$ and $\textbf A_3$, equivalently by all solutions to the quantization condition
\begin{equation}
\label{eq:genIsoQC}
\text{det}_{k,\ell,m, \textbf f} \big [1 - \textbf K_{\df,3}(E^\star) \, \textbf  F_3(E, \boldsymbol P, L) \big ] =0 \,,
\end{equation}
where the subscript $\textbf f$ indicates that the determinant additionally runs over
 flavor space. Note that this expression will give the spectra of all three-particle quantum numbers simultaneously and is therefore not useful in practice. 
 In the section~\ref{sec:blockdiag} below we discuss how to project this result into the various sectors of definite total isospin.

\subsection{Generalized relation to the three-particle scattering amplitude \label{sec:genKtoM}}

First, however, we present the isospin generalizations of
eqs.~\eqref{eq:Ddisc}-\eqref{eq:MLinfty} above, 
thus providing the relation between $\textbf K_{\df,3}$ and the physical scattering amplitude. 
One first defines the modified finite-volume correlator:
\begin{align}
\label{eq:M3LdefgenIso}
 \textbf M_{3,L}(P) & \equiv \mathcal S \Big [  \textbf M^{(u,u)}_{3,L}(P)  \Big  ] \,, \\
 \textbf M^{(u,u)}_{3,L}(P) & \equiv     \textbf D^{(u,u)}  +  \textbf L^{(u)}_L    \frac{1}{1 - \textbf K_{\df,3} \textbf F_3}     \textbf K_{\df,3}    \textbf R^{(u)}_L \,,
 \label{eq:eq:M3LuudefgenIso}
\end{align}
 where 
\begin{align}
\label{eq:Duudef}
\textbf D^{(u,u)}  &   \equiv \textbf F^{-1}  \textbf F_3 \textbf F^{-1} - \textbf D^{(u,u)}_{\text{disc}} \,, 
 & \textbf D^{(u,u)}_{\text{disc}}  &  \equiv   \textbf F^{-1}  \bigg [ \frac{\textbf F}{3} +   \textbf F     \textbf M_{2,L} \textbf F      \bigg ] \textbf F^{-1} \,, 
\\
  \textbf L^{(u)}_L  &  \equiv \textbf F^{-1} \textbf F_3 \,, &
   \textbf R^{(u)}_L & \equiv \textbf F_3 \textbf F^{-1}\,. 
   \label{eq:RdefgenIso}
\end{align}
$\mathcal S$ now denotes a symmetrization procedure in the multi-flavor system, an extension that introduces some additional complications as we discuss in the following paragraphs.
 As in the case of a single channel, an ordered double limit of $\textbf M_{3,L}$ gives a set of integral equations relating $\textbf K_{\df,3}$ to the physical scattering amplitude, denoted $\textbf M_3$,
\begin{equation}
\label{eq:MLinftygenIso}
\textbf M_3(E, \boldsymbol P) = \lim_{\epsilon \to 0^+} \lim_{L \to \infty}  \textbf M_{3,L}(E + i \epsilon, \boldsymbol P) \,.
\end{equation} 
It is straightforward to write out the resulting integral equations explicitly,
as done for identical particles in ref.~\cite{\KtoM}, 
but they are not enlightening and we do not do so here.
This concludes the path from finite-volume spectrum, through $\textbf K_{\df,3}$, to the scattering amplitude $\textbf M_3$.

As in the single-channel case, implicit in this procedure is a conversion from the $k,\ell,m$ index space to a function of the incoming and outgoing three-momenta. This conversion is performed simultaneously with a symmetrization procedure.
We stress that symmetrization is needed even for non-identical particles, 
to ensure that all diagrams are included, i.e.~that the proper definition of the infinite-volume amplitude is recovered. 

At this point, it remains only to specify the symmetrization procedure, encoded in the operator $\mathcal S$, for the case of general pion flavors. 
To do so, we begin by defining 
\begin{equation}
X^{(u,u)}(\boldsymbol k', \boldsymbol a'; \boldsymbol k, \boldsymbol a)   \equiv 4 \pi Y^*_{\ell' m'}(\hat {\boldsymbol a}'^\star_{2,k'}) \, X^{(u,u)}_{k' \ell' m', k \ell m} \,     Y_{\ell m}(\hat {\boldsymbol a}_{2,k}^\star) \,,
\label{eq:Xuumomfunc}
\end{equation}
where $X^{(u,u)}_{k' \ell' m', k \ell m}$ stands for a generic, unsymmetrized quantity,
e.g.~$\mathcal M_{3,L}^{(u,u)}$ in the identical-particle case or an entry of $ \textbf M^{(u,u)}_{3,L} $ in flavor space.  
Here $\hat {\boldsymbol a}^\star_{2,k}$ is the spatial direction of $(\omega_a^\star, \boldsymbol a_{2,k}^\star)$, the four-vector reached by boosting $(\omega_a, \boldsymbol a)$ with velocity $\boldsymbol \beta = - (\boldsymbol P - \boldsymbol k)/(E - \omega_k)$.
In other words $\hat {\boldsymbol a}^\star_2$ gives the direction of back-to-back momenta of the non-spectator pair, which have momenta $\boldsymbol a$ and $\boldsymbol P - \boldsymbol k - \boldsymbol a$ in their two-particle \CMF. The same holds for $\hat {\boldsymbol a}'^\star_{2,k'}$ with $\boldsymbol a \to \boldsymbol a'$ and $\boldsymbol k \to \boldsymbol k'$. Contracting the spherical harmonic indices, as shown on the 
right-hand side of eq.~\eqref{eq:Xuumomfunc}, leads to a function of momenta whose
argument can be take as $\boldsymbol k, \hat {\boldsymbol a}^\star_{2,k}$ 
or, equally well, as $\boldsymbol k, \boldsymbol a$. Here we choose the latter convention, i.e.~specifying all momenta in the finite-volume frame, as this makes the symmetrization procedure more transparent. 

We begin with the case of a single channel of identical particles, where the symmetrization procedure, first introduced in ref.~\cite{\KtoM}, is given by
\begin{gather}
\label{eq:symmX}
X (\boldsymbol k', \boldsymbol a', \boldsymbol b'; \boldsymbol k, \boldsymbol a, \boldsymbol b)   \equiv \mathcal S[X^{(u,u)}_{k' \ell' m', k \ell m}]  \equiv \sum_{\{\boldsymbol p_3', \boldsymbol p_1' \} \in \mathcal P_3'}  \sum_{\{\boldsymbol p_3, \boldsymbol p_1 \} \in \mathcal P_3} X^{(u,u)}(\boldsymbol p_3', \boldsymbol p_1'; \boldsymbol p_3, \boldsymbol p_1)    \,. \end{gather}
The sums here run over the sets 
\begin{equation}
\label{eq:P3choices}
\mathcal P_3 = \big \{ \{ \boldsymbol k, \boldsymbol a \}, \{ \boldsymbol a, \boldsymbol b \}, \{ \boldsymbol b, \boldsymbol k \}   \big \}
\ \ {\rm and} \ \ 
\mathcal P_3' = \big \{ \{ \boldsymbol k', \boldsymbol a' \}, \{ \boldsymbol a', \boldsymbol b' \}, \{ \boldsymbol b', \boldsymbol k' \}   \big \}\,, 
\end{equation}
with $\boldsymbol b \equiv \boldsymbol P - \boldsymbol a - \boldsymbol k$ 
and $\boldsymbol b' \equiv \boldsymbol P - \boldsymbol a' - \boldsymbol k'$. 
 As discussed in ref.~\cite{\KtoM}, this step is necessary to reach the correct
 definition of $\mathcal M_3$, a quantity that is invariant under the exchange 
 of any two incoming or outgoing momenta. 
 The essential point is that the sum
runs over all assignments of the spectator momentum
for both incoming and outgoing particles in $X^{(u,u)}$.

\begin{figure}
\begin{center}
\includegraphics[width=\textwidth]{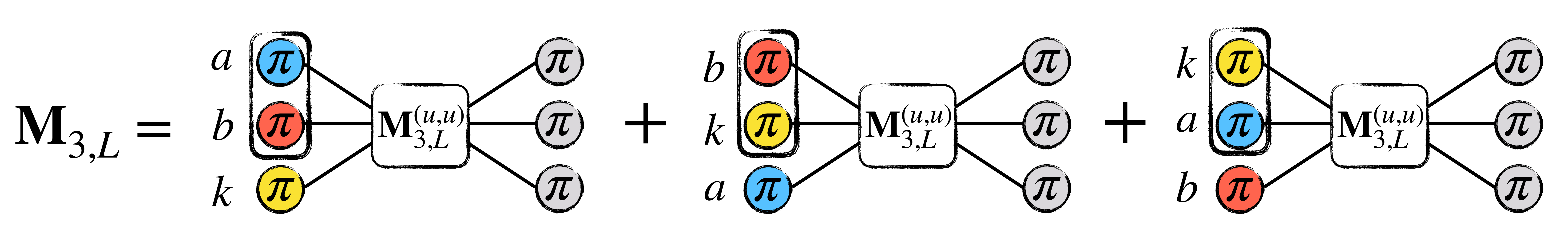}
\caption{ Representation of the symmetrization procedure applied to the outgoing particles. Colors indicate different flavors. \label{fig:symm}}
\end{center}
\end{figure}

To generalize this to non-trivial flavors, 
we first note that the identical-particle prescription,
i.e. simply summing $\textbf M_{3,L}^{(u,u)}$ over all permutations
of the momenta, is clearly incorrect. The issue is that, for
example, the $\pi_0 \, \pi_+ \pi_- \to \pi_0 \, \pi_+ \pi_-$
scattering amplitude is not, in general, invariant under
permutations of either the incoming or the outgoing momenta. Instead,
the required property is that amplitudes must be invariant under the
simultaneous exchange of flavor and momentum labels.
Summing over such exchanges ensures that the all choices of
the spectator pion flavor are included, as illustrated in fig.~\ref{fig:symm}.

To express this we introduce matrices that rearrange flavors 
in accordance with a given momentum permutation. 
For example, the second element in the set $\mathcal P_3$ corresponds to
 $\boldsymbol k \to \boldsymbol a$, 
$\boldsymbol a \to \boldsymbol b$, $\boldsymbol b \to \boldsymbol k$, 
and should be matched with the following flavor rearrangement: 
\begin{equation}
\label{eq:symMat}
\textbf R_{\boldsymbol k \to \boldsymbol a} \equiv   \begin{pmatrix}
 \square \, & \, \square  \, & \,  \square \, & \,  \square \, & \, \blacksquare  \, & \,  \square \, & \, \square     \\
\square  \, & \,  \square \, & \,   \square \, & \,  \square \, & \,  \square \, & \, \square  \, & \,   \blacksquare  \\
\square  \, & \, \blacksquare  \, & \, \square  \, & \,  \square \, & \,  \square \, & \, \square  \, & \,  \square   \\
\square  \, & \, \square  \, & \, \square  \, & \, \blacksquare  \, & \,  \square \, & \,  \square \, & \,  \square   \\
\square  \, & \,  \square \, & \,  \square \, & \, \square  \, & \,  \square \, & \, \blacksquare  \, & \,   \square  \\
\blacksquare  \, & \, \square   \, & \, \square  \, & \, \square  \, & \, \square  \, & \,  \square \, & \,  \square   \\
\square  \, & \, \square  \, & \, \blacksquare  \, & \, \square  \, & \, \square  \, & \, \square  \, & \,      \square
\end{pmatrix}
 \,, \qquad \qquad \square = 0 \,, \ \ \blacksquare = 1 \,,
\end{equation}
We additionally define $\textbf R_{\boldsymbol k \to \boldsymbol k} \equiv \mathbb I$ 
(the identity) and 
$\textbf R_{\boldsymbol k \to \boldsymbol b} \equiv 
\textbf R_{\boldsymbol k \to \boldsymbol a}^2$.{The matrices $\textbf R_{\boldsymbol k \to \boldsymbol b}$, $\textbf R_{\boldsymbol k \to \boldsymbol a}$, and $\textbf R_{\boldsymbol k \to \boldsymbol k}$ are reducible representations of elements $(231)$, $(312)$, and $(1)$ of $\mathcal{S}_3$ [see again appendix \ref{app:A}]. }
This then allows us to succinctly express the generalization of eq.~\eqref{eq:symmX} 
to the space of all possible three-pion flavors
\begin{align}
\label{eq:symmXgenIso}
\textbf X_{\textbf f{\,'}, \textbf f}(\boldsymbol k', \boldsymbol a', \boldsymbol b'; \boldsymbol k, \boldsymbol a, \boldsymbol b) &  \equiv \mathcal S[\textbf X^{(u,u)}_{\textbf f{\,'} k' \ell' m', \textbf f\, k \ell m}]  \,, \\[5pt]
& \equiv \sum_{\{\boldsymbol p_3', \boldsymbol p_1' \} \in \mathcal P_3'}  \sum_{\{\boldsymbol p_3, \boldsymbol p_1 \} \in \mathcal P_3}  \textbf R^T_{\boldsymbol k' \to \boldsymbol p_3'} \cdot \textbf X^{(u,u)}(\boldsymbol p_3', \boldsymbol p_1'; \boldsymbol p_3, \boldsymbol p_1) \cdot  \textbf R_{\boldsymbol k  \to \boldsymbol p_3 }   \,.
\end{align}

Note that the symmetrization also converts us from the index space to
the momentum coordinates 
$(\boldsymbol k', \boldsymbol a', \boldsymbol b'; 
\boldsymbol k, \boldsymbol a, \boldsymbol b)$, 
and thus leads to
the proper dependence for the three-body scattering amplitude. In
fact, the scattering amplitude does not depend on this full set of
vectors, but rather on the subset built from the eight possible
Poincaré invariants that can be built from six on-shell
four-vectors. This statement holds regardless of whether or not the
particles are identical.

 We conclude this subsection by commenting that, as for the quantization condition in eq.~\eqref{eq:genIsoQC}, the relation  \eqref{eq:MLinftygenIso} is in the basis of three-pion states labeled by individual pion flavors. The conversion to definite three-pion isospin, and the resulting block diagonalization, will be addressed in section~\ref{sec:blockdiagM3}.

\subsection{Block diagonalization in isospin: quantization condition}
\label{sec:blockdiag}

We now project the above expressions onto definite two- and three-pion isospin. To achieve this we require a matrix $\mathcal C$ such that
\begin{equation}
 \left (
\begin{array}{c}
\!\!\!{\phantom{\big\rangle}}_{3\!} \big \langle (\pi \pi)_2 \pi \big \vert \\[5pt]
\!\!\!{\phantom{\big\rangle}}_{2\!} \big \langle (\pi \pi)_2 \pi  \big \vert \\[5pt]
\!\!\!{\phantom{\big\rangle}}_{2\!} \big \langle \rho \pi \big \vert \\[5pt]
\!\!\!{\phantom{\big\rangle}}_{1\!} \big \langle (\pi \pi)_2\pi  \big \vert \\[5pt]
\!\!\!{\phantom{\big\rangle}}_{1\!} \big \langle \rho \pi  \big \vert \\[5pt]
\!\!\!{\phantom{\big\rangle}}_{1\!} \big \langle \sigma \pi  \big \vert \\[5pt]
\!\!\!{\phantom{\big\rangle}}_{0\!} \big \langle \rho \pi \big \vert \\[5pt]
\end{array}
\right )
=  \mathcal C \cdot  \left (
\begin{array}{c}
\big \langle {\pi}_{-} \,, \, {\pi}_{0} \,, \, {\pi}_{+} \big \vert  \\[5pt]
\big \langle {\pi}_{0} \,, \, {\pi}_{-} \,, \, {\pi}_{+} \big \vert \\[5pt]
\big \langle {\pi}_{-} \,, \, {\pi}_{+} \,, \, {\pi}_{0} \big \vert \\[5pt]
\big \langle {\pi}_{0} \,, \, {\pi}_{0} \,, \, {\pi}_{0} \big \vert \\[5pt]
\big \langle {\pi}_{+} \,, \, {\pi}_{-} \,, \, {\pi}_{0} \big \vert \\[5pt]
\big \langle {\pi}_{0} \,, \, {\pi}_{+} \,, \, {\pi}_{-} \big \vert \\[5pt]
\big \langle {\pi}_{+} \,, \, {\pi}_{0} \,, \, {\pi}_{-} \big \vert
\end{array}
\right ) \,,
\label{eq:Cdef}
\end{equation}
where the subscripts on the bras on the left-hand side indicate the
total isospin, $I_{\pi\pi\pi}$,
and we have indicated the isospin of the first two pions with the shorthand
 $(\pi \pi)_2$ for $I_{\pi \pi} = 2$, 
$\rho$ for $I_{\pi \pi} = 1$ and $\sigma$ for $I_{\pi \pi} = 0$. 
This notation and some related results are discussed further in appendix~\ref{app:A}.
A simple exercise using Clebsch-Gordon coefficients shows that the 
result is given by the orthogonal matrix
 \begin{equation}
\mathcal C =  \left(
\begin{array}{ccccccc}
 \frac{1}{\sqrt{10}} & \frac{1}{\sqrt{10}} & \frac{1}{\sqrt{10}} & \sqrt{\frac{2}{5}} & \frac{1}{\sqrt{10}} & \frac{1}{\sqrt{10}} & \frac{1}{\sqrt{10}} \\
 -\frac{1}{2} & -\frac{1}{2} & 0 & 0 & 0 & \frac{1}{2} & \frac{1}{2} \\
 -\frac{1}{2 \sqrt{3}} & \frac{1}{2 \sqrt{3}} & -\frac{1}{\sqrt{3}} & 0 & \frac{1}{\sqrt{3}} & -\frac{1}{2 \sqrt{3}} & \frac{1}{2 \sqrt{3}} \\
 \frac{\sqrt{\frac{3}{5}}}{2} & \frac{\sqrt{\frac{3}{5}}}{2} & -\frac{1}{\sqrt{15}} & -\frac{2}{\sqrt{15}} & -\frac{1}{\sqrt{15}} & \frac{\sqrt{\frac{3}{5}}}{2} & \frac{\sqrt{\frac{3}{5}}}{2} \\
 \frac{1}{2} & -\frac{1}{2} & 0 & 0 & 0 & -\frac{1}{2} & \frac{1}{2} \\
 0 & 0 & \frac{1}{\sqrt{3}} & -\frac{1}{\sqrt{3}} & \frac{1}{\sqrt{3}} & 0 & 0 \\
 -\frac{1}{\sqrt{6}} & \frac{1}{\sqrt{6}} & \frac{1}{\sqrt{6}} & 0 & -\frac{1}{\sqrt{6}} & -\frac{1}{\sqrt{6}} & \frac{1}{\sqrt{6}} \\
\end{array}
\right) \,.
\label{eq:Cnumerical}
\end{equation}
The block-diagonalized finite-volume correlator is then given by
\begin{equation}
\mathcal C \cdot \textbf C_L(P) \cdot \mathcal C^T = \mathcal C \cdot  \bigg [ \textbf  C_\infty(P)   -    \textbf A_3' \textbf F_3 \frac{1}{1 - \textbf K_{\df,3} \textbf F_3} \textbf A_3 \bigg ] \cdot \mathcal C^T \,.
\end{equation}

To further reduce these expressions one can insert $\mathcal C^{T} \cdot \mathcal C = 1$ between all adjacent factors, so that every matrix is replaced according to $\textbf X \to \mathcal C \cdot \textbf X \cdot \mathcal C^T$. This transformation block diagonalizes $\textbf F$, $\textbf K_2$, $\textbf G$ and $\textbf K_{\df,3}$ so that the final quantization condition factorizes into four results, one each for the four possibilities of total three-pion isospin, $I_{\pi \pi \pi }=0,1,2,3$. For example, starting with eq.~\eqref{eq:Ggendef} above,  one finds
(with blank entries vanishing)
 \begin{equation}
\mathcal C \cdot \textbf G \cdot \mathcal C^T =  i \frac{1}{2 \omega L^3}   G \begin{pmatrix}
1 &  &  &  &  &  &  \\
  & -\frac{1}{2} & - \frac{\sqrt{3}}{2} &  &  &  &  \\
  &- \frac{\sqrt{3}}{2} & \frac{1}{2} &  &  &  &  \\
  &  &  & \frac{1}{6} &   \frac{\sqrt{15}}{6} & \frac{\sqrt{5}}{3} &  \\
  &  &  &  \frac{\sqrt{15}}{6} & \frac{1}{2} & - \frac{1}{\sqrt{3}} &  \\
  &  &  & \frac{\sqrt{5}}{3} & - \frac{1}{\sqrt{3}} & \frac{1}{3} &  \\
  &  &  &  &  &  & -1 \\
\end{pmatrix} \,.   \end{equation}
 We introduce the shorthand $\textbf G^{[I]}$ to indicate the block within $\mathcal C \cdot \textbf G \cdot \mathcal C^T$ corresponding to a given total isospin. See table~\ref{tab:QCsummary} for the explicit definitions. {It is interesting to note that $\textbf G^{[3]}$, $\textbf G^{[0]}$, and $\textbf G^{[2]}$ each correspond to the element $(13)$, as it is defined, respectively, in the trivial, sign and standard irreps of $\mathcal{S}_3$. In addition $\textbf G^{[1]}$ is this same element in a reducible representation, the direct sum of the trivial and the standard irreps.}
  
  For the two-particle K matrix, $\textbf K_2$, the change of basis gives an exact diagonalization, with each total-isospin block populated by the possible two-pion subprocesses, as illustrated in fig.~\ref{fig:pions}. 
The quantity $\textbf F$ is trivial under the change of basis, since it is proportional to the identity matrix. 
 Finally, the exchange properties of the pions within $\textbf K_{\df,3}$ 
 (which are the same as those of $\textbf M_{3,L}$ and $\textbf M_3$)
 are enough to show that it too block diagonalizes, but now with all elements non-zero 
 in a given total-isospin sector. We conclude that the quantization condition
  divides into four separate relations, compactly represented by 
 adding superscripts $[I]$ to all quantities. 
 The resulting forms of $\textbf K_2^{[I]}$ and $\textbf F^{[I]}$ as well as the 
 corresponding quantization conditions,
  are summarized in table~\ref{tab:QCsummary}.
 One noteworthy result is the change in the sign of the $G$ term for
$I_{\pi\pi\pi}=0$ compared to that for $I_{\pi\pi\pi}=3$, which is a consequence of
the antisymmetry of the isospin wavefunction in the former case.

 \begin{table}
 \setlength\extrarowheight{10pt}
 \begin{equation*}
 \boxed{\ \ \ 
\text{det} \big [1 - \textbf K^{[I]}_{\df,3}(E^\star) \, \textbf  F^{[I]}_3(E, \boldsymbol P, L) \big ] =0 \ \ \ 
}
\end{equation*}\\[-15pt]
  \begin{equation*}
  \boxed{ \ \ \ 
\textbf  F^{[I]}_3 \equiv \frac{\textbf F^{[I]}}{3} + \textbf F^{[I]}   \frac{1}{1 - \textbf M^{[I]}_{2,L} \textbf G^{[I]} } \textbf M^{[I]}_{2,L}  \textbf F^{[I]}  \ \  \ \ \ \ \ \  \textbf M^{[I]}_{2,L} \equiv \frac{1}{\textbf K_2^{[I]-1} - \textbf F^{[I]}} \ \ \  }
\end{equation*}
 \begin{center}
 \begin{tabular}{c | c | c | c }
 $I$ & $\textbf F^{[I]}$ & $\textbf K_2^{[I]}$ & $\textbf G^{[I]}$   \\ \hline \hline 3 & $\dfrac{i F}{2 \omega L^3} $ & $i [2 \omega L^3] \mathcal K_{(\pi \pi)_2}$ & $ i \dfrac{1}{2 \omega L^3}   G$ \\[10pt]
2 & $\dfrac{i F}{2 \omega L^3} {  \setlength\extrarowheight{0pt} \! \begin{pmatrix} 1 & 0 \\[0pt] 0 & 1 \end{pmatrix}  }$ & $i [2 \omega L^3]  \!  {  \setlength\extrarowheight{0pt}  \begin{pmatrix} \mathcal K_{(\pi \pi)_2} & 0 \\ 0 & \mathcal K_\rho \end{pmatrix}} $   &  $ i \dfrac{1}{2 \omega L^3}   G {  \setlength\extrarowheight{0pt} \begin{pmatrix}   -\frac{1}{2} &- \frac{\sqrt{3}}{2}   \\
   - \frac{\sqrt{3}}{2} & \frac{1}{2}  
\end{pmatrix}} $ \\[12pt]
 1& $\dfrac{i F}{2 \omega L^3} \!  {  \setlength\extrarowheight{0pt}  \begin{pmatrix}   1 \ & \ 0 \ & \ 0 \ \\   0 \ & \ 1 \ & \ 0 \ \\   0 \ &  \ 0 \ & \ 1 \ \end{pmatrix} } $ & $i [2 \omega L^3]  \!  {  \setlength\extrarowheight{0pt}  \begin{pmatrix} \mathcal K_{(\pi \pi)_2} & 0 & 0 \\ 0 & \mathcal K_\rho & 0 \\ 0 & 0 & \mathcal K_\sigma \end{pmatrix} } $   &    {  \setlength\extrarowheight{0pt} $  i \dfrac{1}{2 \omega L^3}   G  \begin{pmatrix} \frac{1}{6} &   \frac{\sqrt{15}}{6} & \frac{\sqrt{5}}{3}    \\
   \frac{\sqrt{15}}{6} & \frac{1}{2} & -\frac{1}{\sqrt{3}}    \\
   \frac{\sqrt{5}}{3} & -\frac{1}{\sqrt{3}} & \frac{1}{3} \end{pmatrix}$ } \\[15pt]
0 & $\dfrac{i F}{2 \omega L^3}$ &  $i [2 \omega L^3] \mathcal K_\rho$  & $-  i \dfrac{1}{2 \omega L^3}   G$
\end{tabular}
\end{center}
\caption{Summary of quantization conditions for all allowed values
of the total isospin $I=I_{\pi\pi\pi}$.
\label{tab:QCsummary}}
\end{table}

\subsection{Block diagonalization in isospin: relation to $\textbf M_3$ \label{sec:blockdiagM3}}

To conclude our construction of the general isospin formalism, it remains only to express the relations between $\textbf K_{\df,3}$ and the scattering amplitude, $\textbf M_3$, described in section~\ref{sec:genKtoM}, in the definite-isospin basis. Exactly as with the quantization condition, the approach is to  left- and right-multiply the finite-volume correlator, $\textbf M_{3,L}(P)$, by $\mathcal C$ and $\mathcal C^T$ respectively
\begin{multline}
\mathcal C \cdot \textbf M_{3,L}(P) \cdot \mathcal C^T   = \mathcal C \cdot \mathcal S \Big [  \textbf M^{(u,u)}_{3,L}(P)  \Big  ]  \cdot \mathcal C^T   = \sum_{\{\boldsymbol p_3', \boldsymbol p_1' \} \in \mathcal P_3'}  \sum_{\{\boldsymbol p_3, \boldsymbol p_1 \} \in \mathcal P_3}   \\[5pt]
\times \mathcal C \cdot  \textbf R^T_{\boldsymbol k' \to \boldsymbol p_3'} \cdot \mathcal C^T \cdot \mathcal C \cdot \textbf M_L^{(u,u)}(\boldsymbol p_3', \boldsymbol p_1'; \boldsymbol p_3, \boldsymbol p_1) \cdot \mathcal C^T \cdot \mathcal C \cdot  \textbf R_{\boldsymbol k  \to \boldsymbol p_3 } \cdot \mathcal C^T \,.
\end{multline}
One can then verify that the change of basis block diagonalizes the various $\textbf R_{\boldsymbol k \to \boldsymbol p_3}$ as well as $\textbf M_L^{(u,u)}$. In other words, the symmetrization does not mix the different total isospin so that we can write
\begin{equation}
\textbf M^{[I]}_{3,L}(P) =   \sum_{\{\boldsymbol p_3', \boldsymbol p_1' \} \in \mathcal P_3'}  \sum_{\{\boldsymbol p_3, \boldsymbol p_1 \} \in \mathcal P_3}  
   \textbf R^{[I]\,T}_{\boldsymbol k' \to \boldsymbol p_3'}   \, \textbf M_L^{[I](u,u)}(\boldsymbol p_3', \boldsymbol p_1'; \boldsymbol p_3, \boldsymbol p_1) \,  \textbf R^{[I]}_{\boldsymbol k  \to \boldsymbol p_3 } \,,
\end{equation}
where each object on the right-hand is reached by identifying a specific block after the change of basis. The symmetrizing matrices are defined as follows: $\textbf R_{\boldsymbol k' \to \boldsymbol k'} = \textbf R_{\boldsymbol k \to \boldsymbol k} = \mathbb I$, $\textbf R_{\boldsymbol k \to \boldsymbol b}^{[I]} = \textbf R_{\boldsymbol k' \to \boldsymbol b'}^{[I]} = \big ( \textbf R_{\boldsymbol k \to \boldsymbol a}^{[I]} \big )^2$, and $\textbf R_{\boldsymbol k \to \boldsymbol a}^{[I]} = \textbf R_{\boldsymbol k' \to \boldsymbol a'}^{[I]}$ are given in table~ \ref{tab:KtoMsummary}. {For $I_{\pi\pi\pi}=0,\,2$, and $3$,  $\textbf R_{\boldsymbol k \to \boldsymbol a}^{[I]}$ coincides with the element $(312)$ in the irreps of $\cS_3$, see eqs.~\eqref{eq:perm1d} and \eqref{eq:perm2d}.     }

To conclude we only need the isospin specific definitions for the building blocks entering $ \textbf M^{[I]}_{3,L}(P)$. These are natural generalizations of eqs.~\eqref{eq:eq:M3LuudefgenIso}-\eqref{eq:RdefgenIso} but we repeat the expressions here for convenience: 
\begin{align}
 \textbf M^{[I](u,u)}_{3,L}(P) & \equiv     \textbf D^{[I](u,u)} +   \textbf L^{[I](u)}_L    \frac{1}{1 - \textbf K^{[I]}_{\df,3} \textbf F_3^{[I]}}     \textbf K^{[I]}_{\df,3}    \textbf R^{[I](u)}_L \,,
\end{align}
 where
 \begin{align}
 \begin{split}
 \textbf D^{[I](u,u)}_{\text{disc}}   &\equiv   {\big (}\textbf  F^{[I]}{ \big )}^{-1}  \bigg [ \frac{\textbf F^{[I]}}{3} +   \textbf F^{[I]}     \textbf M_{2,L} \textbf F^{[I]}      \bigg ]  {\big (}\textbf  F^{[I]}{ \big )}^{-1} \,,  \\[5pt]\textbf D^{[I](u,u)}    &\equiv \textbf {\big (}\textbf  F^{[I]}{ \big )}^{-1}  \textbf F_3^{[I]}  {\big (}\textbf F^{[I]}{ \big )}^{-1} - \textbf D^{[I](u,u)}_{\text{disc}} \,, \\[5pt]
  \textbf L^{[I](u)}_L   &\equiv  {\big (}\textbf  F^{[I]}{ \big )}^{-1} \textbf F_3^{[I]} \,,  \\[5pt]
   \textbf R^{[I](u)}_L  &\equiv \textbf F_3^{[I]} {\big (} \textbf F^{[I]}{ \big )}^{-1}\,.  
 \end{split}
 \end{align}
\begin{table}
 \setlength\extrarowheight{10pt}
 \begin{center}
 \begin{tabular}{c | c }
 $I$  & $\textbf R_{\boldsymbol k \to \boldsymbol a}^{[I]}$   \\ \hline \hline  \\[-30pt]
3  &  1 \\[0pt]
2  &  $  {  \setlength\extrarowheight{0pt}
\begin{pmatrix}
 -\frac{1}{2} & -\frac{\sqrt{3}}{2} \\
\frac{\sqrt{3}}{2} & -\frac{1}{2} 
 \end{pmatrix}
} $ \\[15pt]
 1 &    {  \setlength\extrarowheight{0pt} $    \begin{pmatrix} 
 \frac{1}{6} & \frac{\sqrt{\frac{5}{3}}}{2} & \frac{\sqrt{5}}{3} \\
- \frac{\sqrt{\frac{5}{3}}}{2} & -\frac{1}{2} & \frac{1}{\sqrt{3}} \\
 \frac{\sqrt{5}}{3} & -\frac{1}{\sqrt{3}} & \frac{1}{3} \end{pmatrix}$ } \\[0pt]
0 &   1
\end{tabular}
\end{center}
\caption{Summary of the symmetrization matrices entering the relation between the scattering amplitude and $\textbf K_{\df,3}^{[I]}$.
\label{tab:KtoMsummary}}
\end{table} 
\section{Parametrization of $\textbf K_{\df,3}$ in the different isospin channels}
\label{sec:param}

In order to use the quantization condition detailed in the previous section, $\Kdf$ must be parametrized in a manner that is consistent
with its symmetries. In the ideal situation, only a few free parameters will be needed
describe $\Kdf$ in the kinematic range of interest, such that one can overconstrain the system with many finite-volume energies and thereby extract reliable predictions for the three-particle scattering amplitude. There are two regimes in which this is expected
to hold: near the three-particle threshold and in the vicinity of a three-particle resonance. 
In this section we describe the parametrizations in these two regimes.

An important property of $\Kdf$ that has been left implicit heretofore is that it can
be chosen real.\footnote{%
This assumes that, as is the case for QCD,
 the underlying theory is invariant under T, or equivalently CP,
so that coupling constants in the effective field theory can be chosen to be real.}
This applies when $\Kdf$ is expressed as a function of
momenta, using Eqs.~(\ref{eq:Xuumomfunc}) and (\ref{eq:symmX}),
rather than in the $\{k\ell m\}$ basis.\footnote{%
In the $\{k\ell m\}$ basis, $\Kdf$ becomes complex due to the spherical
harmonics in the decomposition (\ref{eq:Xuumomfunc}). This applies also
to $F$, $G$ and $\mathcal K_2$. The key point, however, is that each of these objects, and thus any symmetric product built from them, is a hermitian matrix on the $\{k \ell m\}$ space. The determinant of any such matrix, in particular the determinant defining the quantization condition, must then be a real function. Similarly, since $\mathcal M^{(u,u)}_{3,L}$ is hermitian, one recovers a real function upon contracting this with spherical harmonics.
This subtlety be avoided by using real spherical harmonics, as we do in our
numerical implementation below.
}
The reality of $\Kdf$ in the case of identical scalars arises
in the derivation of Ref.~\cite{\LtoK} from the use of the PV prescription to define
integrals over poles.
The same argument applies here, except that, in addition, one must choose
the relative phases between different flavor channels to be real.
This additional condition is relevant for the multichannel cases, $I=1$ and $2$.

\subsection{Threshold expansion of $\textbf K_{\df,3}$}
\label{sec:threshold}

Although in the discussion above $\textbf K_{\df,3}$ appears in the finite-volume quantization condition, it is
important to remember that it is an infinite-volume quantity. 
In addition, like the physical scattering amplitude, it is a Poincare-invariant function
(equivalently a Lorentz-invariant and momentum-conserving function)
of the six on-shell momenta.
It also inherits from $\textbf M_3$ invariance under the simultaneous exchange of particle species and momenta in both the initial and final state,
as well symmetry under charge conjugation (C), parity (P) and time-reveral (T)
transformations~\cite{\BHSQC}.

To make this final point clear it is useful to introduce $\mathcal K_{\df,3}$ (representing here a generic entry of the flavor matrix $\textbf K_{\df,3}$) as a function of six three-vectors, in direct analogy to the left-hand side of eq.~\eqref{eq:symmXgenIso}.
Working in the basis of definite individual pion flavors allows us to readily express the consequences of various symmetries. For example, the exchange symmetry can be written as
\begin{multline}
\mathcal K_{\text{df},3;  [\pi^+ \pi^0 \pi^- \ \leftarrow \ \pi^+ \pi^0 \pi^-]}(\boldsymbol p_1', \boldsymbol p_2', \boldsymbol p_3'; \boldsymbol p_1, \boldsymbol p_2, \boldsymbol p_3) = \\ \mathcal K_{\text{df},3;  [\pi^+ \pi^0 \pi^- \ \leftarrow \ \pi^+ \pi^- \pi^0 ]}(\boldsymbol p_1', \boldsymbol p_2', \boldsymbol p_3'; \boldsymbol p_1, \boldsymbol p_3, \boldsymbol p_2) \,,
\label{eq:exchange1}
\end{multline}
where we have swapped the second and third species and momenta on the 
in-state.\footnote{%
This property may seem obvious, but we stress that it does not hold for individual Feynman diagrams. Because the definition for $\Kdf$ is built up diagrammatically,
the exchange invariance does not hold for various 
intermediate quantities entering the original derivation and only emerges in the final definition.
This point is discussed in more detail in appendix~\ref{app:derivation}.}
Using T invariance then implies the following relation,
\begin{multline}
\mathcal K_{\text{df},3;  [\pi^+ \pi^0 \pi^- \ \leftarrow \ \pi^+ \pi^0 \pi^-]}(\boldsymbol p_1', \boldsymbol p_2', \boldsymbol p_3'; \boldsymbol p_1, \boldsymbol p_2, \boldsymbol p_3) = \\ \mathcal K_{\text{df},3;  [\pi^+ \pi^0 \pi^- \ \leftarrow \ \pi^+ \pi^0 \pi^-]}(-\boldsymbol p_1, -\boldsymbol p_2, -\boldsymbol p_3; -\boldsymbol p_1', -\boldsymbol p_2', -\boldsymbol p_3')\,.
\end{multline}
Combining with parity implies that
$\mathcal K_{\df,3}$ is unchanged when the initial- and final-state momenta triplets are swapped:
\begin{multline}
\mathcal K_{\text{df},3;  [\pi^+ \pi^0 \pi^- \ \leftarrow \ \pi^+ \pi^0 \pi^-]}(\boldsymbol p_1', \boldsymbol p_2', \boldsymbol p_3'; \boldsymbol p_1, \boldsymbol p_2, \boldsymbol p_3) = \\ \mathcal K_{\text{df},3;  [\pi^+ \pi^0 \pi^- \ \leftarrow \ \pi^+ \pi^0 \pi^-]}(\boldsymbol p_1, \boldsymbol p_2, \boldsymbol p_3; \boldsymbol p_1',  \boldsymbol p_2',  \boldsymbol p_3')\,.
\end{multline}
This result holds  for all theories that are PT invariant.

As proposed in ref.~\cite{\BHSnum}, and worked out in ref.~\cite{\dwave} for three identical bosons,
one can expand $\Kdf$ (which in the present case is replaced with the matrix $\textbf K_{\df,3}$) about the three-particle threshold in a consistent fashion,
and use the symmetries to greatly restrict the number of terms that appear.
The results of ref.~\cite{\dwave} apply to the $I_{\pi\pi\pi}=3$ three-pion system; here we generalize
them to the $I_{\pi\pi\pi}=0$, $1$ and $2$ channels. The new feature is the need to include isospin indices
in the particle interchange transformations.

For the parametrizations, we use the same building blocks as in ref.~\cite{\dwave},
\begin{equation}
    \Delta \equiv \frac{s - 9m^2}{9m^2}\,, \ \
    \Delta_i \equiv \frac{s_{jk} - 4m^2}{9m^2} \,, \ \
     \quad \Delta_i' \equiv \frac{s_{jk}' - 4m^2}{9m^2}\,, \ \
    \widetilde{t}_{ij} \equiv \frac{t_{ij}}{9m^2}\,,
    \label{eq:Deltas}
\end{equation}
with generalized Mandelstam variables defined as
\begin{equation}
    s \equiv  E^2 \,, \ \
    s_{ij} \equiv (p_i+p_j)^2=s_{ji}, \ \ s_{ij}' \equiv (p_i'+p_j')^2 =s'_{ji}\,, \ \ 
    t_{ij} \equiv (p_i-p_j')^2\,.
    \label{eq:Mandelstams3}
\end{equation}
The power counting scheme for the expansion will be
\begin{equation}
\Delta \sim \Delta_{ij} \sim  \Delta'_{ij} \sim \wt{t}_{ij}  \,. \end{equation}
As discussed in ref.~\cite{\dwave}, only eight of the sixteen quantities
in eq.~(\ref{eq:Deltas}) are independent---the overall CMF energy, and seven
angular variables. The relations between the quantities will be used to
simplify the threshold expansions.

In the following, we work out the leading two or three terms
in the parametrizations of $\textbf K_{\df,3}$ in each of the isospin channels. 
A summary of key aspects of the results is given in table~\ref{tab:thrsumm}.
The presence of even or odd values of $\ell$ is determined by whether
the states in the isospin decomposition are given by $\ket{(\pi\pi)_2\pi}$ and $\ket{\sigma \pi}$,
leading to even angular momentum in the first two pions, or else
$\ket{\rho\pi}$, leading to odd angular momenta.\footnote{We stress that the notation $\ket{\rho\pi}$ indicates only that the first two pions are
combined into an isotriplet. This implies that their relative angular momentum must be odd,
but does not restrict the pions to be in a $p$-wave.}
The fact that only small values of angular momentum appear in the table ($\ell,\ell'\le 2$)
is due to our consideration of only the lowest few terms in the threshold expansion. 
Only a few cubic-group irreps appear for the same reason.
All values of $\ell$ and $\ell'$, as well as all cubic-group irreps, will appear at some order
in the expansion.

\begin{table}[h!]
\centering
\begin{tabular}{c c c c}
$I_{\pi\pi\pi}$ & term & $(\ell',\ell)$ & irreps  \\ \hline\hline
3 & $\Kiso$ & $(0,0)$ & $A_1^-$ \\ \hline
3 & $\cK_{\df,3}^{(2,A)}$\vphantom{\bigg(} & $(0,0), (0,2),(2,0)$ & $A_1^-$ \\ \hline
3 & $\cK_{\df,3}^{(2,B)}$\vphantom{\bigg(} & $(0,0), (0,2), (2,0), (2,2)$ &$A_1^-, E^-,T_2^-, T_1^+$ \\ \hline
0 & $\cK_{\df,3}^{(\rm AS)}$ \vphantom{\bigg(}& (1,1) & $T_1^-, T_1^+$ \\ \hline
0 & $\cK_{\df,3}^{(\rm AS,2)}$ \vphantom{\bigg(}& (1,1) & $T_1^-$ \\ \hline
2 & $\cK_{\df,3}^{T}$
&  \vphantom{$\begin{pmatrix} 0 \\ 0 \\ 0 \end{pmatrix}$} $\begin{pmatrix} (0,0) & (0,1)\\(1,0) & (1,1) \end{pmatrix}$ &$A_1^-, T_1^+$ \\ \hline
2 & $\cK_{\df,3}^{T,2}$ 
&  \vphantom{$\begin{pmatrix} 0 \\ 0 \\ 0 \end{pmatrix}$} $\begin{pmatrix} (0,0) & (0,1)\\(1,0) & (1,1) \end{pmatrix}$ &$A_1^-$ \\ \hline
2 & $\cK_{\df,3}^{T,3} $
&  \vphantom{$\begin{pmatrix} 0 \\ 0 \\ 0 \end{pmatrix}$} $\begin{pmatrix} (0,0),(0,2),(2,0)\ &\ (0,1),(2,1)\\(1,0),(1,2) & (1,1) \end{pmatrix}$ 
&$A_1^-, T_1^+$ \\ \hline
2 & $\cK_{\df,3}^{T,4}$ 
&  \vphantom{$\begin{pmatrix} 0 \\ 0 \\ 0 \end{pmatrix}$} $\begin{pmatrix} (0,0),(0,2),(2,0),(2,2) &\ (0,1),(2,1)\\(1,0),(1,2) & (1,1) \end{pmatrix}$ 
&$A_1^-,E^-,T_2^-, T_1^+$ \\ \hline
1 & $\cK_{\df,3}^{\rm SS}$ 
& $\begin{pmatrix} (0,0) & - & - \\ - & - & -\\ - & - & - \end{pmatrix}$ 
&$A_1^-$ \\ \hline
1 & $\cK_{\df,3}^{\rm SD}$ 
& $\begin{pmatrix} -& (0,0) & (0,1) \\ (0,0) & - & -\\ (1,0) & - & - \end{pmatrix}$ 
&$A_1^-$ \\ \hline
1 & $\cK_{\df,3}^{\rm DD}$ 
& $\begin{pmatrix} - & - & - \\ - & (0,0) & (0,1)\\ - &(1,0) & (1,1) \end{pmatrix}$ 
&$A_1^-, T_1^-$ \\ \hline
\end{tabular}
\caption{Properties of low-order terms in the threshold expansion of $\textbf K_{\df,3}$.
 The terms
are specificed by their coefficients in eqs.~(\ref{eq:termsKdf}), (\ref{eq:KdfI0}), (\ref{eq:KdfI2}),
and (\ref{eq:KdfI1}).
The values of $(\ell',\ell)$ are obtained by decomposing the expessions into the $k\ell m$
basis, following the method of ref.~\cite{\dwave}.
The matrix structure corresponds to the isospin decomposition of appendix~\ref{app:A},
which is also used in the aforementioned equations.
The final column lists the cubic-group irreps that are present 
in finite volume when one considers the rest frame, $\boldsymbol P=0$.
The superscript gives the parity, which includes the intrinsic negative parity of the three-pion state.
The irreps are determined by first working out which $J^P$ values are present,
and then subducing to the cubic group. 
Results for $I_{\pi\pi\pi}=3$ are taken from ref.~\cite{\dwave}.}
\label{tab:thrsumm}
\end{table}

\subsubsection{$I_{\pi\pi\pi}=3$}

This is the simplest channel, and has been analyzed previously
in ref.~\cite{Blanton:2019igq}, from which we simply quote the results. 
The $I_{\pi\pi\pi}=3$ state is fully symmetric in isospin, so the momentum-dependent
part of $\textbf K_{\df,3}^{[I=3]} $ must be symmetric under
particle interchanges. In the charge neutral sector, there is only a single $I_{\pi\pi\pi}=3$ state, and thus no
isospin indices are needed. $\textbf K_{\df,3}^{[I=3]} $ is therefore a function only of the momenta, and, 
through quadratic order, there are only five independent terms that can appear:
\begin{align}
    m^2 \textbf K_{\df,3}^{[I=3]}  &= \K^{\iso}
    + \KA\Delta^{(2)}_A + \KB \Delta^{(2)}_B + \mc{O}(\Delta^3)\,, 
    \label{eq:termsKdf}
    \\[5pt]
    \K^{\iso} &= \Kiso + \Kisoone\Delta + \Kisotwo\Delta^2
    \label{eq:Kisoterm}
    \\
    \Delta^{(2)}_A &= \sum_{i=1}^3 (\Delta_{i}^2 + \Delta_{i}'^{\,2}) - \Delta^2, \label{eq:K3Aterm} \\ 	
    \Delta^{(2)}_B &= \sum_{i,j=1}^3 \widetilde{t}_{ij}^{\;2} - \Delta^2 \,.
    \label{eq:K3Bterm}
\end{align}
Here $ \Kiso ,  \Kisoone,  \Kisotwo, \KA$ and $\KB$ are numerical constants. 
An extensive study of how these terms affect the finite-volume
spectrum has been performed in ref.~\cite{Blanton:2019igq}.

\subsubsection{$I_{\pi\pi\pi}=0$}

The three-pion state with $I_{\pi\pi\pi}=0$ is totally antisymmetric under the permutation of isospin indices, 
as shown explicitly by the last row of $\mathcal C$ in eq.~(\ref{eq:Cdef}). 
Thus, to satisfy the exchange symmetry exemplified by eq.~(\ref{eq:exchange1}),
the momentum-dependent part of $\textbf K_{\df,3}^{[I=0]} $ 
must also be totally antisymmetric under particle exchange,
in order that the full three-pion state remains symmetric. Again, no explicit isospin indices are needed,
as there is only one $I_{\pi\pi\pi}=0$ state.

It is straightforward to see that the leading completely antisymmetric term that can appear in the
momentum-dependent part of $\textbf K_{\df,3}^{[I=0]} $ is of quadratic order in the threshold expansion:
\begin{equation}
\textbf K_{\df,3}^{[I=0]}  \supset \KAS  \sum_{\substack{ijk \\ mnr}} \epsilon_{ijk} \epsilon_{mnr} t_{im} t_{jn}  \equiv \KAS \Delta_{\text{AS}}^{(2)} \,.
\end{equation} 
At next order two new structures arise and the full form can be written
\begin{equation}
\textbf K_{\df,3}^{[I=0]}   = \left(\KAS + \KASone \Delta  \right)   \Delta_{\text{AS}}^{(2)} 
+ \KAStwo \Delta_{\text{AS}}^{(3)} + O(\Delta^4),
\label{eq:KdfI0}
\end{equation}
with
\begin{equation}
\Delta_{\text{AS}}^{(3)} \equiv \sum_{\substack{ijk \\ mnr}} \epsilon_{ijk} \epsilon_{mnr} t_{im} t_{jn} t_{kr} \,.
\label{eq:DeltaAS3}
\end{equation}

\subsubsection{$I_{\pi\pi\pi}=2$}

As discussed in the previous section, 
and summarized in table~\ref{tab:QCsummary},
the isotensor channel involves a two-dimensional flavor space.
This space can be understood in terms of the permutation group $S_3$,
as described in appendix~\ref{app:A}.
The two isospin basis vectors, $\ket{\chi_1}_2 = \ket{(\pi\pi)_2 \pi}_2$ 
and $\ket{\chi_2}_2 = \ket{\rho \pi}_2$,
also given in eqs.~(\ref{eq:chi12}) and (\ref{eq:chi22}),
transform in the standard irrep of $S_3$.
To satisfy the exchange relations exemplified by eq.~(\ref{eq:exchange1}),
the combined transformation of isospin indices and momenta 
must lie in the trivial irrep of $S_3$.
This requires combining the isospin doublet with a
momentum-space doublet also transforming in the standard irrep.
At linear order, there are three momenta, and these decompose
into a symmetric singlet ($p_1+p_2+p_3$) and the standard-irrep doublet
\begin{equation}
{\xi_1}=\frac{1}{\sqrt{6}} ( 2p_3 -p_1-p_2)\ \ {\rm and} \ \   {\xi_2} =\frac{1}{\sqrt{2}}(  p_2 -p_1)\,.
\label{eq:xis}
\end{equation}
There is an analogous doublet, $\xi'_i$, built from final-state momenta.
 The symmetric combinations are then
\begin{align}
\ket{\psi_{\text{sym}}}  &= \xi_1 \ket{\chi_1}_2 + \xi_2 \ket{\chi_2}_2  
\equiv \begin{pmatrix} \xi_1 \\ \xi_2 \end{pmatrix} \equiv \vec \xi  \,,
\\
\ket{\psi'_{\text{sym}}} &= \xi'_1 \ket{\chi_1}_2 + \xi'_2 \ket{\chi_2}_2
\equiv \begin{pmatrix} \xi'_1 \\ \xi'_2 \end{pmatrix}\equiv \vec \xi^{\,\,\prime}\,,
\end{align}
where the last two forms introduce a convenient column vector notation.
 The leading term in $\textbf K_{\df,3}^{[I=2]} $ then becomes 
\begin{align}
\begin{split}
\textbf K_{\df,3}^{[I=2]} & \supset \cK^{\text{ST}}_{\df,3}  \ket{\psi'_{\text{sym}}} \cdot \bra{\psi_{\text{sym}}}  
\equiv  \cK_{\df,3}^{\text{ST}}   \begin{pmatrix}
 \xi'_1 \cdot \xi_1 & \xi'_1 \cdot \xi_2 \\
 \xi'_2 \cdot \xi_1 & \xi'_2 \cdot \xi_2
\end{pmatrix} 
\equiv \cK_{\df,3}^{\text{ST}} \  \vec \xi^{\,\,\prime\mu} \otimes \vec \xi_\mu
\,, \\
& \hspace{-15pt} =   \frac{\cK_{\df,3}^{\text{ST}} }{6}  \begin{pmatrix}
 ( 2p'_3 -p'_1-p'_2) \cdot  ( 2p_3 -p_1-p_2) \ \ \ & \ \ \   \sqrt{3} ( 2p'_3 -p'_1-p'_2) \cdot (  p_2 -p_1) \\
\sqrt{3}(  p'_2 -p'_1) \cdot  ( 2p_3 -p_1-p_2) \ \ \ & \ \ \ 3 (  p'_2 -p'_1) \cdot(  p_2 -p_1)
\end{pmatrix}  \,,
\label{eq:isotensorKdf} 
\end{split}
\end{align}
where $ \cK^{\text{ST}}_{\df,3}$ is a constant.
Note that this is of linear order in $\Delta$, since 
the inner products $\xi_i\cdot \xi'_j$ can  be written as linear combinations of the $t_{ij}$.
There are no terms of $\cO(\Delta^0)$.

At next order, there are three sources of contributions. First, 
one can multiply the term in eq.~(\ref{eq:isotensorKdf}) by $\Delta$.
Second,
one can build additional basis vectors transforming as doublets, but of higher order in momentum.
Third, one can form Lorentz singlets in more than one way.
We discuss the latter two issues in turn.

To proceed systematically, we begin by classifying 
objects quadratic in momenta, of the general form $p_i^\mu p_j^\nu$. 
The nine such objects contain three standard-irrep doublets:
\begin{equation}
\xi(S)_i^{\mu\nu} = \xi_i^\mu P^\nu + \mu\leftrightarrow \nu \,,\quad
\xi(A)_i^{\mu\nu} = \xi_i^\mu P^\nu - \mu\leftrightarrow \nu \,,
\end{equation}
and
\begin{equation}
\vec \xi(\bar{S})^{\mu\nu} \equiv
\left(\xi(\bar{S})_1^{\mu\nu},\xi(\bar{S})_2^{\mu\nu}\right)
= \left(\xi_2^\mu\xi_2^\nu - \xi_1^\mu \xi_1^\nu, \ \xi_1^\mu\xi_2^\nu + \xi_2^\mu\xi_1^\nu\right)\,.
\end{equation}
The latter is the standard irrep that results from the direct product of $\vec \xi$ with itself.
Each of these doublets can be combined with the isospin-space doublet to make 
fully symmetric objects out of both initial- and final-state momenta. These are then
combined as in eq.~(\ref{eq:isotensorKdf}) to give a contribution to $\Kdf$. When Lorentz
contractions are included, as discussed below, symmetric doublets ($\xi(S)$ and $\xi(\bar{S})$)
must be combined with other symmetric objects, and similarly for the antisymmetric doublet
$\xi(A)$. Taking into account also CPT symmetry, there are then four possible 
combinations, schematically given by
\begin{equation}
\xi(S)' \xi(S)\,, \ \
\xi(S)' \xi(\bar S) + \xi(\bar S)'\xi(S)\,, \ \
\xi(\bar{S})' \xi(\bar{S})\ \ {\rm and}\ \ 
\xi(A)'\xi(A)\,.
\end{equation}

Lorentz indices can be contracted in three ways:
\begin{equation}
(i) \ g_{\mu \nu} g_{\mu' \nu'}\,,\ \
(ii)\ g_{\mu \mu'} g_{\nu \nu'}\ \ {\rm and} \ \
(iii)\ \epsilon_{\mu \nu\mu'\nu'}\,.
\end{equation}
The first two can be used only for the symmetric objects, while the last two can be used for
the antisymmetric objects.
We begin with the Lorentz contractions of type $(i)$. Here it turns out that all three symmetric
combinations lead to the same result, namely the outer product
\begin{equation}
\textbf K_{\df,3}^{[I=2]}  \supset \vec \xi^{\,\,\prime(2)} \otimes \vec \xi^{\,\,(2)}\,,
\end{equation}
where
\begin{equation}
\vec \xi^{\,\,(2)} = \left(\frac{2\Delta_3-\Delta_1-\Delta_2}{\sqrt6}, \frac{\Delta_2-\Delta_1}{\sqrt{2}}\right)
\propto \left(\xi_1\cdot P, \xi_2 \cdot P\right)\,,
\label{eq:xi2}
\end{equation}
with $P=p_1+p_2+p_3=p'_1+p'_2+p'_3$.
Next we consider Lorentz contractions of type $(ii)$. Here we find only two
combinations lead to new structures, namely,
\begin{equation}
\textbf K_{\df,3}^{[I=2]}  \supset  \vec \xi(\bar S)'^{\mu\nu} \otimes \vec \xi(S)_{\mu\nu}
+ \vec \xi(S)'^{\mu\nu} \otimes \vec \xi(\bar S)_{\mu\nu}  \,,
\end{equation}
and
\begin{equation}
\textbf K_{\df,3}^{[I=2]}  \supset 
\vec \xi(\bar S)'^{\mu\nu} \otimes \vec \xi(\bar S)_{\mu\nu}\,.
\end{equation}
Finally, the contraction of type $(iii)$ leads to
\begin{equation}
\textbf K_{\df,3}^{[I=2]}  \supset \epsilon_{\mu\nu\rho\sigma} \
 \vec \xi(A)'^{\mu\nu} \otimes \vec \xi(A)^{\rho\sigma}\,,
 \end{equation}
 which vanishes identically.
 
 Thus, at this stage, we have found four terms of $\cO(\Delta^2)$.
 A further potential source of such terms is to combine contributions linear in $\xi$ with those
 cubic in $\xi'$, and {\em vice versa}. Carrying out an analysis similar to that above, we find,
 however, that all such terms can be written in terms of those already obtained.
 Thus the final form of $\textbf K_{\df,3}^{[I=2]} $ is
 \begin{multline}
\textbf K_{\df,3}^{[I=2]}  =
\left( \cK_{\df,3}^{\text{T}} +  \cK_{\df,3}^{\text{T},1}  \Delta \right)
\vec \xi^{\,\,\prime\mu} \otimes \vec \xi_\mu 
+ \cK_{\df,3}^{\text{T},2} \ \vec \xi^{\,\,\prime(2)} \otimes \vec \xi^{\,\,(2)} +
\\
+  \cK_{\df,3}^{\text{T},3} \left(
 \vec \xi(\bar S)'^{\mu\nu} \otimes \vec \xi(S)_{\mu\nu}
 + \vec \xi(S)'^{\mu\nu} \otimes \vec \xi(\bar S)_{\mu\nu} \right)
+  \cK_{\df,3}^{\text{T},4}\ \vec \xi(\bar S)'^{\mu\nu} \otimes \vec \xi(\bar S)_{\mu\nu}
+ \cO(\Delta^3)\,,
\label{eq:KdfI2}
\end{multline}
where the superscript T refers to isotensor.

\subsubsection{$I_{\pi\pi\pi}=1$}
\label{sec:thrI1}

Lastly, we consider the parametrization of $\textbf K_{\df,3}^{[I=1]}$. 
Here the isospin subspace is three-dimensional and in section \ref{sec:derivation} we used a basis with definite two-pion isospin, \begin{equation}
\left\{ \ket{(\pi \pi)_2 \pi }_1, \ \ket{\rho \pi}_1,\ \ket{\sigma \pi}_1 \right\}\,.
\end{equation}
In this section we find it convenient to use a different basis, consisting of a singlet transforming
in the trivial irrep of $S_3$ and a doublet in the standard irrep. 
The relation between bases  is shown explicitly in eqs.~(\ref{eq:chi11})--(\ref{eq:chis1})
 and, in the matrix notation that follows, we order the basis vectors such that the singlet comes first:
\begin{equation}
\left\{ \ket{\chi_s}_1, \ \ket{\chi_1}_1,\ \ket{\chi_2}_1 \right\}\,.
\label{eq:chibasis}
\end{equation}

The presence of two irreps implies a greater number of options for building a fully symmetric object.
In particular, the analysis for the symmetric singlet component is identical to that for the $I_{\pi\pi\pi}=3$ sector,
with the leading two terms being of $\cO(\Delta^0)$ and $\cO(\Delta)$, respectively.
 Combining a final-state singlet with an initial-state doublet,
an overall singlet of $\cO(\Delta)$ is obtained using the 
Lorentz-scalar doublet $\vec \xi^{\,\,(2)}$ of eq.~(\ref{eq:xi2}).
An analogous term is obtained by interchanging initial and final states. 
At this same order,  initial- and final-state doublets can be combined as in eq.~(\ref{eq:isotensorKdf}).
In total, enforcing CPT invariance, we end up with
\begin{align}
\begin{split}
\textbf K_{\df,3}^{[I=1, \vert \chi \rangle]} = \left(\cK_{\df,3}^{\text{SS} }+  \cK_{\df,3}^{\text{SS},1}\Delta\right)&
\begin{pmatrix}
1 & 0 & 0 \\
0 & 0 & 0 \\
0 & 0 & 0 
\end{pmatrix}
+\cK^{\text{SD}}_{\df,3}
\begin{pmatrix}
0 & \xi^{(2)}_1 & \xi^{(2)}_2  \\
\xi'^{(2)}_1  & 0 & 0 \\
\xi'^{(2)}_2 & 0 & 0 
\end{pmatrix}  \\&
+\cK^{\text{DD}}_{\df,3}  
\begin{pmatrix}
0 & 0 & 0 \\
0 & \xi'_1 \cdot \xi_1 & \xi'_1 \cdot \xi_2 \\
0 & \xi'_2 \cdot \xi_1 & \xi'_2 \cdot \xi_2
\end{pmatrix} + O(\Delta^2)\,,
\end{split}
\label{eq:KdfI1}
\end{align}
where the $\vert \chi \rangle$ superscript on the left-hand side emphasizes that we are using the new basis, introduced in \eqref{eq:chibasis}. The SS and DD superscripts on the right-hand side refer to singlet and doublet irreps.

\subsection{Three-particle resonances}
\label{sec:resonance}

The threshold expansion derived in the previous section plays a similar role for three-particle
interactions as the effective-range expansion does for the two-particle K matrix. It provides a
smooth parametrization of the interaction, valid for some range around threshold, that respects
the symmetries. However, we expect that the convergence of the series is limited by the singularities in $\Kdf$ closest to the three-particle threshold, just as the expansion for $\cK_2$ is limited either by the nearest poles, possibly associated with a two-resonance, or else by the $t$-channel cut.  As studying three-particle
resonances is one of the major goals behind the development of the three-particle quantization condition,
it is important to determine appropriate forms of $\Kdf$ in the channels that contain such resonances.
This is the task of the present section.

We begin by listing, in table~\ref{tab:resonances}, the total $J^P$ and isospin for the resonant
channels observed in nature that couple to three pions~\cite{Tanabashi:2018oca}.
We include only cases where the coupling is allowed in isosymmetric QCD.
Resonances are present only for $I_{\pi\pi\pi}=0$ and $I_{\pi\pi\pi}=1$. We note the absence of the $J^P=0^+$,
$I_{\pi\pi\pi}=1$, $a_0(980)$, for which no three-pion coupling is possible that is simultaneously consistent
with angular momentum and parity conservation.
For each resonance, we also note the corresponding subduced cubic group irreps.
The cubic symmetry group including parity (also called the achiral or full octahedral group) defines the symmetry of the system provided that the total momentum is set to zero.
In a lattice QCD calculation, one can project the three-pion states onto definite
cubic-group irreps by choosing appropriate three-pion interpolating operators, as discussed
in appendix~\ref{app:B}.
Note that, for the values of $J^P$ arising in the table, a finite-volume irrep can always be identitifed that does not couple to any other listed values.
The final column in the table gives the lowest three-pion orbit that couples to the irrep(s) for
the corresponding state. The ordering of the orbits is described in appendix~\ref{app:B};
see in particular table~\ref{tab:restirreps}.

\begin{table}
\centering
\begin{tabular}{ccccc}
Resonance & $I_{\pi\pi\pi}$ &$J^P$ & Irrep ($\boldsymbol P =0$)  & $3\pi$ orbit \\ \hline\hline
$\omega(782)$ & 0  & $1^-$ & $T_1^-$   & 4 \\ \hline
$h_1(1170)$ & 0  & $1^+$ &  $T_1^+$  & 2 \\ \hline
$\omega_3(1670)$ & 0  & $3^-$ & $A_2^-$ & 4  \\ \hline \hline
  $\pi(1300)$ & 1  & $0^-$ &  $A_1^-$  & 1 \\ \hline
 $a_1(1260)$ & 1  & $1^+$ &  $T_1^+$ & 2 \\ \hline
 $\pi_1(1400)$ & 1  & $1^-$ &  $T_1^-$  & 4 \\ \hline
 $\pi_2(1670)$ & 1  & $2^-$ &  $E^-$ and $T_2^-$ & 2 \\ \hline
 $a_2(1320)$ & 1  & $2^+$ &  $E^+$ and $T_2^+$ & 3 \\ \hline
 $a_4(1970)$ & 1  & $4^+$ &  $A_1^+$ & 16 \\ \hline
\end{tabular}
\caption{Lowest lying resonances with negative G-parity, and which couple to three pions,
in the different isospin and $J^P$ channels. 
The fourth column shows the cubic group irreps that are subduced from the rotation group irreps, 
assuming that the resonance is at rest ($\boldsymbol P=0$).
The final column gives the lowest three-pion momentum orbit that contains the
corresponding cubic group irrep, again assuming $\boldsymbol P=0$. 
}
\label{tab:resonances}
\end{table}

In the remainder of this section we determine the forms of the entries of $\textbf K_{\df,3}$ that couple to three pions
having each of the quantum numbers listed in table~\ref{tab:resonances}. 
We stress that, as in the previous section, this is an infinite-volume exercise. When using the
resulting forms for $\textbf K_{\df,3}^{[I]}$ in the quantization condition, 
one must covert the forms given here to the $k \ell m$ index set introduced above.
This is a straightforward exercise that we do not discuss further here.

By analogy with the two-particle case, we expect that a three-particle resonance can be represented
by a pole in the part of $\textbf K_{\df,3}^{[I]}$ with the appropriate quantum numbers~\cite{\BHSnum}, i.e.
\begin{equation}
\textbf K_{\df,3}^{[I, \vert \chi\rangle]} =   \cK^X_{\df,3} \frac{ c_X}{s - M_X^2} + \mathcal O \big [ (s - M_X^2)^0 \big ]  \,, \label{eq:reso1}
\end{equation}
where the superscript $\vert \chi \rangle$ on the left-hand side emphasizes that we work in the basis of definite symmetry states for $I_{\pi \pi \pi}=1$ (see also appendix \ref{app:A}). On the right-hand side,$X$ labels the quantum numbers, 
$M_X$ is close to the resonance mass (at least in the case of narrow resonances),
the real constant $c_X$ is related to the width of the resonance,
and $\cK^X_{\df,3}$ carries the overall quantum numbers.
The precise relationship of $c_X$ and $M_X$ to the resonance parameters in $\cM_3$
is not known analytically, since determining $\cM_3$ requires solving the non-trivial integral
equations discussed above. 

We stress that, once a form for $\cK^X_{\df,3}$ is known,
only one sign of $c_X$ will lead to a resonance pole with the physical sign for the residue.
The correct choice can be identified by requiring that the finite-volume correlator $C_L$ has a single pole with
the correct residue~\cite{\BHSnum,\dwave}.
In the limit $c_X\to 0$, one recovers an additional decoupled state in the finite-volume spectrum
at energy $E=M_X$ (assuming $\boldsymbol P=0$), corresponding to a stable would-be resonance.
The form in eq.~(\ref{eq:reso1})
was proposed in ref.~\cite{\BHSnum} for the case of identical scalars (which is equivalent to
the $I_{\pi\pi\pi}=3$ channel here) for which $\cK^X_{\df,3}$ is a constant. 
As noted above, however, there are no resonances in nature in the $I_{\pi\pi\pi}=3$ or $I_{\pi\pi\pi}=2$ channels,
so the example given in ref.~\cite{\BHSnum} is for illustrative purposes only.
In the following we determine forms for $\cK^X_{\df,3}$ that can be used
for all the resonant channels listed in table~\ref{tab:resonances}.

We also enforce an additional requirement on $\cK^X_{\df,3}$, namely that it  has 
a factorized form in isospin space. This is motivated by the fact that 
the residues of resonance poles in $\cM_2$ and $\cM_3$ do factorize, and it was argued in
ref.~\cite{\BHSK} that this carries over to poles in $\cK_2$ evaluated at off-shell momenta. Here we assume that this holds
also for resonance poles in $\Kdf$. We view this as plausible, but leave the proof to future work.

Before turning to the detailed parametrizations, we comment on the range of validity for the quantization condition. All the resonances in table \ref{tab:resonances} have, in principle, additional decay channels, such as $5\pi$ or $K \bar{K}$. One must consider on a case by case basis whether neglecting these is justified, based on the the couplings between the resonance of interest to the neglected channels, as well as the target precision of the calculation. Another possibility is to work at unphysically heavy pion masses, such that some of the neglected channels are kinematically forbidden. While the procedure for including additional two-particle channels should be given by a straightforward generalization of ref.~\cite{\BHSQC}, rigorously accommodating the 5$\pi$ state would be a significant formal undertaking.

\subsubsection{Isoscalar resonances}

The symmetry requirements for the $\KdfX{X}$ are exactly as in the threshold expansion. For $I_{\pi\pi\pi}=0$, this means
complete antisymmetry under particle exchange. Useful building blocks are the following objects:
\begin{align}
V^\alpha &= P_\mu \sum_{ijk} \epsilon_{ijk}\, p_j^\mu p_k^\alpha  \ \ \ \   \xrightarrow{ \ \text{\CMF} \ }  \ \ \ \
\frac{E}{2}\left(0, -3 \omega_- \boldsymbol p_3 - \boldsymbol p\,^- [E - 3 \omega_3 ]\right)\,,
\label{eq:Vdef}\\
\begin{split}
A^\alpha &= \epsilon_{\alpha \beta \gamma \delta}\, p_1^\beta p_2^\gamma p_3^\delta  \ \ \ \ \ \ \ \hspace{2pt}
\xrightarrow{ \ \text{\CMF} \ }  \ \ \ \ E \left(0, \, \boldsymbol p_1 \times \boldsymbol p_2\right)
= E \left(0, \, \boldsymbol p_2 \times \boldsymbol p_3\right) \,, \\ & \hspace{220pt}  = E \left(0, \, \boldsymbol p_3 \times \boldsymbol p_1\right)\,,
\end{split}
\label{eq:Adef}
\end{align}
where $p^{-\mu} = p_1^\mu-p_2^\mu = (\omega^-,\boldsymbol p\,^-)$,
$p_3^\mu=(\omega_3,\boldsymbol p_3)$, etc.
The quantities
$V^\alpha$ and $A^\alpha$ are fully antisymmetric under particle exchange, and describe a vector and axial vector, respectively, as can be seen from their forms in the \CMF.
 In particular, the vanishing of the temporal components in this frame shows the absence of
scalar and pseudoscalar contributions (with the respect to the three-dimensional rotation group).

Taking into account the negative parity of the pion,
the momentum-space amplitude for the $J^P=1^-$ $\omega(782)$ to decay to three pions
must transform as an axial vector. This leads to the following form for $\Kdf$,
\begin{equation}
\cK^\omega_{\df,3} = A^{\prime\mu}  A_\mu  \,,
\label{eq:Kdfomega}
\end{equation}
where $A^{\prime\mu}$ has the same form as $A^\mu$ but expressed in terms of final-state
momenta. The expression (\ref{eq:Kdfomega}) is manifestly Lorentz and CPT invariant.
We have checked explicitly that, when reduced to the $ k \ell m$ basis
used in the quantization condition, this expression transforms purely as a $T_1^-$ under
the cubic group. 
 Indeed, it turns out to be proportional to the operator
$\Delta_{\rm AS}^{(3)}$, given in eq.~(\ref{eq:DeltaAS3}),
that arises in the threshold expansion.
Furthermore, we note from table~\ref{tab:restirreps} in appendix~\ref{app:B} that the lowest three-pion
state in a cubic box that transforms in the $T_1^-$ irrep lies in the fourth orbit and
has momenta $(1,1,0)$, $(-1,0,0)$ and $(0,-1,0)$
(or a cubic rotation thereof) in units of $2\pi/L$.
This can be understood from the fact that,
in the \CMF, $A^\mu$ vanishes if {\em any} of the three pion momenta vanish,
as  can be seen from eq.~(\ref{eq:Adef}).

These results have implications for a practical study of the $\omega$ resonance.
As is known from the study of two-particle resonances, to map out the resonant structure
(e.g.~the rapid rise in the phase shift) requires many crossings between 
the finite-volume resonance level and those of weakly-interacting multi-particle states.
Since the lowest, non-interacting three-pion state with the quantum numbers of the $\omega$
lies in the fourth orbit, it occurs at relatively high energy.
Thus for small to moderate volumes, 
the finite-volume level corresponding to the $\omega$ will be the lowest lying state 
and there will be no avoided level crossings.
Only by going to larger boxes will the level-crossings needed to constrain $\Kdf$
in detail be present.
For physical pion masses the constraint is not too strong---an avoided level crossing
requires $mL \gtrsim 4.6$. However, if working with heavier-than-physical pions,
such as in the example presented in section~\ref{sec:toy}, larger values of $mL$ are needed
($mL \gtrsim 6.5$ in the toy model).
These constraints apply, however, only in the overall rest frame.
It is likely that moving frames, for which the constraints will be relaxed, will
play an important role in any detailed investigation of the $\omega$ resonance.

For the $J^P=1^+$ $h_1(1170)$, the momentum-space
decay amplitude must transform as a vector, leading to
\begin{equation}
\cK^{h_1}_{\df,3} = V^{\prime\mu} V_\mu \,. \label{eq:Kdfh1}
\end{equation}
Only two momenta need to be nonzero for $V^\mu$ to be nonvanishing, and indeed the
lowest momentum configuration transforming as the required $T_1^+$ lies in the second orbit and has
momenta $(1,0,0)$, $(-1,0,0)$ and $(0,0,0)$ (see table~\ref{tab:restirreps}). Applying the same estimate as above based on the non-interacting energy, the first \CMF~avoided-level crossing 
for physical pion masses is already expected for $m L \gtrsim 1.8$. Thus, for all volumes where the neglected $e^{- m L}$ is a reasonable approximation (typically requiring $m L \gtrsim 4$), we expect to recover useful constraints on the $h_1$ width in all finite-volume frames.

Finally, for the $J^P=3^-$ $\omega_3(1670)$, the momentum-space amplitude must transform
as $J^P=3^+$. One possible form is
\begin{equation}
\cK^{\omega_3}_{\df,3} =  ( A_\mu A'\,^\mu  )^3 - \frac{3}{5} (A^2)(A'\,^2)(A_\mu A'\,^\mu )\,,
\end{equation}
where the second term is required to project against a $J^P=1^+$ component.
The corresponding cubic-group irrep, $A_2^-$, appears first in the same  three-pion orbit as
for the $\omega$, for then the axial current $A_\mu$ is nonzero.

\subsubsection{Isovector resonances}

We turn now to parameterizations of $\mathcal K_{\df,3}^X$ in the three-dimensional isovector case, working always in the $\chi$-basis of \eqref{eq:chibasis} [defined explicitly in eqs.~\eqref{eq:chi11}-\eqref{eq:chis1}].

Beginning with the $J^P=0^-$ $\pi(1300)$, the simplest case in this sector,
we note that these quantum numbers can be obtained from three pions at rest, so that no momentum 
dependence is required in $\KdfX{\pi}$. However, as we have seen in section~\ref{sec:thrI1},
momentum-independence is possible only for the component connecting 
permutation-group-singlets in the initial and final states. For the other components momentum
dependence is needed to obtain a form that is fully symmetric under permutations.
Using results from our discussion of the threshold expansion,
we find the following possible form\footnote{%
We stress that we are not here doing an expansion in momenta, but rather writing
a simple form that has the appropriate symmetries.
More complicated expressions consistent with the desired quantum numbers are certainly possible.}
\begin{align}
\KdfX{\pi} = 
\begin{pmatrix}
s_\pi \\
d_\pi \,\xi'^{(2)}_1 \\
d_\pi \,\xi'^{(2)}_2 \\
\end{pmatrix}
\otimes
\begin{pmatrix}
s_\pi, &  d_\pi \,\xi^{(2)}_1, & d_\pi\, \xi^{(2)}_2 
\end{pmatrix}\,.
\label{eq:Kdfpi}
\end{align}
Here $s_\pi$ and $d_\pi$ are real constants, corresponding to couplings
to the singlet and doublet components, respectively. 
The outer product structure is necessary due to the factorization of the residue at the
K-matrix pole.
We stress that the components of the two vectors in the outer product must be
Lorentz scalars in order that $\KdfX{\pi}$ couples to $J^P=0^-$.
Thus, for example, $\xi^{(2)}_1$ cannot be replaced by $\xi_1^\mu$.
We also note that we do not expect the momentum-dependent parts of this expression
to be suppressed relative to the momentum-independent ones, since we are far from threshold.

We can use the properties of the physical $\pi(1300)$ resonance to guide our expectations
concerning $s_\pi$ and $d_\pi$. In particular, the resonance has been observed to have both
$\sigma\pi$ and $\rho\pi$ final states~\cite{Tanabashi:2018oca}.
Recalling from appendix~\ref{app:A} that the
first two entries of the vector space are linear combinations of the states $\ket{(\pi\pi)_2\pi}_1$
and $\ket{\sigma\pi}_1$, while the third is $\ket{\rho \pi}_1$, we see that $s_\pi$ describes
the coupling to the former two states, while $d_\pi$ couples to all three.
Thus $d_\pi$ must be nonzero to describe the physical resonance, with its $\rho\pi$ decay,
while the importance of $s_\pi$ depends on the details of the amplitude.

Next we turn to the $J^P=1^+$ $a_1(1260)$. Taking into account the intrinsic parity of the pion,
the decay amplitude must transform as a vector. A possible form is thus
\begin{equation}
\KdfX{a_1} = g^P_{\mu\nu}
\begin{pmatrix}
s_{a_1} V_S'^\mu \\
d_{a_1} \,\xi'^\mu_1 \\
d_{a_1} \,\xi'^\mu_2 \\
\end{pmatrix}
\otimes
\begin{pmatrix}
s_{a_1} V_{S}^{\nu}, &  d_{a_1} \,\xi_{1}^{\nu}, & d_{a_1}\, \xi_{2}^{\nu} 
\end{pmatrix}\,,
\label{eq:Kdfa1}
\end{equation}
where
\begin{equation}
V_S^\nu = \xi_1^\nu \xi_1^{(2)} + \xi_2^\nu \xi_2^{(2)} \,,
\end{equation}
is a vector that is symmetric under permutations,
and
\begin{equation}
g^P_{\mu\nu} =(g_{\mu\nu}-P_\mu P_\nu/P^2) \,,
\label{eq:gP}
\end{equation}
is the projector that arises from the sum over polarizations of $\epsilon_\mu \epsilon_\nu^*$.
It projects against $P^\mu$, and in the CM frame it picks out the the spatial part, ${\boldsymbol V}_{\!\!S}$,
which transforms as a vector, while removing the $J^P=0^+$ quantity, $V_S^0$.
We are forced to use a form for $V_S^\nu$ that is cubic in momenta because the only
 symmetric vector linear in momenta is simply $P^\mu$, which vanishes when contracted
 with $g^P$.
In contrast to the form for the $\pi(1300)$, eq.~(\ref{eq:Kdfpi}), the doublet portion of 
the amplitude in eq.~(\ref{eq:Kdfa1})
has a simpler momentum-dependence than the singlet part.
The real constants $s_{a_1}$ and $d_{a_1}$ play the same role as for the 
$\pi(1300)$, and again $d_{a_1}$ must be nonzero since $\rho\pi$ and $\sigma \pi$ decays
are observed. 

Next we turn to the $J^P=1^-$ $\pi_1(1400)$. It is not possible to construct a fully symmetric
axial vector from three momenta, and thus the decay amplitude of the symmetric component
vanishes. For the doublet part, a nonzero amplitude can be obtained by combining the completely
antisymmetric axial vector $A^\mu$ [eq.~(\ref{eq:Adef})] with the doublet $\vec \xi^{\,\,(2)}$
in the appropriate manner. This leads to
\begin{equation}
\KdfX{\pi_1} = A'^\mu g^P_{\mu\nu} A^\nu \
\begin{pmatrix}
0 \\
-\xi'^{(2)}_2 \\
\xi'^{(2)}_1\\
\end{pmatrix}
\otimes
\begin{pmatrix}
0, & -\xi_{2}^{(2)}, & \xi_{1}^{(2)} 
\end{pmatrix} \,.
\label{eq:Kdfpi1}
\end{equation}

To parametrize the $J^P=2^-$ $\pi_2(1670)$ requires a tensor composed of momentum vectors,
with the appropriate symmetry properties. Using the constructions from the previous section,
we find the following form:
\begin{equation}
\KdfX{\pi_2} =  \left(g^P_{\rho\mu} g^P_{\sigma\nu}-\tfrac13 g^P_{\rho\sigma} g^P_{\mu\nu}\right)
\begin{pmatrix}
s_{\pi_2} T'^{\rho\sigma} \\
d_{\pi_2}\,\xi(\bar S)'^{\rho\sigma}_1\\
d_{\pi_2}\,\xi(\bar S)'^{\rho\sigma}_2
\end{pmatrix}
\otimes
\begin{pmatrix}
s_{\pi_2} T^{\mu\nu},
& d_{\pi_2} \,\xi(\bar S)_{1}^{\mu\nu}, 
& d_{\pi_2}\, \xi(\bar S)_{2}^{\mu\nu}
\end{pmatrix} \,,
\label{eq:Kdfpi2}
\end{equation}
where 
\begin{equation}
T^{\mu\nu} = \xi_1^\mu\xi_1^\nu + \xi_2^\mu\xi_2^\nu
\,,
\end{equation}
is a Lorentz tensor that is an $S_3$ singlet.
The tensor containing $g^P$ projects out the $J=2$ part in the CM frame.

For the $J^P=2^+$ $a_2(1320)$ we need to construct a pseudotensor from momentum vectors.
The simplest form that we have found is
\begin{equation}
\KdfX{a_2} =  \left(g^P_{\rho\mu} g^P_{\sigma\nu}-\tfrac13 g^P_{\rho\sigma} g^P_{\mu\nu}\right)
\begin{pmatrix}
s_{a_2} A'^\rho V'^\sigma \\
-d_{a_2} A'^\rho \xi'^\sigma_2\\ 
d_{a_2} A'^\rho \xi'^\sigma_1
\end{pmatrix}_{\!\!\text{sym}}
\!\!\!\!\!\!\!\!\otimes
\begin{pmatrix}
s_{a_2} A^\mu V^\nu,
&-d_{a_2} A^\mu \xi_2^\nu,
& d_{a_2} A^\mu \xi_1^\nu
\end{pmatrix}_\text{sym} \,,
\label{eq:Kdfa2}
\end{equation}
where the subscript ``sym'' indicates symmetrizing the tensors.

Finally, for the $J^P=4^+$ $a_4(1970)$, we need to construct an $\ell = 4$ pseudotensor from
momentum vectors. 
One possible form is
\begin{align}
\KdfX{a_4} &= \left(g^P_{\mu'\mu}g^P_{\nu'\nu}g^P_{\rho'\rho}g^P_{\sigma'\sigma}
-\tfrac67 g^P_{\mu'\nu'}g^P_{\mu\nu}g^P_{\rho'\rho}g^P_{\sigma'\sigma}
+\tfrac3{35} g^P_{\mu'\nu'}g^P_{\rho'\sigma'}g^P_{\mu\nu}g^P_{\rho\sigma}\right)
T'^{\mu'\nu'\rho'\sigma'}_4 \otimes T_4^{\mu\nu\rho\sigma}\,,
\label{eq:Kdfa4}
\\
T_4^{\mu\nu\rho\sigma} &=
\begin{pmatrix}
s_{a_4} (A^\mu A^\nu A^\rho V^\sigma), 
&-d_{a_4} (A^\mu A^\nu A^\rho \xi_2^\sigma),
& d_{a_4} (A^\mu A^\nu A^\rho \xi_1^\sigma)
\end{pmatrix}_{\rm sym} \,,
\\
T'^{\mu\nu\rho\sigma}_4 &=
\begin{pmatrix}
s_{a_4} (A'^\mu A'^\nu A'^\rho V'^\sigma) \\
-d_{a_4} (A'^\mu A'^\nu A'^\rho \xi'^{2\sigma})\\ 
d_{a_4} (A'^\mu A'^\nu A'^\rho \xi'^{1\sigma})
\end{pmatrix}\,.
\end{align}
An alternative form replaces two of the axial vectors with vectors (in either or both the initial and
final states).

\section{Toy model: spectrum in $I_{\pi\pi\pi}=0$ channel}
\label{sec:toy}

The goal of this section is to present an example of the implementation 
of the new quantization conditions derived in this paper.
We choose the $I_{\pi\pi\pi}=0$ channel, which is the simplest
of the new results,
since the quantization condition is one-dimensional in isospin space.
The extension of the implementation to the other channels is, however, 
straightforward.

The $I_{\pi\pi\pi}=0$ channel is of direct phenomenological relevance, 
due to the presence of two 
(relatively) light three-particle resonances, the $\omega(782)$ and the $h_1(1170)$. 
In particular, at physical pion masses, the $\omega$ lies
only slightly above the five-pion inelastic threshold, and the isospin-violating
couplings to two and four pions are weak, 
so that the three-particle quantization condition is likely to provide a good description.
Indeed, at somewhat heavier-than-physical pion masses (e.g. $M_\pi \sim 200$ MeV),
the $\omega$ should lie between the three- and five-pion thresholds.
If, in addition, one has exact isospin symmetry, 
there will be no coupling to channels with an even number of pions.
This example can thus be explored in a rigorous way using 
the quantization condition derived in this work,
and is an excellent candidate for the first lattice QCD study of a three-particle resonance.

Another feature of interest in these examples is the presence of the $\rho$ resonance in
two-particle subchannels. Although the decay $\omega\to\rho\pi$ is kinematically forbidden,
we expect, given the width of the resonance, that it will have a significant impact on the energy
levels in the vicinity of the $\omega$ mass. 
For the $h_1$, the $\rho\pi$ decay is allowed (and seen
experimentally), and thus the system provides an example in which the full complication of
cascading resonant decays, $h_1\to\rho\pi\to3\pi$, occurs.
We also note that, away from the three-particle resonance energy, the dominant effect on the
three-pion spectrum arises from pairwise interactions, and thus this spectrum provides an alternative source of information on the $\rho$ resonance. Indeed, the effect on the 
three-particle spectrum is enhanced relative to that for two pions due to the presence of three pairs.

The implementation of the isoscalar three-particle quantization condition requires only minor
generalizations of the $I_{\pi\pi\pi}=3$ case implemented previously in
refs.~\cite{Briceno:2018mlh,Blanton:2019igq,Romero-Lopez:2019qrt,Blanton:2019vdk}.
Specifically, appendices A and B of  ref.~\cite{Blanton:2019igq} provide a summary of all necessary results.
The new features here are two-fold:~(i) the expression for $F_3$ contains a relative minus sign for $G$ compared
to that for $I_{\pi\pi\pi}=3$ (see table~\ref{tab:QCsummary}), which is trivial to implement;
(ii) the angular momentum indices $\ell, m$ of the interacting pair contain only odd partial waves.
Concerning the latter point, in our illustrative example we restrict to the 
lowest allowed partial wave, namely $\ell=1$. 
While odd two-particle partial waves have not previously been implemented in
the three-particle quantization condition, this requires only a simple generalization from
the work in ref. \cite{Blanton:2019igq}, where $\ell=0$ and $2$ were considered.
In particular, we follow that work in using real spherical harmonics, and in the method of projection onto different irreps of the cubic group.

We now describe how the resonances are included in our example.
We stress at the outset that the parameters we choose are not intended to
be close to those for the physical particles, but rather are choices
that allow certain features of the resulting spectrum to be clearly seen.
For the $\rho$, we use the Breit-Wigner parametrization:
\begin{equation}
\left(\frac{k}{M_\pi}\right)^3\cot \delta_1 = \frac{M^2_\rho - E^2 }{E M_\pi} \frac{6\pi}{g^2} \frac{E^2}{M_\rho^2},
\end{equation}
with $g=1$ and $M_\rho = 2.8 M_\pi$.\footnote{Our chosen value of $M_\rho/M_\pi$ corresponds to a theory with $M_\pi \sim 320$ MeV
(see ref.~\cite{Werner:2019hxc}). Our choice of the coupling $g$ is, however, significantly
smaller than the observed value (corresponding to a narrower-than-physical decay width).}
As explained in ref.~\cite{Romero-Lopez:2019qrt}, in order for the three-particle quantization
condition to remain valid in the presence of two-particle resonances, we
must use a  modified principal value prescription.
This requires the following changes to $\wt F$ and $\wt\K_2$:
\begin{align}
[ F]_{k\ell' m';p\ell m} &\to [ F]_{k\ell' m';p \ell m} 
+ \delta_{kp} \delta_{\ell'\ell}\delta_{m' m}
 H(\boldsymbol k)  \frac{I_\PV^{(\ell)}(q_{2,k}^{\star2})}{32\pi}\,,
\label{eq:Fellshift}
\\
\big [  (  \K_2)^{-1} \big]_{k\ell' m'; p \ell m} &\to\
\big [   (\K_2)^{-1} \big]_{k\ell' m'; p \ell m} 
- \delta_{kp} \delta_{\ell'\ell}\delta_{m' m}
 H(\boldsymbol k)  \frac{I_\PV^{(\ell)}(q_{2,k}^{\star2})}{32\pi}\,,
\label{eq:Kellshift}
\end{align}
where $\ell$ and $\ell'$ are odd, and in this case $\ell=\ell'=1$. 
We find that $I_\PV^{(\ell=1)}(q) = C/q^2$, 
with $C=100 M_\pi^2$ is enough to accommodate any resonance in the region $M_\rho < 5 M_\pi$. 

For the three-particle resonances, we use the general form given in eq.~(\ref{eq:reso1})
for $\Kdf$, with the specific momentum-dependent expressions for
$\cK^\omega_{\df,3}$ and $\cK^{h_1}_{\df,3}$ given in
eqs.~(\ref{eq:Kdfomega}) and (\ref{eq:Kdfh1}), respectively.
We set $M_\omega = 4.3 M_\pi, \, M_{h_1} = 4.7 M_\pi$, and $c_\omega = 0.007, \, c_{h_1} = 0.42$. 
This choice is motivated by the hierarchy of the resonance parameters known from
experiment, i.e., $M_{h_1} > M_\omega$, $\Gamma_{h_1} > \Gamma_{\omega}$. 
We stress, however, that we do not at present know how to relate the parameters 
$c_X$ to the physical width, and that these values are chosen only for illustrative purposes.

\begin{figure}
\centering 
\subfigure[ $\omega$ channel. \label{fig:omega} ]{\includegraphics[width=.49\textwidth]{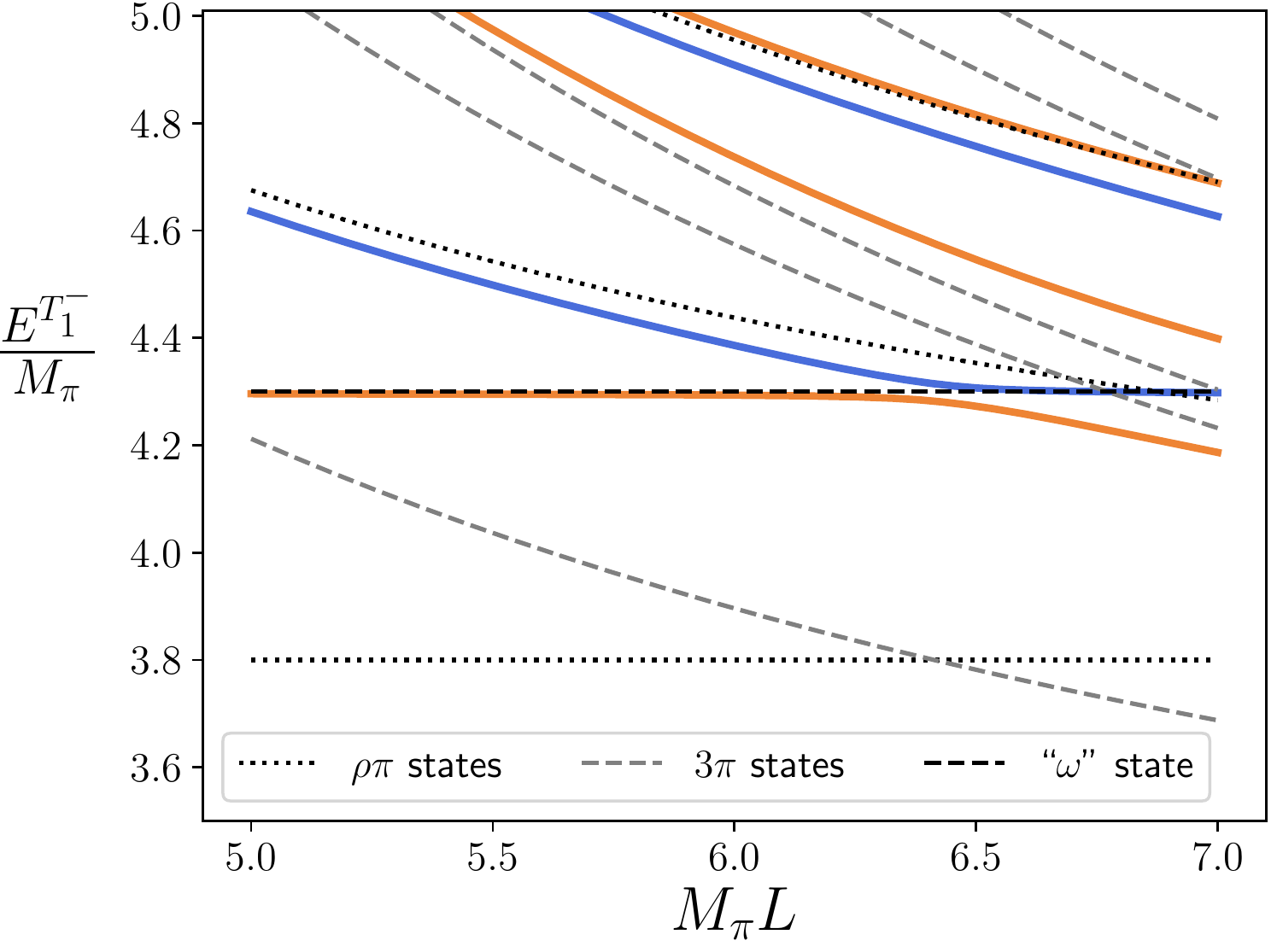}}
\hfill
\subfigure[ $h_1$ channel. \label{fig:h1} ]{\includegraphics[width=.49\textwidth]{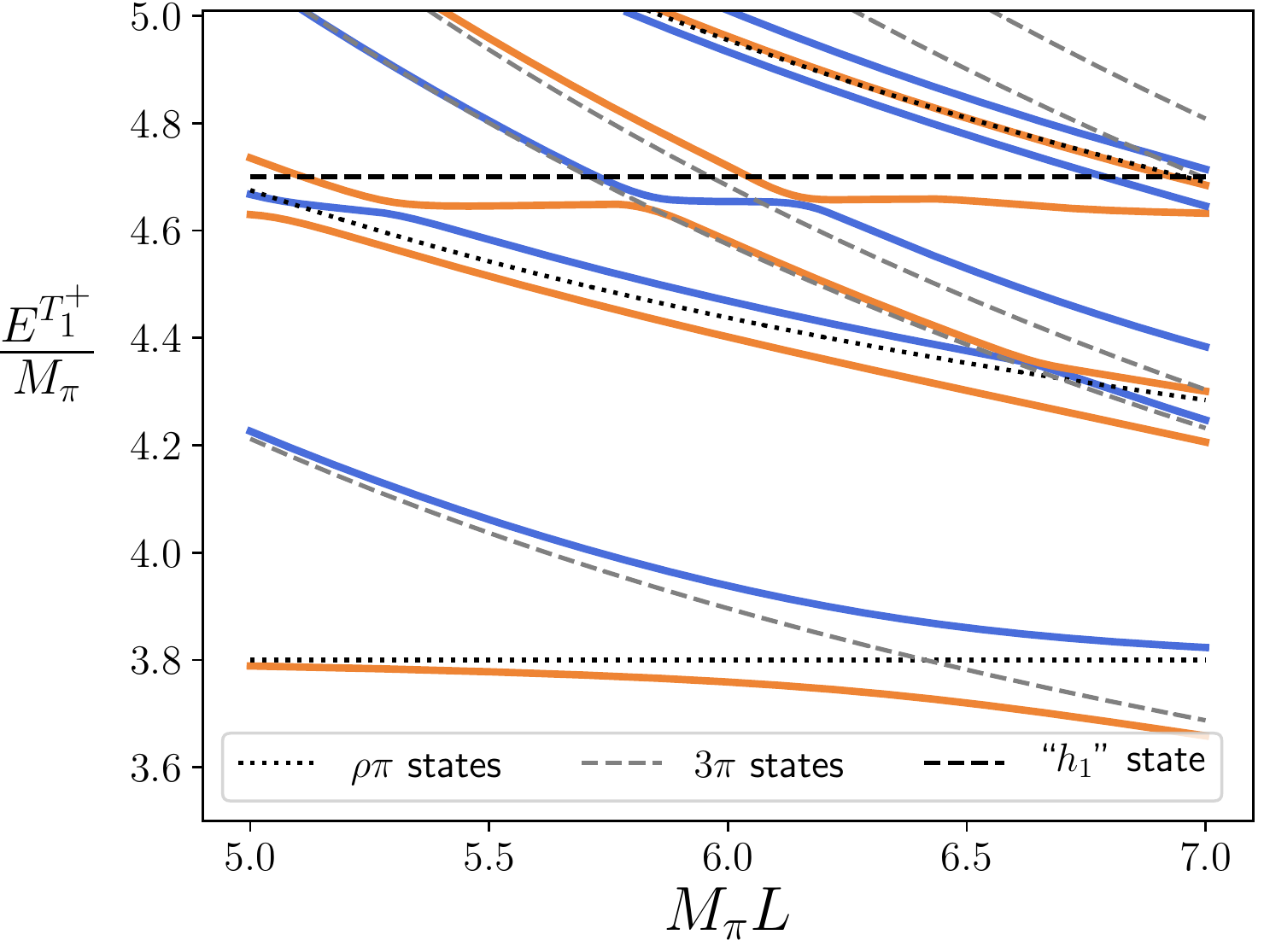}}
\hfill
\caption{\label{fig:I0spect}  Illustrative finite-volume spectra for three pions with $I_{\pi\pi\pi}=0$
and irreps (a) $T_1^-$ and (b) $T_1^+$, plotted versus $M_\pi L$. 
The interacting spectrum is shown by
solid lines, with the alternating orange and blue colors only used to distinguish adjacent
levels. Dashed and dotted grey lines show the comparison with different noninteracting levels.
The parameters used for $\cK_2$ and $\Kdf$ are described in the text.
}
\end{figure}

The resulting three-pion spectra for two different irreps, $T_1^\mp$, are shown in figure~\ref{fig:I0spect}
as a function of $M_\pi L$. 
As described in section~\ref{sec:resonance},
these irreps couple to resonances with $J^P=1^\mp$, i.e.~to the $\omega$ and $h_1$
channels, respectively.
For comparison, we include noninteracting energies for the 
finite-volume $3\pi$, $\rho\pi$, and $\omega$/$h_1$ states.
The actual spectral lines show significant shifts from the noninteracting levels,
as well as the usual pattern of avoided level crossings.
For our choice of parameters of the $\omega$ and $h_1$, the 
avoided level crossings are quite narrow. 
This could be a result of the resonance being narrow, or a volume suppression of the gap in the avoided level crossings.
Moreover, the finite-volume state related to the toy $h_1$ is significantly shifted with respect to the position of the pole in $\Kdf$. Further investigation is needed to identify to what extent this is driven by a difference between the pole in $\Kdf$ and the real part of the $\mathcal M_3$ pole position (i.e.~between the parameter $M_R$ and the physical three-particle resonance mass) and to what extent it is an effect of the finite volume. The fact that there are fewer levels in the $T_1^-$ plot can be understood in terms of the
antisymmetry of the momentum wavefunctions---as discussed in appendix~\ref{app:B}.
Indeed, one can understand precisely the counting of levels in both plots.

\section{Conclusion}
 \label{sec:conc}

This work constitutes the first extension of the finite-volume three-particle formalism to include nonidentical particles. We have focused on the description of a generic three-pion system in QCD with exact isospin symmetry. The main difference with the original quantization condition
of refs.~\cite{\LtoK, \KtoM} is that there are different subchannels for pairwise interactions ($I_{\pi\pi}=0,1,2$) that must be taken into account. The new three-particle quantization condition, and the infinite-volume three-particle integral equations, look formally identical to those for identical particles, but live in an enlarged matrix space with additional flavor indices.
The central point of this work is to give the explicit forms of all building blocks in this enlarged space, and to outline a strategy for extracting three-pion scattering amplitudes, in both weakly-interacting and resonant systems, for all possible quantum numbers.  

  As described in section~\ref{sec:derivation}, to carry out the derivation
it is convenient to first generalize the quantization condition 
using the basis with definite individual pion flavors. 
The final result is then block-diagonalized by performing a standard change of basis 
in flavor space, with the resulting blocks labeled by the three-pion isospin
$I_{\pi \pi \pi}=0-3$, and the elements within each block labeled by the allowed values of incoming and outgoing two-pion isospin $I_{\pi \pi}$. In this way, the three-pion quantization condition turns into a set of four independent expressions, to be applied separately to finite-volume energies with the corresponding quantum numbers. 
The $I_{\pi\pi\pi}=3$ quantization condition is the same as 
that for three identical (pseudo-)scalars derived in refs.~\cite{\LtoK,\KtoM},
while those for $I_{\pi\pi\pi}=0,1,2$ are new.
The implementation of the new quantization conditions is of similar complexity to
the $I_{\pi\pi\pi}=3$ case, where there have been extensive previous
studies~\cite{\BHSnum,\dwave,\largera,Blanton:2019vdk}. 
They do, however, exhibit some new features, such as the presence of odd partial waves and different relative signs between the finite-volume objects involved. 

In section~\ref{sec:param}, we also have addressed the parametrization of $\Kdf$ in a general isospin channel, which is a crucial point for the extraction of three-particle scattering amplitudes from lattice QCD. First, we have extended the threshold expansion of $\Kdf$ to
all values of $I_{\pi\pi\pi}$. 
This is a series expansion about threshold based on symmetry properties of $\Kdf$: 
Lorentz invariance, CPT and particle exchange. We have worked out the first few terms for all isospin channels. In addition, we propose parametrizations of $\Kdf$ to describe  all three-particle resonances present in the $I_{\pi\pi\pi}=0$ and $1$ channels. These generate an additional state in the spectrum, which decouples in the limit of zero coupling.

Given these results, 
all ingredients are now available for lattice studies of resonances with three-particle decay channels, such as the $\omega(782)$ and the $h_1(1170)$. These two $I_{\pi\pi\pi}=0$
resonances are particularly good candidates for a first study, 
as they lie below the $5M_\pi$ threshold for slightly heavier-than-physical pions. 
In section~\ref{sec:toy} we
 use the new quantization condition to determine the finite-volume spectrum 
for these two channels in a toy model
motivated by the experimentally observed hierarchies of masses and widths.
Our exploration suggests that, in practice, moving frames will be needed to gain insight in the nature of the resonances, especially in the case of the $\omega(782)$. 
We stress, however, not yet established how the parameters of $\Kdf$ relate to the physical  masses and widths of the resonances and thus more investigation is needed. 

Going forward, the next steps fall into three basic categories.
First, it would be instructive to study various limiting cases, 
in order to provide useful crosschecks and gain insights into the structure of the new quantization conditions. One concrete example would be to study the $I_{\pi \pi \pi} = 2$ expressions, continued to parameters such that the $\rho$ resonance becomes a stable particle. In this case one can restrict to the energy regime $M_\rho + M_\pi < \sqrt{s} < 3 M_\pi $, and the result should coincide with the two-particle, finite-volume formalism for vector-scalar scattering \cite{\RaulSpin}, already used to analyze finite-volume energies in 
 ref.~\cite{Woss:2018irj}.
Second, it is necessary to further generalize the formalism, so as to
  describe all possible systems of two- and three-particles with generic interactions, quantum numbers, and degrees of freedom. Specific cases, ranked from most straightforward to most difficult, include three pseudoscalar particles in $SU(N_f)$-symmetric QCD, three-nucleon systems (i.e.~the inclusion of spin) and, by far the most challenging, $N \pi \to N \pi \pi$ transitions in the Roper channel (requiring spin, $2 \to 3$ transitions, and non-identical and non-degenerate particles).
Finally, and most importantly, the application of this formalism to three-pion resonances using lattice QCD is now well within reach. This will represent the achievement of a long-standing milestone on the way towards unlocking the exotic excitations of the strong force. 
 
\acknowledgments

We thank Raúl Briceño for helpful comments and many fruitful discussions. We also thank Mattia Bruno, Christopher Thomas, David Wilson, and Antoni Woss 
for useful discussions.
FRL acknowledges the support provided by the European projects H2020-MSCA-ITN-2015/674896-ELUSIVES, H2020-MSCA-RISE-2015/690575-InvisiblesPlus, the Spanish project FPA2017-85985-P, and the Generalitat Valenciana grant PROMETEO/2019/083. The work of FRL also received funding 
from the European Union Horizon 2020 research and innovation program 
under the Marie Sk{\l}odowska-Curie grant agreement No. 713673 and ``La Caixa'' Foundation  (ID 100010434, LCF/BQ/IN17/11620044). The work of SRS is supported in part by the United States Department of Energy (USDOE)
grant No. DE-SC0011637, in part by the DOE.

\appendix

\section{Further details of the derivation}\label{app:derivation}

In this appendix we provide more details of the derivation of the 
result for the generalized finite-volume correlator,
 eq.~(\ref{eq:CLdecomgenIso}).
As noted in the main text, most of the steps in the original derivation of ref.~\cite{\LtoK}
go through, with the only change being the need to generalize the
core quantities $F$, $G$ and $\mathcal K_2$  in the presence of flavor
[using the definitions of eqs.~(\ref{eq:Fgendef}), (\ref{eq:K2def1}) and (\ref{eq:Ggendef})].
In other words, almost all of the equations in ref.~\cite{\LtoK} can be taken over unchanged
as long as one adds flavor indices and uses the new definitions.
There is, however, one step in the derivation that needs further generalization, as we now explain.

The most challenging part of the derivation of ref.~\cite{\LtoK} is to show that
$\mathcal K_{\df,3}$ has the appropriate symmetry. 
Since the symmetrization procedure must be generalized here,
as described in section~\ref{sec:genKtoM},
a natural question is whether the derivation of the quantization condition in
the presence of flavor leads to the appropriately symmetrized version of
$\mathcal K_{\df,3}$, denoted $\textbf K_{\df,3}$.
A second aim of this appendix is to explain why this is indeed the case.

For the sake of brevity, we assume that the reader has a copy of ref.~\cite{\LtoK} in front
of them and we do not repeat equations from that work. We refer to equations from 
ref.~\cite{\LtoK} as (HS1), (HS2), etc.%
\footnote{Some aspects of the derivation of ref.~\cite{\LtoK} were streamlined in ref.~\cite{\BHSK},
which generalized the derivation to include a K-matrix pole.
We do not refer to the latter work, however, since the notation therein is quite involved,
as there is an additional channel needed for the  K-matrix pole, which is not relevant here.
In any case, our aim  is not to repeat the derivation, but rather to describe how
it can be taken over wholesale. The more pedestrian approach of ref.~\cite{\LtoK} is
adequate for this purpose.}

The first place in ref.~\cite{\LtoK} where the discussion does not generalize in a simple way
is in the discussion between (HS140) and (HS146). This concerns the introduction
of quantities with a superscript $(s)$, e.g. $A'^{(1,s)}$ in (HS140).
These are to be contrasted with quantities having a superscript $(u)$,
such as $\textbf D^{(u,u)}$ in eq.~(\ref{eq:Duudef}).
For the latter quantities, the matrix index $k$ corresponds to the spectator momentum,
while for quantities with superscript $(s)$, $k$ labels the momentum of one
of the nonspectator pair.
To be more precise, in the symmetrization described in eq.~(\ref{eq:symmX}),
the choice $\mathcal P_3=\{ \boldsymbol k, \boldsymbol a \}$ from
eq.~(\ref{eq:P3choices}) corresponds to a $(u)$ quantity,
while that with $\mathcal P_3= \{ \boldsymbol a, \boldsymbol b \}$ corresponds to
an $(s)$ quantity.
The third choice, $\mathcal P_3= \{ \boldsymbol b, \boldsymbol k \} $, leads
to quantities denoted by $(\tilde s)$ in ref.~\cite{\LtoK}.
These three choices are illustrated in fig.~13b of ref.~\cite{\LtoK}.

We choose our flavor generalizations of $A'^{(1,u)}$ and $A'^{(1,s)}$ such that
(HS140) maintains its form, becoming\footnote{The numerical superscripts indicate the order in an expansion in numbers of
``switch states". The details, described in ref.~\cite{\LtoK},
 are not important for the present discussion.}
\begin{equation}
\textbf A_L'^{(2,u)} = \textbf A'^{(2,u)}
+ 2 \textbf A'^{(1,s)}\; \textbf F\; \textbf K_2
\,.
\end{equation}
With this choice, the coupling of flavor and momentum labels is automatically maintained.
For example, in the product $[\textbf A'^{(1,s)}]_{ij} [\textbf F]_{jl}$, if $j=2$, 
corresponding to $\widetilde \pi_0(a) \widetilde \pi_-(b) \widetilde \pi_+(k)$,
then the spectator attaching to the endcap has momentum $\bm a$
and is a neutral pion.
Thus no additional permutation matrix is needed.
With this choice the symmetrized endcap is simply given by\footnote{A potentially confusing issue is why there are only three terms
in the symmetrization sums, as opposed to six, the number of permutations of
the three momenta. In other words, why is it sufficient to have one contribution
from each of the different choices of spectator momenta, while the order of
the nonspectator momenta is irrelevant? In the case of three neutral pions
($j=4$) this is because the amplitude is symmetric under exchange of the
nonspectator pair. For other choices of the flavor index $j$,
the two pions in the nonspectator pair have different charges, and their
order  has no meaning in the context of a Feynman diagram, as long as
we associate a given momentum label always with a given flavor, as is the case here.
}
\begin{equation}
\textbf A' = \textbf A'^{(u)}+ \textbf A'^{(s)} + \textbf A'^{(\tilde s)}\,.
\end{equation}
Here we are considering endcaps obtained by summing to all orders in perturbation 
theory, and thus there is no numerical superscript. In this notation the complete
endcap appearing in the main text  is $\textbf A'_3 = \bm \sigma+ \textbf A'$
[see, e.g., eq.~(\ref{eq:CLdecomgenIso})].

Now we come to the core issue of this appendix. The derivation of
ref.~\cite{\LtoK} produces, in many places,\footnote{Strictly speaking, these quantities should have a common numerical superscript
indicating the order in the expansion in switch states, but this plays no role in the
present derivation, so we drop it for the sake of brevity.}
the combination $\textbf A'^{(u)}+ 2 \textbf A'^{(s)}$, 
rather than the desired symmetric quantity $\textbf A'$.
The key results needed to allow symmetrization generalize here to
\begin{align}
\left\{ \textbf A'^{(u)}  + 2 \textbf A'^{(s)}\right\} \textbf F \textbf A^{(u)} 
= \textbf A' \textbf F \textbf A^{(u)} \ \ &\Leftrightarrow \ \
\textbf A'^{(s)} \textbf F \textbf A^{(u)} 
= \textbf A'^{(\tilde s)} \textbf F \textbf A^{(u)}\,,
\label{eq:key1}
\\
\left\{ \textbf A'^{(u)}  + 2 \textbf A'^{(s)}\right\} \textbf F\textbf  A 
= \textbf A' \textbf F \textbf A
\ \ &\Leftrightarrow \ \
\textbf A'^{(s)}\textbf F \textbf A 
= \textbf A'^{(\tilde s)} \textbf F \textbf A\,.
\label{eq:key2}
\end{align}
In each line, the two forms are algebraically equivalent, and we will demonstrate the
second forms. The argument for (the ungeneralized form of) these results given
in ref.~\cite{\LtoK} applies only for identical particles. Here we give the generalization.

In both eqs.~(\ref{eq:key1}) and (\ref{eq:key2}) there is an implicit sum over the
flavor indices. The matrix $\textbf F$ is diagonal in flavor [see eq.~(\ref{eq:Fgendef})],
so the right-hand flavor index of  the left endcap and the left-hand flavor index of
the right endcap are the same, and we call this common index $j$.
In the all-neutral case, $j=4$, the arguments of ref.~\cite{\LtoK} hold and demonstrate
the equalities.
For other choices, the equalities hold only after summing over the pairs of values of
$j$ that are related by interchanging the first two pions, i.e.
$j=\{1,2\}$, $\{3,5\}$ and $\{6,7\}$.
For each of these pairs, we denote the two values as $j_1$ and $j_2$.
The new results that are needed are
\begin{align}
(\textbf A^{(u)}_{j_1 i})_{k\ell m} &= (-1)^\ell (\textbf A^{(u)}_{j_2 i})_{k\ell m}\,,
\label{eq:key3}
\\
(\textbf A'^{(s)}_{ij_1})_{k\ell' m'} &= (-1)^{\ell'} (\textbf A'^{(\tilde s)}_{i j_2})_{k\ell' m'}\,,
\label{eq:key4}
\end{align}
as well as a result derived in ref.~\cite{\LtoK},
\begin{align}
(-1)^{\ell'} F_{k'\ell' m'; k\ell m} (-1)^{\ell} &= F_{k'\ell' m';k\ell m}\,,
\label{eq:key5}
\end{align}
using which it is simple to derive eqs.~(\ref{eq:key1}) and (\ref{eq:key2}).

We discuss eqs.~(\ref{eq:key3})-(\ref{eq:key5}) in turn. Note that in
the first two of these equations, the flavor label $i$ plays no role.
What eq.~(\ref{eq:key3}) states is that, if we interchange the momenta $\bm a$
and $\bm b$, and interchange the flavors $j_1$ and $j_2$,
then we obtain the same amplitude. The factor of $(-1)^\ell$
arises because we are decomposing into spherical harmonics with respect to
$\widehat{ \boldsymbol a}^\star$ on the left-hand side and 
$\widehat{ \boldsymbol b}^\star$ on the right-hand side,
corresponding to a parity flip in the CMF of the nonspectator pair.
The same explanation holds for eq.~(\ref{eq:key4}), except here there is the
additional feature that interchanging $\bm a$ and $\bm b$ also 
interchanges $(s)$ and $(\tilde s)$.
Finally, eq.~(\ref{eq:key5}) encodes the statement that $F$ vanishes 
(up to exponentially suppressed corrections) unless $\ell+\ell'$ is even.

The remainder of the derivation in ref.~\cite{\LtoK} generalizes step
by step in the presence of flavor. Each equation holds when the original
quantities are replaced by their flavored (bold faced) generalizations (taking into
account the factors of $i$ and $2\omega L^3$ absorbed into the bold faced
definitions). No new results are needed. For example, the key result given in
(HS196)-(HS198), which is also crucial to allow symmetrization, carries over
verbatim for each choice of flavor indices. Also, the complicated steps
in (HS213)-(HS239), which result in a symmetrized $\mathcal K_{\df,3}$,
carry over and (using the key results given above) lead to a
$\textbf K_{\df, 3}$ with exactly the generalized symmetry properties
described in section~\ref{sec:genKtoM}.
Finally we note that the inclusion of the generalized
three-particle Bethe-Salpeter kernel, $\textbf B_3$,
also follows the same steps as in section~IV.E of ref.~\cite{\LtoK},
because $\textbf B_3$ has the same symmetry properties as $\boldsymbol \sigma$,
namely those of $\textbf M_3$.

\section{Building blocks of the quantization condition} \label{app:QC}

This appendix provides a self-contained collection of all necessary definitions to implement the three-particle quantization condition.

First, we define  the cutoff function:
\begin{align}
    H(\boldsymbol{k}) &= J(z)\,, \qquad
    z = \frac{E_{2,k}^{\star2} - (1+\alpha_H)m^2}{(3-\alpha_H)m^2}\,, 
        \label{eq:H_k}
    \\
    J(z) &=
    \begin{cases}
        0, & z\leq 0 \,, \\
        \exp{\left( -\frac{1}{z}\exp{\left[ -\frac{1}{1-z} \right]} \right)}, & 0<z<1 \,, \\
        1, & 1\leq z \,,
    \end{cases}
\end{align}
where $E_{2,k}^{\star2} = (E-\omega_k)^2  - (\boldsymbol P- \boldsymbol k)^2$ and $\alpha_H\in [-1,3)$ a constant that sets the scheme for $\Kdf$ but does not affect the relation between finite-volume energies and the physical amplitude. We typically choose $\alpha_H=-1$, corresponding to the highest cutoff. 

For $G$ we use the relativistic form described in ref.~\cite{\BHSQC}, 
\begin{equation}
  {G}_{p\ell'm';k\ell m} (E, {\boldsymbol P},L) \equiv \frac{1}{L^3}   \frac{H(\boldsymbol{p}) H(\boldsymbol{k})}{b^2-m^2}
    \frac{4\pi \mc{Y}_{\ell'm'}(\boldsymbol{k}^\star) \mc{Y}^*_{\ell m}(\boldsymbol{p}^{\star}) }{q_{2,p}^{\star\ell'}\,q_{2,k}^{\star\ell} } \frac{1}{2\omega_k}\,, 
    \label{eq:defG}
\end{equation}
where $b= P-p - k$ is the momentum of the exchanged particle and $q_{2,k}^{\star2} = E_{2,k}^{\star2} /4 - m^2$ is the squared back-to-back momentum of the non-$b$ pair in its \CMF.
We have also used the two-particle \CMF~quantities $\boldsymbol{p}^{\star}$ and $\boldsymbol{k}^{\star}$, defined via
\begin{align}
\begin{split}
    &\boldsymbol{p}^{\star} = (\gamma_k-1)\left(\boldsymbol{p}\cdot(\widehat{\boldsymbol k \!\! - \!\! \boldsymbol P})\right)(\widehat{\boldsymbol k \!\! - \!\! \boldsymbol P}) + \omega_p\gamma_k \beta_k(\widehat{\boldsymbol k \!\! - \!\! \boldsymbol P}) + \boldsymbol{ p}, \\ & \quad \quad \quad \quad
    {\beta}_k = \frac{\vert \boldsymbol P-  \boldsymbol{k} \vert}{E-\omega_k}, \quad \gamma_k = (1-\beta_k^2)^{-1/2}
\,,
\end{split}
\label{eq:pstardef}
\end{align}
where $\hat {\boldsymbol x} = \boldsymbol x / \vert   \boldsymbol x \vert$. 
The definition for $\boldsymbol k^\star$ is given by 
exchanging $\boldsymbol p \leftrightarrow \boldsymbol k$ everywhere.
Finally, $\mathcal Y_{\ell m}(\boldsymbol k)$ are harmonic polynomials,
\begin{equation}
    \mathcal{Y}_{\ell m}(\boldsymbol{k}) \equiv \vert \boldsymbol{k} \vert^\ell Y_{\ell m}(\hat{\boldsymbol k})\,,
\end{equation}
where $Y_{\ell m}$ are the spherical harmonics. In practice, it is more convenient to use the {real} spherical harmonics, as discussed in ref.~\cite{\dwave}.

Next, 
\begin{equation}
  {F}_{k'\ell'm';k\ell m} (E, {\boldsymbol P},L) \equiv \delta_{k' k} F_{\ell' m', \ell m}(\boldsymbol k) \,,
\end{equation}
where $ F(\boldsymbol k)$ is a sum-integral difference that 
is proportional to the zeta functions that appear in the two-particle
quantization condition~\cite{Luscher:1986n2,Luscher:1991n1}. This object also depends on $(E, {\boldsymbol P},L)$ but we leave this implicit, focusing on the role of the spectator momentum.
$ F(\boldsymbol k)$ requires ultraviolet (UV) regularization, and can be written in various forms that are
equivalent up to exponentially-suppressed corrections. The original form, presented in ref.~\cite{\HSQCa}, uses a product of $H$ functions as  a UV regulator. 
Here, we give a different form that is simpler to evaluate numerically.
Following 
ref.~\cite{Kim:2005gf}, we write
\begin{align}
 {F}_{\ell'm';\ell m} (\boldsymbol{k})   &= 
    \frac{1}{16\pi^2 L   (E\!-\!\omega_k)} 
     \left[ \sum_{\boldsymbol{n}_a} \!- {\rm PV}\!\int d^3 {\boldsymbol n}_a \right]
    \frac{e^{\alpha(x^2-r^2)}}{x^2-r^2} 
    \frac{4\pi \mathcal{Y}_{\ell'm'}(\boldsymbol{r}) \mathcal{Y}^*_{\ell m}(\boldsymbol{r})}{x^{\ell'+\ell}}\,,
    \label{eq:Ft1}
\end{align}
where $\boldsymbol n_a= \boldsymbol a L/(2 \pi) $, $x=q_{2,k}^\star L/(2\pi)$, and  \begin{align}
    \boldsymbol{r}(\boldsymbol{n}_k,\boldsymbol{n}_a) &= \boldsymbol{n}_a + \boldsymbol{n}_{kP} \left[
    \frac{\boldsymbol{n}_a\cdot\boldsymbol{n}_{kP}}{n_{kP}^2}
    \left(\frac{1}{\gamma_k}-1\right) + \frac{1}{2\gamma_k}
    \right]\,,
    \label{eq:rdef}
\end{align}
with $\boldsymbol k-\boldsymbol P = \boldsymbol n_{kP} (2\pi/L)$, and $\gamma_k$ as in eq.~\eqref{eq:pstardef}.
The UV regularization is now provided by the exponential in the integrand with $\alpha > 0$. The $\alpha$ dependence is exponentially suppressed in $L$ but can become numerically significant if $\alpha$ is taken too large. We find that $\alpha \lesssim 0.5$ is usually sufficient. In this regularization, the integral can be performed analytically, as explained in appendix B of ref.~\cite{\dwave}.

Finally, we turn to $\cK_2$, which is a diagonal matrix:
\begin{align}
	&\left[\frac{1}{\mc{K}_2}\right]_{p\ell'm';k\ell m} = \delta_{pk}\delta_{\ell'\ell}
	\delta_{m'm} \frac{1}{\mc{K}_{2;k}^{(\ell)}}\,,
	\label{eq:K2mat} \\
&\frac{1}{\mc{K}_{2;k}^{(\ell)}} = \frac{1}{16\pi E_{2,k}^\star}
	\left\{  q_{2,k}^{\star}  \cot \delta_\ell(q_{2,k}^{\star}) + |q_{2,k}^\star| [1-H(\boldsymbol{k})] \right\}\,,
	\label{eq:K2ellwave}
\end{align}
where  $ \delta_\ell (q_{2,k}^{\star})$ is the two-particle phase-shift in the $\ell$th partial wave.

\section{Three-pion states} \label{app:A}

{We collect in this appendix some additional details concerning the basis we use
for the neutral three-pion states.}
The first two pions are combined into a state of definite isospin.
The $I_{\pi\pi}=2$, $1$ and $0$ states are denoted $(\pi\pi)_2^q$,  $\rho^q$,
and $\sigma$, respectively, with $q$ the charge.
The two-pion state is then combined with the remaining pion to create a state of total isospin $I_{\pi\pi\pi}$
(denoted by a subscript on the kets listed below).
{This leads to}
\begin{align}
\ket{(\pi\pi)_2 \pi}_3 =& \frac{1}{\sqrt{5}} \left(  \ket{(\pi\pi)^+_2 \pi_-} + \sqrt{3}  \ket{(\pi\pi)^0_2 \pi_0} + \ket{(\pi\pi)^-_2 \pi_+}  \right)  \,,
\\
\ket{(\pi\pi)_2 \pi}_2 =&  \frac{1}{\sqrt{2}} \left(  \ket{(\pi\pi)^+_2 \pi_-} - \ket{(\pi\pi)^-_2 \pi_+}  \right)   \,,
\\
\ket{\rho \pi}_2 =& \frac{1}{\sqrt{6}} \left(  \ket{\rho^+ \pi_-} + 2  \ket{\rho^0 \pi_0} + \ket{\rho^- \pi_+}  \right)  \,,
 \\
\ket{(\pi\pi)_2 \pi}_1 =&  \frac{1}{\sqrt{10}} \left( \sqrt{3} \ket{(\pi\pi)^+_2 \pi_-} -2 \ket{(\pi\pi)^0_2 \pi_0} + \sqrt{3} \ket{(\pi\pi)^-_2 \pi_+}  \right)   \,,
\\
\ket{\rho \pi}_1 =&  \frac{1}{\sqrt{2}} \left(  \ket{\rho^+ \pi_-} - \ket{\rho^- \pi_+}  \right)    \,,
\\
 \ket{\sigma \pi}_1 =& \ket{\sigma \pi_0}   \,,
\\
\ket{\rho \pi}_0 =&  \frac{1}{\sqrt{3}} \left(  \ket{\rho^+ \pi_-} -  \ket{\rho^0 \pi_0} + \ket{\rho^- \pi_+}  \right)\,.
\label{eq:I0form}
\end{align}
The right-hand sides can be further decomposed into the $\ket{\pi\pi\pi}$ basis
used in the main text, resulting in eqs.~(\ref{eq:Cdef}) and (\ref{eq:Cnumerical}).

We make extensive use of the irreducible representations (irreps) of the symmetry group $S_3$, 
which  describes permutations of three objects. It has 6 elements, divided into three
conjugacy classes as
\begin{equation}
\{(1)\}, \ \{(12), \ (23),\ (13)\} \ \ {\rm and}\ \ \{  (231), \ (312)\}\,.
\end{equation}
The three irreps are as follows.
\begin{enumerate}
\item The trivial representation, with all elements being the identity. States
transforming according this irrep are denoted $\ket{\chi_s}$.
\item The sign or alternating representation:
\begin{align}
\begin{split}
(1), \ (231), \ (312) &\to + 1, \\  (12), \ (23), \ (13) &\to -1. \label{eq:perm1d}
\end{split}
\end{align}
States transforming according to this irrep are denoted $\ket{\chi_a}$.
\item The standard representation, which is two dimensional. A convenient choice
of basis vectors, denoted $\ket{\chi_1}$ and $\ket{\chi_2}$, leads to:
\begin{align}
\begin{split}
&(1) \to \begin{pmatrix}
1 & 0 \\ 0 & 1
\end{pmatrix}, \ (12) \to \begin{pmatrix}
1 & 0 \\ 0 & -1
\end{pmatrix},   \ (13) \to \frac{1}{2} \begin{pmatrix}
-1 & -\sqrt{3} \\ -\sqrt{3} & 1
\end{pmatrix}, \\ 
(23) \to \frac{1}{2}& \begin{pmatrix}
-1 & \sqrt{3} \\ \sqrt{3} & 1
\end{pmatrix}, \
(231) \to \frac{1}{2} \begin{pmatrix}
-1 & \sqrt{3} \\ -\sqrt{3} & -1
\end{pmatrix}, \
(312) \to \frac{1}{2} \begin{pmatrix}
-1 & -\sqrt{3} \\ \sqrt{3} & -1
\end{pmatrix}. \label{eq:perm2d}
\end{split}
\end{align}
\end{enumerate}

The three-pion states listed above can be classified according to their transformations
under permutations. The $I_{\pi\pi\pi}=3$ state transforms in the symmetric irrep, 
the $I_{\pi\pi\pi}=2$ states in the standard irrep,
the $I_{\pi\pi\pi}=1$ states in a direct sum of the symmetric and standard irreps,
and the $I_{\pi\pi\pi}=0$ state in the sign irrep.
The linear combinations that lie in the permutation-group irreps are
(with the subscript on the ket again denoting isospin)
\begin{align}
\ket{\chi_s}_3 &= \ket{(\pi\pi)_2 \pi}_3,  \\
\ket{\chi_1}_2 &=\ket{(\pi\pi)_2 \pi}_2,  
\label{eq:chi12} \\
\ket{\chi_2}_2 &=\ket{\rho \pi}_2,  
\label{eq:chi22} \\
\ket{\chi_1}_1 &=- \frac{\sqrt{5}}{3}\ket{(\pi\pi)_2 \pi}_1 + \frac{2}{3}\ket{\sigma \pi}_1, \\
&=  \frac{1}{\sqrt{12}} \Big( 2 \ket{\pi_+,\pi_-,\pi_0} + 
2 \ket{\pi_-,\pi_+,\pi_0  }  -  \ket{\pi_+,\pi_0,\pi_-  } 
\nonumber \\
&\qquad\qquad\qquad
- \ket{\pi_0,\pi_+,\pi_-  } - \ket{\pi_0,\pi_-,\pi_+ } - \ket{\pi_-,\pi_0,\pi_+  }     \Big),  
\label{eq:chi11} \\
\ket{\chi_2}_1 &=\ket{\rho \pi}_1,  
\label{eq:chi21} \\
\ket{\chi_s}_1 &= \frac{2}{3}\ket{(\pi\pi)_2 \pi}_1 + \frac{\sqrt{5}}{3}\ket{\sigma \pi}_1,  \\ 
&= \frac{1}{\sqrt{15}} \Big( 
\ket{\pi_+,\pi_-,\pi_0  }+\ket{\pi_0,\pi_+,\pi_-  }+\ket{\pi_-,\pi_0,\pi_+  }
+\ket{\pi_-,\pi_+,\pi_0  }\nonumber\\
&\qquad\qquad\qquad
 +\ket{\pi_0,\pi_-,\pi_+  }+\ket{\pi_+,\pi_0,\pi_-  }  - 3\ket{\pi_0,\pi_0,\pi_0} \Big),
\label{eq:chis1} \\
 \ket{\chi_a}_0 &= \ket{\rho \pi}_0 .
\end{align}

\section{Group-theoretic results} \label{app:B}

In this appendix we collect some group-theoretic 
results that are relevant for the practical implementation of
the quantization condition described in the main text. 
We restrict our considerations to the overall rest frame, i.e.~we set $\boldsymbol P=0$;
generalizations to moving frames are straightforward but tedious.

We begin by listing the irreps that are created and annihilated by 
operators with $(I_{\pi\pi\pi})_z=0$, having the form of three
noninteracting pions, each with a definite momentum.
Focusing on annihilation operators, we write 
\begin{equation}
\widetilde\pi_i(\boldsymbol a) \widetilde \pi_j (\boldsymbol b) \widetilde \pi_k (\boldsymbol c)\,,
\end{equation}
with $\widetilde\pi$ the Fourier transform of some choice of local pion operator.
The indices $i,j,k$ denote $(I_\pi)_z$, and the constraint that the total operator is neutral
restricts the choices of indices to seven options, as described in appendix~\ref{app:A}.
The momenta are $\boldsymbol a= 2 \pi \boldsymbol m_1/L$, $\boldsymbol b= 2 \pi \boldsymbol m_2/L$,
and $\boldsymbol c =-\boldsymbol a -\boldsymbol b = 2\pi \boldsymbol m_3/L$.
One then projects onto definite isospin using the results given in eqs.~\eqref{eq:Cdef} and \eqref{eq:Cnumerical} and appendix~\ref{app:A}.
Operators of this type are typically
used as part of the variational basis in lattice QCD calculations, and the energies of the
corresponding noninteracting states provide points of comparison for the spectrum of the
interacting theory (see, e.g., fig.~\ref{fig:I0spect}).

Each choice of $\boldsymbol m_1$ and $\boldsymbol m_2$ (which fixes $\boldsymbol m_3=-\boldsymbol m_1- \boldsymbol m_2$) 
is related to some number of other choices by cubic group transformations. We specify the resulting
orbit by giving the values of $m_1^2$, $m_2^2$ and $m_3^2$, which provide a unique
specification for the examples we consider (although not in general).
Each orbit decomposes into irreps of the cubic group, and these are listed
in table~\ref{tab:restirreps} for the operators coupling to the seven lowest-energy states
in the absence of interactions.
We recall that the irreps for the $48$-element cubic group (including parity) are
$A_1^\pm$, $A_2^\pm$, $E^\pm$, $T_1^\pm$ and $T_2^\pm$, with
dimensions of $\{1,1,2,3,3\}$, respectively.
The result from appendix~\ref{app:A} 
that the $I_{\pi\pi\pi}=1$ triplets decompose into a trivial singlet
and a standard irrep doublet under the permutation group $S_3$, leads to the result shown in
the table that the irreps for $I_{\pi\pi\pi}=1$ are simply the sum of those for $I_{\pi\pi\pi}=2$ and $I_{\pi\pi\pi}=3$.

\begin{table}
\begin{tabular}{c c  c   c  c  c  c}
orb. & $\left(\boldsymbol m_1^2,\boldsymbol m_2^2, \boldsymbol m_3^2\right)$ & 
 \ dim.\   & 
$I_{\pi\pi\pi}=0$ & $I_{\pi\pi\pi}=1$ 
& $I_{\pi\pi\pi}=2$ & $I_{\pi\pi\pi}=3$ \\ \hline\hline
1 &(0,0,0) & 2 & --- 
& $R_2^{(o)}+ R_3^{(o)}$ & --- & $A_1^-$ 
\\ \hline
2 &(1,1,0) & 21 & $T_1^+$  
& $R_2^{(o)}+ R_3^{(o)}$&  $A_1^-, E^-, T_1^+$ &  $A_1^-, E^-$ 
\\ \hline
3 &(2,2,0) & 42  & $T_1^+,T_2^+$ 
& $R_2^{(o)}+ R_3^{(o)}$& $A_1^-, E^-, T_2^-, T_1^+, T_2^+$  & $A_1^-, E^-, T_2^-$ 
\\ \hline
4 &(2,1,1) & 84  
& $R_0^{(4)}$ 
& $R_2^{(o)}+ R_3^{(o)}$
& $R_2^{(4)}$ 
& $R_3^{(4)}$ 
\vphantom{\big(}
\\ \hline
5 &(3,3,0) & 28 & $A_2^+,T_1^+$ 
& $R_2^{(o)}+ R_3^{(o)}$&$A_1^-,T_2^-,A_2^+,T_1^+$ & $A_1^-, T_2^- $
\\ \hline
6 &(4,1,1) & 24 & --- 
& $R_2^{(o)}+ R_3^{(o)} $& $ A_1^-, E^-, T_1^+$
& $A_1^-,  E^-, T_1^+$
\\ \hline
7 &(3,2,1) & 168 & $R_3^{(7)} $ 
& $R_2^{(o)}+ R_3^{(o)}$& 2 $R_3^{(7)}$ & $R_3^{(7)}$\vphantom{\bigg(}
\\ \hline
16 & (5,3,2) & 336 &  $R_3^{(16)} $ 
& $R_2^{(o)}+ R_3^{(o)}$& 2 $R_3^{(16)}$ & $R_3^{(16)}$\vphantom{\bigg(}
\end{tabular}
\caption{
Cubic-group irreps for the three-pion operators with $\boldsymbol P=0$ and total charge zero for 
isospin $I_{\pi\pi\pi}=0$, $2$ and $3$. These results include the intrinsic negative parity of the pions.
The operators are those with the lowest seven noninteracting energies for a cubic box with
$m L\approx 4$, together with the lowest-lying orbit having the maximal possible dimension.
The first column gives the orbit number, $o$, the second specifies the orbit, as described in the text,
while the third gives the dimension of the orbit.
The remaining columns list the irreps appearing in the orbit, $R_I^{(o)}$.
As indicated, results for $I_{\pi\pi\pi}=1$ are given by summing the irreps in the 
$I_{\pi\pi\pi}=2$ and $I_{\pi\pi\pi}=3$ columns.
Entries in the $I_{\pi\pi\pi}=3$ column agree with those in table 2 of ref.~\cite{\dwave} {(up to intrinsic parity, which is omitted in the earlier work)}.
The missing entries are $R_0^{(4)}=A_2^-, E^-, T_1^-, T_1^+, T_2^+$,
$R_2^{(4)}= A_1^-,A_2^-,2E^-,T_1^-,T_2^-,2T_1^+,2T_2^+$ ,
$R_3^{(4)}=A_1^-,E^-, T_2^-, T_1^+, T_2^+$,
$R_3^{(7)}=A_1^-, E^-, T_1^-, 2T_2^-, A_2^+, E^+, 2T_1^+, T_2^+$,
and
$R_3^{(16)}=A_1^-, A_2^-, 2E^-, 3T_1^-, 3T_2^-, A_1^+,A_2^+, 2E^+, 3T_1^+, 3T_2^+$.
\label{tab:restirreps} }
\end{table}

We stress that it is always possible to choose particular linear combinations of operators
that pick out each of the irreps in a given orbit. This is very useful in practice as it restricts the
number of terms in $\textbf K_{\df,3}$ that contribute (see section~\ref{sec:threshold}),
and allows one to consider the resonances discussed in section~\ref{sec:resonance} one by one.
We note that certain irreps do not appear until quite high orbits, e.g. $A_2^-$ 
and $T_1^-$ do not appear until the fourth orbit, while $E^+$ and $A_2^+$ do not appear
until the seventh. This leaves only $A_1^+$, which does not appear until the sixteenth  orbit.
This is the lowest ``generic'' orbit, i.e. one in which all nontrivial cubic-group 
transformations have vanishing characters.

In order to interpret the interacting spectra in the presence of narrow two-particle resonances, it is also
useful to determine which irreps are present assuming that the resonance is a stable particle. In practice,
for the energy range of interest, the most important such resonance is the $\rho$, as shown
by the examples in fig.~\ref{fig:I0spect}.
Thus we have determined the irreps created by $\rho \pi$ operators, treating the $\rho$ as
a stable particle with $J^P=1^-$.
There are three isospin combinations with total $(I_{\rho\pi})_z=0$, and these decompose into total
isospin $I_{\rho\pi}=0$, $1$ and $2$. Since the $\rho$ and $\pi$ are different particles, the cubic-group
irreps that appear are the same for all choices of isospin, and the results
for the lowest few momentum orbits are given in table~\ref{tab:rhopi}.
The multiplicities of the $T_1^-$ irrep agree with the results from table~3 of
ref.~\cite{Woss:2018irj}.

We can use the results in tables~\ref{tab:restirreps} and \ref{tab:rhopi}
to understand the level-counting in fig.~\ref{fig:I0spect}, which
shows the spectra for $I_{\pi\pi\pi}=0$ and irreps (a) $T_1^-$ and (b) $T_1^+$.
The energies of the second to the sixth noninteracting $3\pi$ orbits are shown in both panels
(the first orbit, having $E/m=3$, lies below the plotted range),
as well as the first three noninteracting $\rho\pi$ levels.

For $T_1^-$ (the $\omega$ channel), we see
from table~\ref{tab:restirreps} that, for the energy range shown in the figure,
only the fourth orbit contains this irrep.
From table~\ref{tab:rhopi}, we see that the second and third $\rho\pi$ orbits contain the $T_1^-$,
but not the first. In all but one case, there is only a single $T_1^-$ irrep present, the exception
being the third $\rho\pi$ orbit, which contains two such irreps.
These results are consistent with the interacting energies plotted in fig.~\ref{fig:I0spect}(a), which
can be interpreted, for $mL \lesssim 6$, as roughly corresponding to the $\omega$ resonance,
second $\rho\pi$ orbit, fourth $3\pi$ orbit, and a pair of $\rho\pi$ third orbits.

\begin{table}[tb]
\centering
\begin{tabular}{c c c c c}
orb. & $m_\rho^2$ & $m_\pi^2$ & dim. &\text{irreps} \\
 \hline
1 & 0 & 0 &3  &  $ T_1^+$
\\
2 & 1 & 1 & 18 &  $A_1^-,  E^-,  T_1^-, T_2^-, 2T_1^+, T_2^+$
\\
3 &2 & 2 &  36 & $ A_1^-, A_2^-, 2E^-, 2T_1^-, 2T_2^-, A_2^+, E^+, 3T_1^+, 2T_2^+$
\\
4 & 3 & 3 & 24 & $ A_1^-, E^-, T_1^-, 2T_2^-, A_2^+,E^+, 2T_1^+, T_2^+$
\end{tabular}
\caption{Cubic-group irreps contained in $\rho\pi$ states. 
The intrinsic negative parity of the pion and the rho are included.
Orbits are numbered, and specified by the squares of the momenta, with  $\boldsymbol p_\rho = 2\pi \boldsymbol m_\rho/L$
and  $\boldsymbol p_\pi=2\pi \boldsymbol m_\pi/L$. 
The irreps shown are present for each the three allowed isospins,
$I_{\rho\pi}=0$, $1$, and $2$. The dimensions of the orbits apply separately for each choice of isospin.
\label{tab:rhopi}}
\end{table}

The results for the $T_1^+$ irrep, displayed in fig.~\ref{fig:I0spect}(b), can be interpreted in
a similar manner. All the $3\pi$ and $\rho\pi$ orbits shown in the figure contain this irrep,
with unit multiplicities except for the second and third $\rho\pi$ orbits, which
have multiplicities $2$ and $3$, respectively.
This counting, together with the $h_1$ state, matches that seen in the figure.

\bibliographystyle{JHEP}      
\bibliography{ref.bib}

\end{document}